%% file: preprint.tex
\documentclass[12pt]{smo-thesis}
\usepackage{epsf,amssymb,graphics}

\newsavebox{\dash}
\author{Stephen M.\ Ouellette}
\title{SU(3) Chiral Symmetry in\\Non-Relativistic Field Theory}
\address{Lauritsen Laboratory\\
    California Institute of Technology\\
    Pasadena, CA 91125}
\printdate{(January 2, 2001)}

\abstract{%
Applications imposing SU(3)~chiral symmetry on non-relativistic field
theories are considered\@.
The first example is a calculation of the self-energy shifts of the
spin-$\frac{3}{2}$ decuplet baryons in nuclear matter, from the chiral
effective Lagrangian coupling octet and decuplet baryon fields\@.
Special attention is paid to the self-energy of the $\Delta$~baryon near the
saturation density of nuclear matter\@.
We find contributions to the mass shifts from contact terms in the effective
Lagrangian with coefficients of unknown value\@.
As a second application, we formulate an effective field theory with manifest
SU(2)~chiral symmetry for the interactions of $K$ and $\eta$ mesons with
pions at low energy\@.
SU(3)~chiral symmetry is imposed on the effective field theory by a matching
calculation onto three-flavor chiral perturbation theory\@.
The effective Lagrangian for the $\pi{}K$ and $\pi\eta$ sectors is worked out
to order~$Q^4$; the effective Lagrangian for the $KK$ sector is worked out to
order~$Q^2$ with contact interactions to order~$Q^4$\@.
As an application of the method, we calculate the $KK$ \mbox{$s$-wave}
scattering phase shift at leading order and compare with chiral perturbation
theory\@.
We conclude with a discussion of the limitations of the approach and propose
new directions for work where the matching calculation may be useful\@.
}

\begin{document}
\input{include}
\begin{frontmatter}
\input{ackno}
\tableofcontents
\end{frontmatter}

\input{intro}                       
\input{theory}                      
\input{10plet}                      
\input{heavyK}                      

\appendix
\input{codes}                       
\input{recur}                       
\input{biblio}                      
\end{document}

%% file: include.tex
\def\L#1#2{\ensuremath{{\mathcal{L}_{#1}^{#2}}}}
\def\Gchi{\ensuremath{{G_\chi}}}
\def\GchiP{\ensuremath{{G_\chi^{\:\prime}}}}
\def\cpt{\ensuremath{{\chi{\rm PT}}}}
\def\su#1{\ensuremath{{{\rm SU}({#1})}}}
\def\sub#1{\ensuremath{{\mathstrut_{\rm{#1}}}}}

\def\ss#1#2{\ensuremath{{\mathstrut%
    _{{#1}\mathstrut}^{{#2}\mathstrut}}}}
\def\tfrac#1#2{\ensuremath{{\textstyle\frac{#1}{#2}}}}
\def\tr#1{\ensuremath{{{\rm Tr}[{#1}]}}}
\def\xform#1{\ensuremath{\mathrel{\stackrel{#1}{\longrightarrow}}}}
\def\sss{\scriptscriptstyle}
\def\mathsp{\mbox{\ \ \ \ \ }}
\def\ms{\ensuremath{\mathstrut}}
\def\MS{\ensuremath{\vphantom{\ss{}{}}}}
\def\vp#1{\ensuremath{\vphantom{#1}}}
\def\vpf#1#2{\ensuremath{\vphantom{%
    \frac{\displaystyle{#1}}{\displaystyle{#2}}}}}
\def\vc#1{\ensuremath{\vcenter{\hbox{$\displaystyle{#1}$}}}}

\def\QCD{{\rm QCD}}
\def\Dslash{{\setbox0=\hbox{$D$}\rlap{\hbox{\kern.24\wd0$/$}}\box0}}
\def\Mq{\ensuremath{{\mathcal{M}_q}}}
\def\Lchi{\ensuremath{{\Lambda_\chi}}}
\def\MSbar{\ensuremath{\overline{\rm MS}}}
\def\ket#1{\ensuremath{{\,|{#1}\rangle}}}
\def\bra#1{\ensuremath{{\langle{#1}|\,}}}

\def\lpartial{{\setbox0=\hbox{$\partial$}\wd0=.80\wd0\box0}}
\def\dmu{\ensuremath{{\lpartial\ss{\mu}{}}}}
\def\dum{\ensuremath{{\partial\ss{}{\mu}}}}
\def\dnu{\ensuremath{{\lpartial\ss{\nu}{}}}}
\def\dun{\ensuremath{{\partial\ss{}{\nu}}}}

\def\D{{\setbox0=\hbox{$D$}\wd0=.80\wd0\box0}}
\def\V{{\setbox0=\hbox{$V$}\wd0=.65\wd0\box0}}
\def\Dmu{\ensuremath{{\D\ss{\mu}{}}}}
\def\Vmu{\ensuremath{{\V\ss{\mu}{}}}}
\def\Amu{\ensuremath{{A\ss{\mu}{}}}}
\def\spin#1{\mbox{spin-\tfrac{#1}{2}}}
\def\vslash{{\setbox0=\hbox{$v$}\rlap{\hbox{\kern.05\wd0$/$}}\box0}}

\def\pFp{\ensuremath{{p_F^{\sss(p)}}}}
\def\pFn{\ensuremath{{p_F^{\sss(n)}}}}
\def\C{\ensuremath{\mathcal{C}}}
\def\tmu{\ensuremath{\tilde{\mu}}}
\def\pse#1{\ensuremath{{\Sigma_{\rm{#1}}}}}
\def\E#1#2{\ensuremath{{\mathcal{E}_{\rm{#1}}^{\sss{#2}}}}}
\def\dE#1#2{\ensuremath{{\delta E_{\rm{#1}}^{\sss{#2}}}}}
\def\rate#1#2{\ensuremath{{\Gamma_{\rm{#1}}^{\sss{#2}}}}}
\def\g#1{\ensuremath{{g_{#1}}}}
\def\G#1{\ensuremath{{\mathcal{G}_{#1}}}}
\def\vbar#1{\ensuremath{\setbox0=\hbox{$|$}%
    \rlap{\raise#1\ht0\copy0}\raise-#1\ht0\box0}}

\def\v{{\setbox0=\hbox{$\mathcal{V}$}\wd0=.65\wd0\box0}}
\def\vv{\ensuremath{\mathcal{V}}}
\def\Mbar{\ensuremath{\bar{M}}}
\def\KKKK#1#2#3#4{\ensuremath{{({#1}\:{#2}|{#3}\:{#4})}}}
\def\KKOKK#1#2#3#4#5{\ensuremath{{({#1}\:{#2}|{#3}|{#4}\:{#5})}}}
\def\coeff#1#2#3{\ensuremath{{{#1}_{#3}^{\sss\pi{#2}}}}}
\def\A#1#2{\ensuremath{{\mathcal{A}_{#1}^{#2}}}}

\def\redefname{\ensuremath{\phi\rightarrow{}\mathcal{F}[\tilde{\phi}]}}

\def\piC{\ensuremath{{\pi^{\sss\pm}}}}
\def\piP{\ensuremath{{\pi^{\sss+}}}}
\def\piM{\ensuremath{{\pi^{\sss-}}}}
\def\piZ{\ensuremath{{\pi^{\sss0}}}}
\def\kC{\ensuremath{{K^{\sss\pm}}}}
\def\kP{\ensuremath{{K^{\sss+}}}}
\def\kM{\ensuremath{{K^{\sss-}}}}
\def\kZ{\ensuremath{{K^{\sss0}}}}
\def\kbar{\ensuremath{{\mathchoice%
    {{\setbox0=\hbox{$\displaystyle K$}\rlap{\hbox{\kern.2\wd0%
        \rule[1.15\ht0]{.75\wd0}{.05\ht0}}}\box0}}%
    {{\setbox0=\hbox{$\textstyle K$}\rlap{\hbox{\kern.2\wd0%
        \rule[1.15\ht0]{.75\wd0}{.05\ht0}}}\box0}}%
    {{\setbox0=\hbox{$\scriptstyle K$}\rlap{\hbox{\kern.2\wd0%
        \rule[1.2\ht0]{.75\wd0}{.05\ht0}}}\box0}}%
    {{\setbox0=\hbox{$\sss K$}\rlap{\hbox{\kern.2\wd0%
        \rule[1.25\ht0]{.75\wd0}{.1\ht0}}}\box0}}%
    }}}
\def\kbarZ{\ensuremath{{\kbar^{\sss0}}}}
\def\eZ{\ensuremath{\eta}}

\def\sigP{\ensuremath{{\Sigma^{\sss+}}}}
\def\sigZ{\ensuremath{{\Sigma^{\sss0}}}}
\def\sigM{\ensuremath{{\Sigma^{\sss-}}}}
\def\xiZ{\ensuremath{{\Xi^{\sss0}}}}
\def\xiM{\ensuremath{{\Xi^{\sss-}}}}
\def\lmdZ{\ensuremath{\Lambda}}

\def\dltPP{\ensuremath{{\Delta^{\sss++}}}}
\def\dltP{\ensuremath{{\Delta^{\sss+}}}}
\def\dltZ{\ensuremath{{\Delta^{\sss0}}}}
\def\dltM{\ensuremath{{\Delta^{\sss-}}}}
\def\sigSP{\ensuremath{{\Sigma^{*{\sss+}}}}}
\def\sigSZ{\ensuremath{{\Sigma^{*{\sss0}}}}}
\def\sigSM{\ensuremath{{\Sigma^{*{\sss-}}}}}
\def\xiSZ{\ensuremath{{\Xi^{*{\sss0}}}}}
\def\xiSM{\ensuremath{{\Xi^{*{\sss-}}}}}
\def\omgM{\ensuremath{{\Omega^{\sss-}}}}

%% file: ackno.tex
\extrachapter{Acknowledgements}%
I recognize three groups of people to whom I owe a particular debt of
gratitude for the ability to finish the work of this thesis\@.
First and foremost are my partner-in-life \mbox{Heather Frase}, my parents,
and my siblings\@.
Their unconditional love, support, and patience have been the glue which
holds my world together\@.
Also I want to acknowledge my mentors \mbox{Mark Wise}, \mbox{Ryoichi
Seki}, and \mbox{Ubirajara van Kolck}, and my former officemate
\mbox{Iain Stewart} for the encouragement and valuable insights they
offered during our discussions\@.
To those four people I owe everything that I know about the \emph{practice}
of scientific inquiry\@.
Finally I want to thank the people of the Caltech Theoretical High Energy
Physics group and a host of other characters (\mbox{Adam Leibovich},
\mbox{Erik Daniel}, \mbox{Torrey Lyons}, \ldots) for making my experience at
Caltech enjoyable as well as educational\@.

\clearpage

%% file: intro.tex
\chapter[Introduction]{\\Introduction}%
\label{ch:intro}%
\marginal{[ch:intro]}%
%
I once heard a wise man say ``In the history of scientific endeavor, no
problem has consumed as much of mankind's resources as the understanding of
nuclear forces,'' or something like that\@.
In any case, for all the considerable effort poured into solving the
mysteries of the strong interaction, a number of significant problems
remain\@.
The fundamental theory of the strong interactions, Quantum
Chromodynamics~(QCD), is solidly established as a pillar of the Standard
Model of particle physics and, to the extent that QCD is a renormalizable
gauge theory, is well understood in the perturbative regime\@.
As a theory with asymptotic freedom~\cite{DP:asymp,G-W:asymp}, QCD is
perturbative in the high-energy regime; for low energies the coupling
constant of the theory becomes large and perturbative treatments break
down\@.
Some features of non-perturbative~QCD which are still not fully understood
are quark structure of hadrons, dynamical symmetry breaking, and quark
confinement\@.

The special difficulties accompanying the non-perturbative regime require
special methods for working within the theory\@.
One direct approach is to formulate~QCD on a lattice of space-time points and
use numerical techniques to perform the functional integrals\@.
The lattice~QCD method~\cite{MC:book,DR:latQCD} has great potential for
shedding light on many of the unanswered questions of~QCD and has become an
industry unto itself\@.
A complementary approach is to focus on the long-distance physics, and base
the field theory description of the physics on the directly observed degrees
of freedom\@.
Approaches of the second type are generically called effective field theories
and rely on a two-part foundation\@.
I refer to the first important concept as the Weinberg
Hypothesis~\cite{SW:count}: the only content of quantum field theory (apart
from the choice of degrees of freedom) is analyticity, unitarity, cluster
decomposition, and the assumed \emph{symmetry principles}\@.
As a consequence, we can describe the strong interactions in terms of hadron
degrees of freedom provided we write down the most general Lagrangian
consistent with the symmetries of~QCD\@.
The second key concept is that we must identify some expansion parameter,
typically a small momentum or energy scale, which permits us to calculate to
any given order in the expansion with a finite amount of work\@.
The predictive power of an effective field theory arises from the combination
of the two underlying concepts; the symmetry restricts the parameters of the
theory to a meaningful set and the expansion parameter allows a systematic
framework in which we can include all contributions of a given order and
estimate the size of the contributions we have neglected\@.
For a more detailed discussion of effective field theory in general, see
references~\cite{HG:eft,AM:eft,DK:eft}\@.

In the case of~QCD, the hope is that by experimental determination of the
parameters of the effective Lagrangian, we can learn about the underlying
theory in the non-perturbative regime, possibly through lattice~QCD as an
intermediary\@.
An alternative possibility will also be considered in this work\@.
If the `fundamental' theory at higher-energy is known and calculable, then at
a momentum scale where the theories meet the parameters of the effective
theory can be determined by matching onto the fundamental theory\@.
This is sometimes done because certain calculations are more easily performed
in the effective theory, either because of additional approximate symmetry in
the low-energy limit or because a non-relativistic framework may be used\@.
For instance this sort of matching calculation has been performed and applied
with success in non-relativistic QED~(NRQED)~\cite{C-L:nrqed} and
non-relativistic QCD~(NRQCD)~\cite{B-B-L:nrqcd}\@.

In this thesis we consider two applications of effective field theory
to exploit the \mbox{$\su{3}\sub{L}\times\su{3}\sub{R}$} chiral
symmetry of~QCD\@.
In Chapter~\ref{ch:theory} we cover the theoretical framework upon which the
effective field theories will be built\@.
We discuss the symmetries of~QCD which constrain the effective
Lagrangian, the principles for constructing an effective Lagrangian
for the hadron degrees of freedom, and the power counting schemes that
apply to the sectors of the theory with only light fields, one heavy
field (static case), or more than one heavy field (non-relativistic
case)\@.
In Chapter~\ref{ch:decuplet} we present an effective field theory calculation
of the self-energy shift of \spin{3} decuplet baryons in nuclear matter\@.
With the exception of an expanded discussion of the \mbox{$\Delta$-baryon}
self-energy, the material in this chapter has already been
published~\cite{O-S:delta}\@.
In Chapter~\ref{ch:heavyK} we present new material on an effective
field theory for low-energy interactions of pions with kaons or an eta
meson as an alternative to the standard \su{3}~chiral perturbation
theory\@.
The objective is to achieve better convergence for very low energies by
treating the kaon in a non-relativistic framework\@.
To determine the parameters of the low-energy theory we perform a
matching calculation from the heavy kaon/eta theory onto \su{3}~chiral
perturbation theory in the spirit of NRQED or NRQCD\@.
Finally, we consider $KK$ scattering in the heavy kaon formalism and discuss
the utility of the matching calculation\@.


%% file: theory.tex
\chapter[Theoretical Background]{\\Theoretical Background}%
\label{ch:theory}%
\marginal{[ch:theory]}%
%
In this chapter chiral perturbation theory~(\cpt) is introduced as the
foundation for describing interactions involving the light pseudoscalar meson
octet\@.
As an effective field theory, \cpt~gets predictive power from the symmetries
of the underlying theory~(QCD) and a consistent scheme for counting powers of
`small' momenta\@.
Section~\ref{sec:symQCD} reviews the relevant symmetries of the
QCD~Lagrangian\@.
In sections~\ref{sec:cpt} and~\ref{sec:matter} we outline the formulation of
\su{3}~\cpt\ and its extension to include heavy fields\@.
The momentum power counting for diagrams with one or two heavy particles is
briefly discussed in section~\ref{sec:counting}\@.

\ifthesisdraft\clearpage\fi%
\section{Symmetries of QCD}%
\label{sec:symQCD}%
\marginal{[sec:symQCD]}%
%
An appropriate starting point is the Lagrangian of QCD,
\marginal{[eq:Lqcd]}%
\begin{equation}
\label{eq:Lqcd}
    \L{\QCD}{} 
    = -\tfrac{1}{4} G\ss{\mu\nu}{A} G\ss{}{A\mu\nu}
        + \overline{q}\ss{L}{} i\Dslash q\ss{L}{} 
        + \overline{q}\ss{R}{} i\Dslash q\ss{R}{}
        - \overline{q}\ss{R}{} \Mq q\ss{L}{} 
        - \overline{q}\ss{L}{} \Mq q\ss{R}{} 
        + \cdots 
\end{equation}
in which the ellipsis denotes gauge-fixing and ghost terms, renormalization
counterterms, and the $\theta$~term\@.
As written \L{\QCD}{}~is invariant under the Poincar\'e group,
\su{3}\sub{C}~gauge transformations, and charge conjugation~(C)\@.
In addition the coefficient of the $\theta$~term is known to be
small~\cite{PH:theta}, so we neglect it throughout; in this approximation
\L{\QCD}{}~is also invariant under parity~(P) and time reversal~(T)\@.

In the limit that~$N$ of the quark masses vanish, the chiral limit,
\L{\QCD}{}~acquires additional symmetries under independent U($N$)~rotations
of the left- and right-handed quark fields;
\marginal{[eq:q-xfm]}%
\begin{equation}
\label{eq:q-xfm}
    q\ss{L}{} \rightarrow L q\ss{L}{} 
, \mathsp
    q\ss{R}{} \rightarrow R q\ss{R}{} 
,
\end{equation}
where~$L$ and~$R$ are unitary matrices restricted to acting in the massless
sector\@.
The U(1)\sub{V}~subgroup of this symmetry corresponds to the conservation of
quark number in the massless flavors\@.
A second subgroup,~U(1)\sub{A}, is broken in the quantum theory by the
axial-vector anomaly\@.
The symmetries of~\L{\QCD}{} relevant for our purposes are the remaining
\mbox{$\su{N}\sub{L}\times\su{N}\sub{R}$} chiral symmetry and the discrete
symmetries~C, P, and~T\@.

The quark masses appearing in~\L{\QCD}{} are non-zero and explicitly break
the chiral symmetry\@.
Even in the limit of massless quarks the vacuum expectation value of the
quark bilinear
\marginal{[eq:qq-vev]}%
\begin{equation}
\label{eq:qq-vev}
    \bra{0} \overline{q}\ss{La}{} q\ss{Rb}{} \ket{0}
    = -\Delta \, \delta\ss{ab}{}
\end{equation}
spontaneously breaks the chiral symmetry
\mbox{$\su{N}\sub{L}\times\su{N}\sub{R}$} down to the vector
subgroup~\su{N}\sub{V}\@.
The scale \mbox{$\Lchi\sim\mbox{1~GeV}$}, associated with spontaneous
chiral symmetry breaking, determines the relative importance of the
quark masses in breaking the chiral symmetry~\cite{HG:book}\@.
Quark masses~$m_q$ which are much less than~\Lchi, specifically
\mbox{$m_{u,d}\lesssim\mbox{10~MeV}$} and
\mbox{$\mbox{75~MeV}\lesssim{}m_s\lesssim\mbox{170~MeV}$}~\cite{PDG:2000},
can be treated as perturbations about the chiral limit
\mbox{$m_q\rightarrow{}0$} by expanding in powers of~\mbox{$m_q/\Lchi$}\@.
For the rest of this chapter we identify the chiral symmetry group as
\mbox{$\Gchi=\su{3}\sub{L}\times\su{3}\sub{R}$} and drop any explicit
reference to the heavy quark flavors\@.

Invariance under~C, P, and~\Gchi\ is imposed on the Lagrangian of the
effective field theory; invariance under~T follows automatically from the
CPT~theorem\@.

\ifthesisdraft\clearpage\fi%
\section{Chiral Perturbation Theory}%
\label{sec:cpt}%
\marginal{[sec:cpt]}%
%
The spontaneous breaking of chiral symmetry to the \su{3}\sub{V}~subgroup
implies the existence of eight massless Goldstone scalar fields\@.
However, because the chiral symmetry is explicitly broken by the quark
masses, these fields acquire (small) finite masses and are commonly referred
to as pseudo-Goldstone bosons\@.
The pseudo-Goldstone bosons of spontaneously broken chiral symmetry are
identified as the light pseudoscalar meson octet of pions~(\piP, \piM,
and~\piZ), kaons~(\kP, \kM, \kZ, and~\kbarZ), and the eta~(\eZ)\@.
Chiral perturbation theory is the effective field theory for describing the
interactions of this meson octet at energies much lower than the chiral
symmetry breaking scale,~\Lchi\@.
For reviews of this subject see
references~\cite{D-H-G:book,B-K-M:review}\@.

The pseudo-Goldstone bosons are represented by a \mbox{$3\times3$} special
unitary matrix, \mbox{$U=e^{i\Phi/F_0}$}, with
\marginal{[eq:U-def]}%
\begin{equation}
\label{eq:U-def}
    \Phi 
    = \phi\ss{a}{} \lambda\ss{a}{}
    = \sqrt{2} \left(
        \begin{array}{ccc}
        \piZ/\sqrt{2} + \eZ/\sqrt{6} & \piP & \kP \\*
        \piM & -\piZ/\sqrt{2} + \eZ/\sqrt{6} & \kZ \\*
        \kM & \kbarZ & -2\eZ/\sqrt{6}
        \end{array} \right)
\end{equation}
and \mbox{$F_0\simeq{}F_\pi\simeq\mbox{93~MeV}$}\@.
The pion decay constant~$F_\pi$ is determined from
\marginal{[eq:Fpi-def]}%
\begin{equation}
\label{eq:Fpi-def}
    \bra{0} J\ss{5}{\mu} \ket{\piM(p)} 
    = i\sqrt{2} F_\pi p\ss{}{\mu}
,
\end{equation}
where \mbox{$J\ss{5}{\mu}=\overline{u}\gamma\ss{}{\mu}\gamma\ss{5}{}d$} is an
octet axial-vector current associated with chiral symmetry\@.
Under the symmetries of section~\ref{sec:symQCD} the field~$U$ transforms as
\marginal{[eq:U-xfm]}%
\begin{equation}
\label{eq:U-xfm}
    U \xform{\Gchi} R U L^\dag 
, \mathsp
    U \xform{P} U^\dag 
, \mathsp
    U \xform{C} U^T 
.
\end{equation}

We introduce the quark mass matrix~\Mq\ through the
field~\mbox{$\chi=2B_0\Mq$}, where the constant~$B_0$ is related to the
vacuum expectation value in equation~(\ref{eq:qq-vev}) and is approximately
\mbox{$B_0\simeq2\Delta/F_0^2\sim\mbox{1300~MeV}$~\cite{B-K-M:review}}\@.
For the purpose of constructing effective Lagrangians $\chi$~is assumed to
transform in a way that preserves the symmetries of the QCD~Lagrangian;
\marginal{[eq:chi-xfm]}%
\begin{equation}
\label{eq:chi-xfm}
    \chi \xform{\Gchi} R \chi L^\dag 
, \mathsp
    \chi \xform{P} \chi^\dag 
, \mathsp
    \chi \xform{C} \chi^T 
.
\end{equation}
In treating~\Mq\ we neglect the quark mass difference~\mbox{$m_u-m_d$} and
replace both~$m_u$ and~$m_d$ with the average
\mbox{$\hat{m}=\frac{1}{2}(m_u+m_d)$}\@.
Because the corresponding \su{2}~subgroup of~\su{3}\sub{V} is isospin
symmetry, this approximation ignores isospin violation in the strong
interaction\@.

Calculations in \cpt\ are organized as expansions in powers of~$m_q$ and~$Q$,
the characteristic momentum scale of the interaction\@.
The terms in the Lagrangian of \cpt~(\L{\cpt}{}) are grouped by the number of
powers of~$m_q$ and~$Q$ each contributes to diagrams when it is present\@.
For a suitable choice of regulator and subtraction scheme, such as
dimensional regularization with modified minimal subtraction~(\MSbar), each
derivative contributes one power of~$Q$\@.
In the expansions of~\L{\cpt}{} (in powers of~$Q$ and~$m_q$ respectively), 
the first terms are given by
\marginal{[eq:Lcpt2]}%
\begin{equation}
\label{eq:Lcpt2}
    \L{\cpt}{} 
    = \frac{F_0^2}{4} \tr{\dmu U \dum U^\dag}
        + \frac{F_0^2}{4} \tr{\chi U^\dag + U \chi^\dag} 
        + \cdots
\end{equation}
The first term yields a canonically normalized kinetic term for the
pseudo-Goldstone bosons~$\phi\ss{a}{}$\@.
The role of the arbitrary coefficient for the second term is played by the
empirically-determined constant~$B_0$\@.
The second term gives the leading contribution to the pseudoscalar masses,
\marginal{[eq:pgb-mass]}%
\begin{eqnarray}
\label{eq:pgb-mass}
    m_\pi^2 & = & 2B_0 \hat{m} + \cdots 
, \\* \nonumber
    m_K^2 & = & B_0 (\hat{m} + m_s) + \cdots 
, \\* \nonumber
    m_\eZ^2 & = & \tfrac{2}{3} B_0 (\hat{m} + 2m_s) + \cdots 
,
\end{eqnarray}
represented collectively as~$m_\Phi^2$\@.

The result for the pseudoscalar masses indicates that the dual expansion
in~$m_q$ and~$Q$ is unnecessary\@.
Since equation~(\ref{eq:pgb-mass}) shows
\mbox{$m_\Phi^2\sim{}B_0m_q\sim\chi$} and we work in the relativistic regime
where~\mbox{$Q^2\sim{}m_\Phi^2$}, one factor of~$m_q$ (i.e.,~$\chi$) counts 
as a contribution of order~$Q^2$\@.
Thus, the dual expansion is replaced with a scheme counting powers of~$Q$
only\@.
The expansion of the Lagrangian is written
\mbox{$\L{\cpt}{}=\L{2}{}+\L{4}{}+\cdots$} where~\L{d}{} contains all terms
of order~$Q^d$ and~\L{2}{} is given by equation~(\ref{eq:Lcpt2})\@.
The conventional parameterization of~\L{4}{} is
\marginal{[eq:Lcpt4]}%
\begin{eqnarray}
\label{eq:Lcpt4}
    \L{4}{} 
    & = & L_1 \tr{\dmu U \dum U^\dag}^2 
        + L_2 \tr{\dmu U \dnu U^\dag} \tr{\dum U \dun U^\dag} 
\\* \nonumber & & \mbox{} 
        + L_3 \tr{\dmu U \dum U^\dag \dnu U \dun U^\dag}
        + L_4 \tr{\dmu U \dum U^\dag} \tr{\chi U^\dag + U \chi^\dag}
\\* \nonumber & & \mbox{} 
        + L_5 \tr{\dmu U \dum U^\dag (\chi U^\dag + U \chi^\dag)} 
        + L_6 \tr{\chi U^\dag + U \chi^\dag}^2 
\\* \nonumber & & \mbox{} 
        + L_7 \tr{\chi U^\dag - U \chi^\dag}^2
        + L_8 \tr{\chi U^\dag \chi U^\dag + U \chi^\dag U \chi^\dag} 
\end{eqnarray}
as first worked out by \mbox{Gasser} and
\mbox{Leutwyler}~\cite{G-L:su2-A,G-L:su2-B,G-L:su2-C,G-L:su3}\@.
Table~\ref{tbl:L4-coeff} presents phenomenological values of the
coefficients~$L_i$ renormalized in \MSbar\@.
In Appendix~\ref{ch:codes} we present \mbox{\textit{Mathematica}} routines to
symbolically expand~\L{\cpt}{} in terms of the meson fields to the order
necessary for this work\@.

\begin{table}[tb]
\centering
\begin{tabular}{|r|r|l|} \hline 
$i$ & $L_i^r(m_\rho)\times10^3$ & source \MS \\* \hline \hline
1 & $0.4\pm0.3$ & $K_{e4}$, $\pi\pi\rightarrow\pi\pi$ \MS \\*
2 & $1.35\pm0.3$ & $K_{e4}$, $\pi\pi\rightarrow\pi\pi$ \\*
3 & $-3.5\pm1.1$ & $K_{e4}$, $\pi\pi\rightarrow\pi\pi$ \\*
4 & $-0.3\pm0.5$ & Zweig rule (large~$N_c$) \\*
5 & $1.4\pm0.5$ & ratio $F_K:F_\pi$ \\*
6 & $-0.2\pm0.3$ & Zweig rule (large~$N_c$) \\*
7 & $-0.4\pm0.2$ & Gell-Mann--Okubo, with $L_5^r$, $L_8^r$ \\*
8 & $0.9\pm0.3$ & ratio $(m_s-\hat{m}):(m_d-m_u)$, \\* 
  & & $(m_{\kZ}^2-m_{\kP}^2)$, with $L_5^r$ \rule[-1.2ex]{0ex}{1ex}\\* \hline
\end{tabular}
\caption{Phenomenological values of the coefficients~$L_i^r(\mu)$,
  renormalized in \MSbar\ at~\mbox{$\mu=m_\rho$}, taken from
  \mbox{J.~Bijnens~\textit{et~al.}~\cite{B-E-G:values}} (also see
  reference~\cite{G-L:su3} for discussion)\@. \drafttext{\protect\\* 
\mbox{\bf[tbl:L4-coeff]}}}
\label{tbl:L4-coeff}
\end{table}

The power counting rules for Feynman diagrams, determined by
\mbox{Weinberg}~\cite{SW:count}, establish the relative importance of
diagrams in an arbitrary process\@.
A diagram with $N_L$~loops, $N_\pi$~meson propagators, and constructed from
$N_d$~vertices derived from~\L{d}{} will contribute at order~$Q^D$ where
\marginal{[eq:cpt-rule]}%
\begin{eqnarray}
\label{eq:cpt-rule}
    D 
    & = & 2 + 2N_L + \sum_d N_d (d - 2) 
\\* \nonumber
    & = & 4N_L - 2N_\pi + \sum_d d N_d 
.
\end{eqnarray}
The leading-order contribution is given by the sum of all tree diagrams
constructed exclusively from operators in~\L{2}{}\@.
At next to leading order one must include all one-loop diagrams constructed
exclusively from operators in~\L{2}{} and all tree diagrams with one vertex
from~\L{4}{} and any number of vertices from~\L{2}{}\@.
Diagrams with more loops and more powers of~$Q^2$ at the vertices must be
included at higher order\@.

On dimensional grounds, the powers of~$Q^2$ which suppress higher-order
contributions must be accompanied by a mass scale squared~($\Lambda^2$) in
the denominator\@.
For powers of~$Q^2$ generated by a loop integral, the factor which appears 
is \mbox{$\Lambda^2\sim(4\pi{}F_0)^2$}\@.
When a factor of~$Q^d$ arises from an operator in~\L{d}{}, the compensating
powers of~$\Lambda$ are implicit in the dimension of the coefficient of the
operator\@.
We can represent the typical size of coefficients appearing in~\L{d}{}
as~\mbox{$C_d=\alpha\Lambda^{4-d}$}, where~$\alpha$ is a dimensionless
constant of `natural' size (discussed below)\@.
Since each successive order in the expansion of~\L{\cpt}{} serves to
approximate physics at short distance better than the preceding order, the
scale relating coefficients in different orders of~\L{\cpt}{} is characteristic
of the short-distance physics approximated\@.
Thus, $\Lambda$~represents scales like the mass of the
$\rho$~meson~($m_\rho$) and the chiral symmetry breaking scale~(\Lchi)\@.
We treat all of the scales which suppress powers of~$Q$ (e.g.,~$4\pi{}F_0$,
\Lchi, $m_\rho$,~\ldots) equally in terms of the power counting and refer to
the common scale as~\mbox{$\Lambda=\Lchi\simeq4\pi{}F_0$}\@.

The convergence of calculations in \cpt\ depends on the assumption of
naturalness, that the coefficients appearing in~\L{\cpt}{} are not much
larger than their natural size\@.
An anomalously large coefficient, by a factor of order~\mbox{$(\Lchi/Q)^2$},
would indicate that the corresponding operator violates the power counting in
equation~(\ref{eq:cpt-rule}) and the systematic expansion breaks down\@.
A priori, there is no reason to expect such large coefficients and the
occurrence of such an anomaly would indicate either the omission of a 
relevant degree of freedom or fine-tuning in the underlying theory\@.
We get an extremely simple estimate of the natural size of the dimensionless
parameter~$\alpha$ by recognizing~\mbox{$C_2\sim{}F_0^2$} or
\mbox{$\alpha\sim{}(F_0/\Lchi)^2\sim(4\pi)^{-2}$}\@.
For instance, we expect the coefficients in~\L{4}{} to be of about the
natural scale, \mbox{$L_i\sim{}C_4=\alpha$}\@.
\mbox{Manohar} and \mbox{Georgi}~\cite{M-G:dimanal} give a more rigorous
derivation of the same estimate based on naive dimensional analysis\@.
The phenomenological values presented in Table~\ref{tbl:L4-coeff} are
slightly smaller than this estimate\@.

\ifthesisdraft\clearpage\fi%
\section{Coupling to Matter Fields}%
\label{sec:matter}%
\marginal{[sec:matter]}%
%
The so-called matter fields are strongly interacting particles with masses
which do not vanish in the chiral limit\@.
Further, in the applications we consider, the masses of the matter fields are
large enough that the fields can be treated in a non-relativistic
formalism\@.
The matter fields do not form representations of the full chiral symmetry
group~\Gchi, but instead form irreducible representations under the
approximate~\su{3}\sub{V} symmetry of the vacuum\@.
In this section we describe a prescription for including the baryon octet
fields (the~$N$, $\Sigma$, and $\Xi$~isomultiplets and the~\lmdZ) and
decuplet fields (the~$\Delta$, $\Sigma^*$, and $\Xi^*$~isomultiplets and
the~\omgM) in the effective Lagrangian\@.
For a review of this subject see reference~\cite{HG:book}\@.

Under a~\Gchi\ transformation~\mbox{$(L,R)$}, fields in the fundamental
representation of~\su{3}\sub{V} transform as \mbox{$\psi\rightarrow{}H\psi$}
where~\mbox{$H\in\su{3}$} is a function of~$L$, $R$, and the pseudo-Goldstone
bosons~$U$\@.
The pure \su{3}\sub{V}~transformations correspond to taking \mbox{$H=L=R$}\@.
The form of~$H$ is determined by specifying the transformation
property of \mbox{$u=\sqrt{U}=e^{i\Phi/2F_0}$} as
\marginal{[eq:u-xfm]}%
\begin{equation}
\label{eq:u-xfm}
    u \xform{\Gchi} R u H^\dag = H u L^\dag
\end{equation}
which, when solved for~$H$, gives
\marginal{[eq:H-def]}%
\begin{equation}
\label{eq:H-def}
    H 
    = \sqrt{R U L^\dag \,} L \sqrt{U^\dag}
    = \sqrt{L U^\dag R^\dag \,} R \sqrt{U} 
.
\end{equation}
(The field~$u$ is an alternative representation for the pseudo-Goldstone
bosons, corresponding to a different choice for the spontaneously broken
generators of~\Gchi\ in the
CCWZ~prescription~\cite{AM:eft,CCSW:splI,CCSW:splII}\@.)
Because the matrix~$H$ is a function of the pseudo-Goldstone boson
field~$U(x)$, $H$~also implicitly depends on the space-time
coordinate~$x\ss{}{\mu}$\@.
To compensate for the \mbox{$x\ss{}{\mu}$-depend}ence of~$H$, we introduce a
vector field
\marginal{[eq:Vmu-def]}%
\begin{equation}
\label{eq:Vmu-def}
    \Vmu = \tfrac{1}{2} (u^\dag \dmu u + u \dmu u^\dag)
\end{equation}
which transforms under~\Gchi\ as
\mbox{$\Vmu\rightarrow{}H\Vmu{}H^\dag-(\dmu{}H)H^\dag$} and construct the
covariant derivative \mbox{$\Dmu\psi=\dmu{}\psi+\Vmu\psi$} such
that~\mbox{$\Dmu\psi\rightarrow{}H(\Dmu\psi)$}\@.

Armed with the \su{3}\sub{V}~formalism for the fundamental representation,
incorporating the \spin{1} octet and \spin{3} decuplet fields is relatively
straight forward\@.
The octet baryons are encoded as a \mbox{$3\times3$} matrix of Dirac fields
\marginal{[eq:B-def]}%
\begin{equation}
\label{eq:B-def}
    B
    = \left(
        \begin{array}{ccc}
        \sigZ/\sqrt{2} + \lmdZ/\sqrt{6} & \sigP & p \\*
        \sigM & -\sigZ/\sqrt{2} + \lmdZ/\sqrt{6} & n \\*
        \xiM & \xiZ & -2\lmdZ/\sqrt{6}
        \end{array} \right)
\end{equation}
and transform in the adjoint representation of~\su{3}\sub{V}, i.e.,\ 
\marginal{[eq:B-xfm]}%
\marginal{[eq:D-on-B]}%
\begin{eqnarray}
\label{eq:B-xfm} & 
    B \xform{\Gchi} H B H^\dag 
, & \\*
\label{eq:D-on-B} & 
    \Dmu B = \dmu B + [\Vmu , B]
.&
\end{eqnarray}
The \spin{3} decuplet baryons form a fully-symmetric \mbox{rank-3} tensor
under~\su{3}\sub{V} characterized by
\marginal{[eq:T-xfm]}%
\marginal{[eq:D-on-T]}%
\begin{eqnarray}
\label{eq:T-xfm} & 
    T\ss{}{abc} \xform{\Gchi} 
        H\ss{}{aa'} H\ss{}{bb'} H\ss{}{cc'} T\ss{}{a'b'c'}
, & \\*
\label{eq:D-on-T} & 
    (\Dmu T)\ss{}{abc} 
    = \dmu T\ss{}{abc}
        + V\ss{\mu}{aa'} T\ss{}{a'bc}
        + V\ss{\mu}{bb'} T\ss{}{ab'c}
        + V\ss{\mu}{cc'} T\ss{}{abc'} 
.&
\end{eqnarray}
The components of~$T$ are Rarita-Schwinger fields subject to the auxiliary
constraint \mbox{$\gamma\ss{\mu}{}T\ss{}{\mu}=0$} and are identified as
follows:
\marginal{[T-def]}%
\begin{equation}
\label{eq:T-def}
    \begin{array}[t]{l@{\mathsp}l@{\mathsp}l}
    T\ss{}{111} = \dltPP , &
    T\ss{}{113} = \sigSP/\sqrt{3} , &
    T\ss{}{133} = \xiSZ/\sqrt{3} , \\*
    T\ss{}{112} = \dltP/\sqrt{3} , &
    T\ss{}{123} = \sigSZ/\sqrt{6} , &
    T\ss{}{233} = \xiSM/\sqrt{3} , \\*
    T\ss{}{122} = \dltZ/\sqrt{3} , &
    T\ss{}{223} = \sigSM/\sqrt{3} , & \\*
    T\ss{}{222} = \dltM , & &
    T\ss{}{333} = \omgM .
    \end{array}
\end{equation}
The minimal-coupling Lagrangian for the~$B$ and~$T$ fields is
\marginal{[eq:L0matter]}%
\begin{equation}
\label{eq:L0matter}
    \L{0}{} 
    = \overline{B} (i\Dslash - m_B) B
        - \overline{T}\ss{\mu}{} (i\Dslash - m_T) T\ss{}{\mu}
\end{equation}
with implied pair-wise summation of the dangling chiral indices\@.

The effective Lagrangian for matter fields will also explicitly include~$U$
and~$\chi$ which transform under~\Gchi\ as \mbox{$O\rightarrow{}ROL^\dag$}\@.
For convenience, we define fields~\mbox{$\widetilde{O}=u^\dag{}Ou^\dag$} such
that \mbox{$\widetilde{O}\rightarrow{}H\widetilde{O}H^\dag$} under
\Gchi~transformations and coupling~$B$ and~$T$ to the new
fields~$\widetilde{O}$ becomes transparent\@.
The result for~$U$ is trivial, \mbox{$\widetilde{U}=u^\dag{}Uu^\dag=1$};
for~\mbox{$\dmu{}U$} we get the more interesting result
\marginal{[eq:Amu-def]}%
\begin{eqnarray}
\label{eq:Amu-def}
    \Amu 
    & = & \tfrac{i}{2} u^\dag \dmu U u^\dag
    \ =\ \tfrac{-i}{2} u \dmu U^\dag u 
\\* \nonumber
    & = & \tfrac{i}{2} (u^\dag \dmu u - u \dmu u^\dag)
,
\end{eqnarray}
where the factor of~$\frac{i}{2}$ is included so~\Amu\ is hermitian\@.
To replace~$\chi$ (and~$\chi^\dag$) we choose the (anti-)hermitian
combinations
\marginal{[eq:chiPM-def]}%
\begin{equation}
\label{eq:chiPM-def}
    \chi_\pm = u^\dag \chi u^\dag \pm u \chi^\dag u 
.
\end{equation}
As an example the \mbox{order-$Q^2$} Lagrangian of \cpt,
equation~(\ref{eq:Lcpt2}), can be rewritten in terms of~\Amu\
and~$\chi_\pm$ as
\mbox{$\L{2}{}=F_0^2\tr{\Amu{}A\ss{}{\mu}}+\frac{1}{4}F_0^2\tr{\chi_+}$}\@.

We briefly mention two relations which help to identify a minimal set of
terms which are needed at higher order, see \mbox{Fearing} and
\mbox{Scherer}~\cite{F-S:orderQ6} for a more complete
account\@.\footnote{When using the external field method to impliment
  \emph{local} chiral symmetry, these relations are modified by additional
  terms involving external (i.e.,~non-propagating) gauge fields\@.}
The first is the chain rule in the form
\marginal{[eq:chain-rule]}%
\begin{equation}
\label{eq:chain-rule}
    u^\dag \dmu O u^\dag
    = \Dmu \widetilde{O} - i\{ \Amu , \widetilde{O} \}
\end{equation}
which allows the replacement of multiple derivatives of~$U$ with factors
of~\Amu\ and covariant derivatives of~\Amu, e.g.,\ 
\mbox{$u^\dag\dmu\dnu{}Uu^\dag=-2i\Dmu{}A\ss{\nu}{}-2\{\Amu,A\ss{\nu}{}\}$}\@.
Two consequences of equation~(\ref{eq:chain-rule}) are
1)~\mbox{$\Dmu{}A\ss{\nu}{}$} is symmetric in~\mbox{$(\mu,\nu)$} and
2)~despite the dependence of~$\chi_\pm$ on~$u(x)$, covariant derivatives
of~$\chi_\pm$ are unnecessary since
\mbox{$\Dmu\chi_\pm=i\{\Amu,\chi_\mp\}$}\@.
The second simplifying relation gives the field strength associated
with~\Vmu\ in terms of the field~\Amu\ 
\marginal{[eq:elim-G]}%
\begin{equation}
\label{eq:elim-G}
    G\ss{\mu\nu}{(V)} 
    = \dmu \V\ss{\nu}{} - \dnu \Vmu + [\Vmu , \V\ss{\nu}{}] 
    = [\Amu , A\ss{\nu}{}] 
.
\end{equation}
Because antisymmetric covariant derivatives result in factors of the field
strength, like
\mbox{$(\Dmu{}\D\ss{\nu}{}-\D\ss{\nu}{}\Dmu)\psi=G\ss{\mu\nu}{(V)}\psi$} in
the fundamental representation, we can treat all multiple covariant
derivatives of any field as implicitly symmetric without loss of
generality\@.
To summarize, the \mbox{\Gchi-invar}iant effective Lagrangian for the baryon
fields~$B$ and~$T$ is written in terms of the building blocks~$B$, $T$, \Amu,
$\chi_\pm$, and fully-symmetric covariant derivatives of them\@.
The field strength~$G\ss{\mu\nu}{(V)}$ and covariant derivatives
of~$\chi_\pm$ may be omitted in favor of terms involving more factors
of~\Amu\@.

\ifthesisdraft\clearpage\fi%
\section{Non-Relativistic Power Counting}%
\label{sec:counting}%
\marginal{[sec:counting]}%
%
The appearance of baryon masses in equation~(\ref{eq:L0matter}) wrecks the
power counting of \cpt\@.
Derivatives of heavy fields would contribute factors of `hard' momenta,
where~\mbox{$p\ss{0}{}\geq{}m_B$}, and loop diagrams could result in explicit
factors of~$m_B$ or~$m_T$ in the numerator\@.
These diagrams are not suppressed relative to `lower order'
diagrams since \mbox{$m_B\sim{}m_T\sim\Lchi$}\@.
We use a non-relativistic approach adapted from the heavy baryon chiral
perturbation theory~(HB\cpt) developed by \mbox{Jenkins} and
\mbox{Manohar}~\cite{J-M:HBcptI,J-M:HBcptII,EJ:masses,J-M:talk} to address
this problem\@.
We start with the velocity-dependent baryon fields of HB\cpt\ then specialize
to working in the rest frame of the baryons\@.

The momentum of a heavy baryon field is decomposed as
\mbox{$p\ss{}{\mu}=m_Bv\ss{}{\mu}+k\ss{}{\mu}$} in HB\cpt, where the residual
momentum~$k\ss{}{\mu}$ is assumed small and reflects how much the baryon is
off mass-shell\@.
The velocity-dependent baryon fields are defined using~$v\ss{}{\mu}$ as
\marginal{[eq:Bv,Tv-def]}%
\begin{eqnarray}
\label{eq:Bv,Tv-def}
    B_v(x) 
    & = & \frac{1 + \vslash}{2} \: e^{i m_B v \cdot x} \, B(x) 
, \\* \nonumber
    T_v(x) 
    & = & \frac{1 + \vslash}{2} \: e^{i m_B v \cdot x} \, T(x) 
,
\end{eqnarray}
where the octet mass~$m_B$ is used in the definition of the decuplet~$T_v$ to
avoid \mbox{$x\ss{}{\mu}$-depend}ent phases in the Lagrangian coupling~$B_v$
to~$T_v$\@.
The factor of~\mbox{$\frac{1}{2}(1+\vslash)$} projects out the particle
components of the Dirac spinors\@.
The anti-particle components are implicitly integrated out of the theory and
the effects of virtual baryon loops are absorbed into terms of the effective
Lagrangian suppressed by powers of~$1/m_B$\@.

This representation of the heavy baryon fields permits a sensible power
counting scheme\@.
Derivatives of the velocity-dependent fields give factors of the small
residual momenta~$k\ss{}{\mu}$ in place of the hard momenta~$p\ss{}{\mu}$\@.
The minimal-coupling Lagrangian corresponding to equation~(\ref{eq:L0matter})
becomes
\marginal{[eq:Lvmatter]}%
\begin{equation}
\label{eq:Lvmatter}
    \L{v}{} 
    = \overline{B_v} (i v \cdot D) B_v
        - \overline{T_v}\ss{\mu}{} (i v \cdot D) T_v\ss{}{\mu}
        + \Delta m \overline{T_v}\ss{\mu}{} T_v\ss{}{\mu} 
        + \cdots
,
\end{equation}
where~\mbox{$\Delta{}m=m_T-m_B\simeq\mbox{300~MeV}$} is considered of
order~$Q$ and the ellipsis denotes higher-order terms, such
as~\mbox{$-\overline{B_v}D^2B_v/2m_B$}, induced by the integration over
anti-particle degrees of freedom\@.
By removing~$m_B$ and~$m_T$ from the baryon propagators, loop diagrams will
not introduce positive powers of the masses in diagrams (except in a case
considered toward the end of the section)\@.

The Dirac structure of the fields can be eliminated in favor of two-component
spinors because we have integrated out the anti-particle components\@.
We choose to explicitly work in the frame~\mbox{$v\ss{}{\mu}=(1,0,0,0)$} and
drop the subscript~$v$ on heavy fields\@.
This choice of frame simplifies the spinor-related notation; for instance,
the auxiliary condition on Rarita-Schwinger fields,
\mbox{$\gamma\ss{\mu}{}T\ss{}{\mu}=0$}, implies constraints which reduce
to~\mbox{$T\ss{}{0}=0$} and \mbox{$\vec{\sigma}\cdot\vec{T}=0$}\@.
In this frame the resulting non-relativistic framework is equivalent to a
Lagrangian formulation of the time-ordered approach discussed by
\mbox{Weinberg}~\cite{SW:nuclI,SW:nuclII}\@.

The power counting for diagrams with a single heavy field generalizes
equation~(\ref{eq:cpt-rule})\@.
Since~\mbox{$m_B\sim{}m_T\sim\Lchi$} we do not need to distinguish between
corrections suppressed by~\mbox{$Q/m_B$} and~\mbox{$Q/\Lchi$}\@.
As before, each meson propagator counts as~$Q^{-2}$ and each loop integration
gives a factor of~$Q^4$\@.
Heavy field propagators each contribute~$Q^{-1}$ as seen from
equation~(\ref{eq:Lvmatter})\@.
We let~$N_i$ represent the number of vertices in a diagram which contribute
$d_i$~factors of~$Q$ and contain $n_i$~heavy fields\@.
A diagram in the single heavy-particle sector with $N_L$~loops, $N_\pi$~meson
propagators, and $N_I$~baryon propagators contributes at order~$Q^D$
where~\cite{SW:nuclI}
\marginal{[eq:stat-rule]}%
\begin{eqnarray}
\label{eq:stat-rule}
    D 
    & = & 4N_L - 2N_\pi - N_I + \sum_i d_i N_i
\\* \nonumber
    & = & 1 + 2N_L + \sum_i N_i (d_i + \tfrac{1}{2} n_i - 2) 
.
\end{eqnarray}
The primary difference from equation~(\ref{eq:cpt-rule}) is that each order
is suppressed by~\mbox{$Q/\Lchi$} rather than~\mbox{$(Q/\Lchi)^2$} relative
to the preceeding order\@.

For diagrams with two or more heavy particles, the power counting is
complicated by infrared divergences in some loop
integrals~\cite{SW:nuclII}\@.
Consider a heavy-particle bubble diagram; for octet fields the loop integral
takes the form
\marginal{[eq:IRdiv-1]}%
\begin{equation}
\label{eq:IRdiv-1}
    \vc{\begin{picture}(50,0)%
        \put(15,0){\line(-1,1){10}}%
        \put(15,0){\line(-1,-1){10}}%
        \put(25,0){\circle{20}}%
        \put(35,0){\line(1,1){10}}%
        \put(35,0){\line(1,-1){10}}%
        \end{picture}}
    \longrightarrow \int\!\! \frac{d^d q}{(2\pi)^d} \;
            \frac{i}{E + q\ss{0}{} + i\epsilon} \;
            \frac{i}{E - q\ss{0}{} + i\epsilon}
    = \frac{i}{2E} \int\!\! \frac{d^{d-1} q}{(2\pi)^{d-1}}
,
\end{equation}
where the incoming particles have energies~$E$ and momenta~${\pm}\vec{p}$\@.
The divergence arises as~\mbox{$E\rightarrow0$} because the
$q\ss{0}{}$~contour is pinched between the poles of the static propagators\@.
By resumming the kinetic energy operator, e.g.,~\mbox{$B^\dag\nabla^2B/2m_B$},
into the heavy field propagator, the infrared divergences are removed\@.
In dimensional regularization the result for the modified loop integral
becomes
\marginal{[eq:IRdiv-2]}%
\begin{eqnarray}
\label{eq:IRdiv-2}
    \lefteqn{\int\!\! \frac{d^d q}{(2\pi)^d} \;
            \frac{i}{E + q\ss{0}{} - |\vec{q}\,|^2 / 2m_B + i\epsilon} \;
            \frac{i}{E - q\ss{0}{} - |{-}\vec{q}\,|^2 / 2m_B + i\epsilon}} 
\hspace*{.5in} \\* \nonumber 
    & = & \int\!\! \frac{d^{d-1} q}{(2\pi)^{d-1}} \;
            \frac{i}{2E - q^2/m_B + 2i\epsilon} 
\\* \nonumber
    & = & \frac{-im_B}{4\pi} \left( 
                \frac{-2 m_B E - i\epsilon'}{4\pi}
                \right)^{\frac{d-3}{2}}
            \Gamma(\tfrac{3-d}{2})
    = \frac{m_B \sqrt{2 m_B E \,}}{4\pi}
\end{eqnarray}
which is well-defined in the infrared\@.
Setting \mbox{$E=p^2/2m_B+\cdots$}, the leading behavior of the loop integral
is~\mbox{${\sim}m_BQ$}, not~\mbox{${\sim}Q^2$} as expected from
equation~(\ref{eq:stat-rule})\@.

By including the kinetic energy in the heavy field propagator we are treating
the operator~\mbox{$\nabla^2/2m_B$} on an equal footing
with~$\lpartial\ss{0}{}$\@.
For consistency, powers of the integration variable~$q\ss{0}{}$ should be
counted as~\mbox{$q\ss{0}{}\sim{}Q^2/m_B$} in loops leading to
nearly-infrared-divergent behavior\@.
Heavy field propagators count as~\mbox{$m_B/Q^2$}, loop integrals now give
\mbox{$dq\ss{0}{}\,d^3q\sim{}Q^5/m_B$}, and meson propagators
contribute~\mbox{$1/Q^2$} as before\@.
Thus the loop integral above counts
as~\mbox{${\sim}(Q^5/m_B)(m_B/Q^2)^2=m_BQ$} which reconciles the power
counting scheme with the result of equation~(\ref{eq:IRdiv-2})\@.
A general Feynman diagram with heavy particles in the initial and final
states may contribute at order~$Q^D$ where
\marginal{[eq:NR-rule]}%
\begin{equation}
\label{eq:NR-rule}
    D = 5N_L - 2N_\pi - 2N_I + \sum_i d_i N_i
\end{equation}
with the same notation as used in equation~(\ref{eq:stat-rule})\@.
If the loop integrals in a diagram are not infrared divergent, as with
crossed pion exchange, the actual contribution will be of higher order
consistent with equation~(\ref{eq:stat-rule})\@.
The systematics of the power counting for diagrams with two or more heavy
particles is developed in
references~\cite{B-B-L:nrqcd,SW:nuclII,L-M:bound,AM:nrqcd,G-R:NRgauge,%
  L-S:nrqcd,L-M-R:RGscale}\@.
The presence of a bound state or resonance near threshold can also complicate
the power counting~\cite{K-S-W:finetune,DK:deuteron,UvK:finetune}\@.
This complication arises in nucleon-nucleon scattering and the power counting
was studied in that context by several authors, in particular \mbox{Kaplan},
\mbox{Savage}, and \mbox{Wise}~\cite{K-S-W:NNintI,K-S-W:NNintII}\@.
In this work we do not consider systems requiring this special treatment and
refer the reader to the literature for a discussion of the relevant power
counting schemes~\cite{M-S:schemeI,M-S:schemeII,JG:scheme}\@.


%% file: 10plet.tex
\chapter[Decuplet Self-Energy in Nuclear Matter]%
    {\\Decuplet Self-Energy in Nuclear Matter}%
\label{ch:decuplet}%
\marginal{[ch:decuplet]}%
%
Strong interaction effects shift the self-energy of hadrons in nuclear matter
from the free-space values\@.
This effect for the \spin{1} octet baryons was studied in an effective field
theory framework by \mbox{Savage} and \mbox{Wise}~\cite{S-W:hyperon}\@.
Using the formalism reviewed in Chapter~\ref{ch:theory}, we calculate the
self-energy shifts of the \spin{3} decuplet baryons in nuclear matter at
leading order\@.
The self-energy shifts of the decuplet baryons, particularly of the
$\Delta$~isomultiplet, are relevant in studies of meson-nucleus
scattering~\cite{E-W:book,E-K:book} and of stellar and neutron star
matter~\cite{S-T:book,RS:delta}\@.
For the $\Delta$~isomultiplet the self-energy shifts have also been examined
in various phenomenological models~\cite{RS:delta,C-D:delta,W-W:delta} and in
QCD sum rules~\cite{XJ:delta}\@.
The work described here differs from the earlier approaches by extending the
calculation to multiplets of chiral~\su{3}\sub{V} and by including contact
diagrams necessary for a consistent and systematic momentum expansion\@.
This chapter makes a minor correction to a prior publication~\cite{O-S:delta}
and includes a more detailed discussion of the self-energy of the
$\Delta$~isomultiplet in nuclear matter near saturation density\@.

The first section discusses how effects of nuclear matter are described in
the effective field theory\@.
In section~\ref{sec:LforNM} we determine what Feynman diagrams contribute to
the self-energy shifts at leading order and construct the relevant effective
Lagrangian\@.
Section~\ref{sec:dEresult} presents the main results of the calculation\@.
Finally, a discussion and interpretation of the results is contained in
section~\ref{sec:NMconcl}\@.

\ifthesisdraft\clearpage\fi%
\section{Effects of Nuclear Matter}%
\label{sec:NMintro}%
\marginal{[sec:NMintro]}%
%
In nuclear matter the propagation of decuplet baryons is effected by
interactions with the background medium\@.
In particular the self-energy~\E{}{}, i.e.,\ the location of the
$k\ss{0}{}$~pole in the decuplet two-point function, is shifted relative to
free space\@.
As an additional consideration, the background medium breaks Lorentz boost
invariance by specifying a unique frame, the zero-momentum frame of the
nuclear matter\@.
The self-energy in free space of a decuplet baryon with
momentum~\mbox{$k=|\vec{k}\,|$} is
\marginal{[eq:Evac-form]}%
\begin{eqnarray}
\label{eq:Evac-form}
    \E{vac}{}(k^2) 
    & = & \sqrt{m_T^2 + k^2 - i m_T \rate{vac}{} \,} - m_B 
\\* \nonumber
    & = & \left( \Delta m + \frac{k^2}{2m_T} + \cdots \right)
        - \frac{i}{2} \rate{vac}{} \left( 
            1 - \frac{k^2}{2m_T^2} + \cdots \right) 
    + \mathcal{O}(\rate{vac}{\scriptsize2})
,
\end{eqnarray}
where~$\rate{vac}{}$ is the free-space decay rate
and~\mbox{$\Delta{}m=m_T-m_B$}\@.
Throughout this chapter, the octet mass~$m_B$ has been implicitly subtracted
whenever we refer to the decuplet self-energy, as discussed in
section~\ref{sec:counting}\@.
Equation~(\ref{eq:Evac-form}) contains only two decuplet parameters; the full
\mbox{$k^2$-depend}ence of the self-energy is determined by Lorentz
invariance\@.
The corresponding expression for a decuplet baryon in nuclear matter can be
written
\marginal{[eq:Enm-form]}%
\begin{equation}
\label{eq:Enm-form}
    \E{nm}{}(k^2) 
    = (\Delta m^* + \alpha k^2 + \cdots)
        - \tfrac{i}{2} \Gamma^* (1 + \beta k^2 + \cdots)
\end{equation}
in which the \mbox{$k^2$-depend}ence (e.g.,~$\alpha$, $\beta$,~\ldots) 
cannot be determined from symmetry arguments alone\@.

Although invariance under Lorentz boosts is lost, the remaining rotational
symmetry constrains the spin-dependence of the self-energy~\E{nm}{}\@.
Because the nuclear medium is rotationally invariant, the only preferred
spacial directions are along the decuplet baryon three-momentum~$\vec{k}$
and spin~$\vec{S}$\@.
As a function of~$\vec{k}$ and~$\vec{S}$, the self-energy depends on only 
the combinations~$k^2$ and~\mbox{$\vec{k}\cdot\vec{S}$}\@.
(\mbox{$S^2=\frac{15}{4}$}~is trivial\@.)
Further, parity invariance of the strong interaction dictates that the
self-energy depends on even powers of~$\vec{k}$, which means
replacing~\mbox{$\vec{k}\cdot\vec{S}$} with
\mbox{$(\vec{k}\cdot\vec{S}\,)^2=k^2h^2$} in terms of the baryon
helicity~$h$\@.
Consequently, the self-energy in nuclear matter is diagonal in the baryon
helicity states and takes the values~$\E{nm}{(1/2)}(k^2)$
for~\mbox{$h=\pm\frac{1}{2}$} and~$\E{nm}{(3/2)}(k^2)$
for~\mbox{$h=\pm\frac{3}{2}$}\@.
In the limit of vanishing momentum, the rotational symmetry is elevated to
full \su{2}~invariance, and the self-energy must be independent of the
decuplet spin projection along any direction\@.
In terms of the parameters in equation~(\ref{eq:Enm-form}),
$\Delta{}m^*$~and~$\Gamma^*$ are helicity-independent, while the coefficients
of powers of~$k^2$ (e.g.,~$\alpha$, $\beta$,~\ldots) depend on~$|h|$\@.

The lowest order in a density expansion for nuclear matter is a Fermi gas of
non-interacting protons and neutrons with Fermi momenta~\pFp\ and~\pFn\ 
respectively\@.
In this framework, the characteristic momenta relevant in the chiral
derivative expansion are~\pFp, \pFn, and~$k$ the decuplet baryon momentum\@.
Since the density of a degenerate Fermi gas is given
by~\mbox{$d_F=p_F^3/3\pi^2$}, the density expansion for nuclear matter is
consistent with the chiral derivative expansion\@.
The static nucleon propagator in nuclear matter with Fermi momentum~$p_F$
is~\cite{S-W:hyperon}
\marginal{[eq:NMprop]}%
\begin{equation}
\label{eq:NMprop}
    \widetilde{S}_{\rm nm} (q\ss{0}{}, \vec{q}\,)
    = \frac{i\Theta(|\vec{q}\,| - p_F)}{q\ss{0}{} + i\epsilon}
        + \frac{i\Theta(p_F - |\vec{q}\,|)}{q\ss{0}{} - i\epsilon}
\end{equation}
at lowest order in the nuclear density\@.
The modified nucleon propagator reflects the presence of the background
medium through two effects, nucleon states inaccessible due to Pauli-blocking
and nucleon-hole intermediate states allowed
for~\mbox{$|\vec{q}\,|\leq{}p_F$}\@.

As the location of the $k\ss{0}{}$~pole in the exact decuplet two-point
function, the self-energy in nuclear matter is given by the solution of
\marginal{[eq:Enm-def]}%
\begin{equation}
\label{eq:Enm-def}
    \E{nm}{}(k^2) - \Delta m - \pse{nm}(\E{nm}{}(k^2) , \vec{k} \,) 
    = 0 
,
\end{equation}
where~\pse{nm} is the \emph{proper} self-energy for nuclear matter,
\mbox{i.e.,~$-i\pse{nm}$} is the sum of connected one-particle-irreducible
diagrams in the two-point function\@.
What we calculate is the self-energy shift,
\mbox{$\delta\E{}{}=\E{nm}{}-\E{vac}{}$}, obtained from
equation~(\ref{eq:Enm-def}) by expanding the proper self-energy for nuclear
matter about the free-space pole~\E{vac}{},
\marginal{[eq:dE-calc]}%
\begin{eqnarray}
\label{eq:dE-calc}
    \delta\E{}{}(k^2) 
    & = & \pse{nm}(\E{vac}{}(k^2), \vec{k} \,)
        - \pse{vac}(\E{vac}{}(k^2), \vec{k} \,)
\\* \nonumber & & \mbox{} 
        + \left\{ \E{nm}{}(k^2) - \E{vac}{}(k^2) \MS \right\}
                \frac{\partial}{\smash{\partial k\ss{0}{}}\ms} 
                    \pse{nm}(k\ss{0}{}, \vec{k} \,)
                    \left. \vpf{\ms}{\ms} \right|_{
                        k\ss{0}{} = \E{vac}{}(k^2)}
        + \cdots
\end{eqnarray}
and in turn expanding~\mbox{$\E{vac}{}(k^2)$} in powers of~$Q$ as shown in
equation~(\ref{eq:Evac-form})\@.
Note that only Feynman diagrams with an internal nucleon propagator
contribute to the difference on the first line\@.
Because the chiral expansion of the proper self-energy~\pse{nm} starts at
order~$Q^2$, successive terms in equation~(\ref{eq:dE-calc}) only contribute
at higher order\@.
(A very important exception to the last point is discussed in detail in
subsection~\ref{sub:dlt-sat}\@.)
The real part of the self-energy shift~$\dE{}{(h)}(k^2)$ modifies the
decuplet baryon energy-momentum dispersion relation from the free-space
form\@.
The change in the decuplet decay rate is given
by~\mbox{$\delta\rate{}{(h)}(k^2)=-2\,\mbox{Im}[\delta\E{}{(h)}(k^2)]$} at
leading order\@.

\ifthesisdraft\clearpage\fi%
\section{The Effective Lagrangian}%
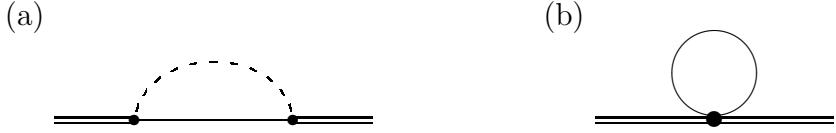
\begin{figure}[tb]
\hspace*{\fill}
\begin{picture}(120,50)%
    \put(0,50){\makebox(0,0)[rt]{(a)~}}%
    \put(0,4){\line(1,0){30}}%
    \put(0,6){\line(1,0){30}}%
    \put(30,5){\circle*{4}}%
    \put(30,5){\line(1,0){60}}%
    \put(90,5){\circle*{4}}%
    \put(90,4){\line(1,0){30}}%
    \put(90,6){\line(1,0){30}}%
    \savebox{\dash}(0,0){%
        \begin{picture}(0,0)%
        \put(30,-1.5){\line(0,1){3}}%
        \end{picture}}%
    \put(60,2){\rotatebox{15}{\usebox{\dash}}}%
    \put(60,.5){\rotatebox{30}{\usebox{\dash}}}%
    \put(60,-.75){\rotatebox{45}{\usebox{\dash}}}%
    \put(60,-1.75){\rotatebox{60}{\usebox{\dash}}}%
    \put(60,-2.5){\rotatebox{75}{\usebox{\dash}}}%
    \put(60,-3){\rotatebox{90}{\usebox{\dash}}}%
    \put(60,-2.5){\rotatebox{105}{\usebox{\dash}}}%
    \put(60,-1.75){\rotatebox{120}{\usebox{\dash}}}%
    \put(60,-.75){\rotatebox{135}{\usebox{\dash}}}%
    \put(60,.5){\rotatebox{150}{\usebox{\dash}}}%
    \put(60,2){\rotatebox{165}{\usebox{\dash}}}%
\end{picture}
\hspace*{\fill}
\begin{picture}(90,50)%
    \put(0,50){\makebox(0,0)[rt]{(b)~}}%
    \put(0,4){\line(1,0){90}}%
    \put(0,6){\line(1,0){90}}%
    \put(45,23){\circle{32}}%
    \put(45,5.5){\circle*{6}}%
\end{picture}
\hspace*{\fill}
\caption{Feynman diagrams for the self-energy shifts at leading order of
  decuplet baryons in nuclear matter, (a)~meson-nucleon loop diagrams and
  (b)~contact diagrams\@.  Double lines represent decuplet baryons, single
  lines represent nucleons, and dashed lines represent pseudo-Goldstone
  bosons\@.  \drafttext{\protect\\*
    \mbox{\bf[fig:NMdiags]}}}
\label{fig:NMdiags}
\end{figure}

\label{sec:LforNM}%
\marginal{[sec:LforNM]}%
%
At leading order in the chiral expansion the self-energy
shifts~$\delta\E{}{}$ coincide with the difference
\mbox{$\pse{nm}(\E{vac}{},\vec{k}\,)-\pse{vac}(\E{vac}{},\vec{k}\,)$}\@.
For diagrams contributing to the proper self-energies~\pse{nm} and~\pse{vac},
the power counting is given by equation~(\ref{eq:stat-rule}) for the single
heavy-particle sector\@.
Because a nucleon propagator is required for a non-zero difference, the
leading contribution arises from one-loop diagrams in which the vertices
satisfy \mbox{$d+\frac{1}{2}n-2=0$} and contributes at order~$Q^3$\@.
Topologically, the one-loop diagrams are constructed from either two
three-leg vertices or a single four-leg vertex; the two possibilities with a
nucleon propagator are shown in Figure~\ref{fig:NMdiags}\@.

The vertices of meson-nucleon loop diagrams are derived from operators
coupling the fields~\mbox{$TB^\dag\Amu$} in the \mbox{order-$Q$}
Lagrangian\@.
The most general Lagrangian of that form, invariant under~\Gchi\ and parity,
is~\cite{J-M:HBcptII,EJ:masses}
\marginal{[eq:NM-La]}%
\begin{equation}
\label{eq:NM-La}
    \L{\rm a}{} 
    = - \C \epsilon\ss{}{abc} \left(
            \vec{A}\ss{}{ad} \cdot 
                [B^\dag\ss{}{be}\, \vec{T}\ss{}{cde}]
            + \vec{A}\ss{}{da} \cdot 
                [\vec{T}^\dag\ss{}{cde} B\ss{}{eb}]
            \right) 
,
\end{equation}
where square brackets denote summation on spinor indices\@.
The value of the coefficient, \mbox{$|\C|\simeq1.53$}, is empirically
determined from \mbox{$T\rightarrow{}B\pi$}~decays~\cite{J-M:talk}\@.
The meson-nucleon loop diagrams contribute to the self-energy shifts of
the~$\Delta$ and $\Sigma^*$~isomultiplets only; the Lagrangian~\L{\rm{}a}{}
does not couple~$\Xi^*$ or $\omgM$~baryons to a nucleon and single
pseudo-Goldstone boson, a consequence of strangeness conservation\@.

For contact diagrams the vertex contributes \mbox{$d=0$}~powers of~$Q$, so
the relevant Lagrangian contains only simple products of field
operators~\mbox{$TB(TB)^\dag$}\@.
To construct the four-baryon operators we start with the spin and chiral
structures of the product~$TB$\@.
The operator product decomposes under rotational~\su{2} as
\mbox{$\bf\frac{3}{2}\otimes\frac{1}{2}=2\oplus1$} and under
chiral~\su{3}\sub{V} as \mbox{$\bf10\otimes8=35\oplus27\oplus10\oplus8$}\@.
By coupling products~$TB$ and~$(TB)^\dag$ to form chiral and rotational
singlets, we find eight linearly-independent four-baryon operators which
contribute to the self-energy shifts at leading order\@.
We choose to write the effective Lagrangian which contains these operators
as\footnote{In our prior publication we used a different convention for the
  pion decay constant, specifically
  \mbox{$f=\sqrt{2}F_0\simeq\mbox{132~MeV}$}\@.}
\marginal{[eq:NM-Lb]}%
\begin{eqnarray}
\label{eq:NM-Lb}
    \L{\rm b}{} \!\!
    & =\!\!\!\! & \mbox{} 
        - \frac{d_1}{2F_0^2} 
            [T^\dag\ss{}{j\,abc}\, T\ss{}{j\,abc}]
            [B^\dag\ss{}{ed} B\ss{}{de}]
        - \frac{d_5}{2F_0^2}
            [T^\dag\ss{}{j\,abc} \sigma\ss{}{k}\, T\ss{}{j\,abc}]
            [B^\dag\ss{}{ed} \sigma\ss{}{k} B\ss{}{de}] 
\\* \nonumber & & \mbox{} 
        - \frac{d_2}{2F_0^2} 
            [T^\dag\ss{}{j\,abc}\, T\ss{}{j\,abd}]
            [B^\dag\ss{}{ed} B\ss{}{ce}]
        - \frac{d_6}{2F_0^2}
            [T^\dag\ss{}{j\,abc} \sigma\ss{}{k}\, T\ss{}{j\,abd}]
            [B^\dag\ss{}{ed} \sigma\ss{}{k} B\ss{}{ce}] 
\\* \nonumber & & \mbox{} 
        - \frac{d_3}{2F_0^2} 
            [T^\dag\ss{}{j\,abc}\, T\ss{}{j\,abd}]
            [B^\dag\ss{}{ce} B\ss{}{ed}]
        - \frac{d_7}{2F_0^2}
            [T^\dag\ss{}{j\,abc} \sigma\ss{}{k}\, T\ss{}{j\,abd}]
            [B^\dag\ss{}{ce} \sigma\ss{}{k} B\ss{}{ed}] 
\\* \nonumber & & \mbox{} 
        - \frac{d_4}{2F_0^2} 
            [T^\dag\ss{}{j\,abc}\, T\ss{}{j\,ade}]
            [B^\dag\ss{}{bd} B\ss{}{ce}]
        - \frac{d_8}{2F_0^2}
            [T^\dag\ss{}{j\,abc} \sigma\ss{}{k}\, T\ss{}{j\,ade}]
            [B^\dag\ss{}{bd} \sigma\ss{}{k} B\ss{}{ce}]
,
\end{eqnarray}
where~\mbox{($j$,$k$)} are vector indices, \mbox{($a$--$e$)}~are chiral
indices, and square brackets indicate sums over spinor indices\@.
Factors of~$F_0^{-2}$ are included in~\L{\rm{}b}{} to make the
coefficients~$d_i$ dimensionless\@.

The values of the eight coefficients~$d_i$ have not yet been experimentally
determined\@.
However, because the baryon helicity is conserved in the self-energy
diagrams, the terms in~\L{\rm{}b}{} of the form
\mbox{$[T^\dag\ss{}{j}\sigma\ss{}{k}\,T\ss{}{j}][B^\dag\sigma\ss{}{k}B]$} do
not generate self-energy shifts and the results are independent
of~\mbox{$d_5$--$d_8$}\@.
Values for the remaining four coefficients~\mbox{$d_1$--$d_4$} are important
for quantitative predictions\@.
Unfortunately, knowledge of the values will likely have to wait until
low-energy octet-decuplet scattering data become available\@.
In section~\ref{sec:NMconcl} we discuss constraints on the coefficients~$d_i$
from the further assumption of \su{6}~spin-flavor symmetry\@.

Two characteristics of the contact diagrams allow some predictions which are
independent of the coefficients~$d_i$\@.
In a contact diagram there is no `intermediate state' which corresponds to an
allowed decay of the decuplet baryon so the~$d_i$ do not appear in the
imaginary part of the self-energy shift\@.
Also the contact diagrams are independent of the momentum on the external
line so the \mbox{$d_i$-depend}ence is restricted to the
\mbox{$k^2$-indep}endent parts of the self-energy shifts\@.
In terms of the parameters in equation~(\ref{eq:Enm-form}), at leading order
only the helicity-independent quantity~\mbox{$\Delta{}m^*$} depends on the
coefficients~$d_i$\@.
Quantities independent of the coefficients~$d_i$, for which we present
quantitative results, are the helicity-splitting of the self-energy shifts
\mbox{$\Delta^{\sss(h)}E=\dE{}{(1/2)}-\dE{}{(3/2)}$} and the decuplet decay
rates in nuclear matter~$\rate{nm}{(h)}=\rate{vac}{}+\delta\rate{}{(h)}$\@.

\ifthesisdraft\clearpage\fi%
\section{Self-Energy Shift Results}%
\label{sec:dEresult}%
\marginal{[sec:dEresult]}%
%
From the effective Lagrangians,
equations~\mbox{(\ref{eq:NM-La},~\ref{eq:NM-Lb})}, we calculate the
self-energy shifts~$\dE{}{(h)}$ and~$\delta\rate{}{(h)}$ of the \spin{3}
decuplet baryons in nuclear matter to leading order~(${\sim}Q^3$) in the
chiral momentum expansion\@.
For convenience we introduce the two `threshold' mass scales
\marginal{[eq:2mu-def]}%
\begin{eqnarray}
\label{eq:2mu-def}
    \mu 
    & = & \sqrt{(m_\Delta - m_N)^2 - m_\pi^2 \MS\,} 
        \ \simeq\ \mbox{255~MeV} 
, \\* \nonumber
    \tmu 
    & = & \sqrt{m_K^2 - (m_{\Sigma^*} - m_N)^2 \MS\,}
        \ \simeq\ \mbox{215~MeV} 
,
\end{eqnarray}
and present the results in terms of the functions
\marginal{[eq:G(h)-def]}%
\begin{eqnarray}
\label{eq:G(h)-def}
    \g{\frac{1}{2}}(k,p,m)
    & = & \frac{1}{16m^3k^3} (m^2 \!-\! p^2 \!+\! k^2) \left\{ 
            (m^2 \!-\! p^2 \!+\! k^2)^2 + 4m^2k^2 \MS \right\}
, \\* \nonumber
    \G{\frac{1}{2}}(k,p,m)
    & = & \frac{-1}{12m^2k^2} \left\{ 3(m^2 \!-\! p^2 \!+\! k^2)^2 
            - 2k^2(p^2 \!+\! 3k^2 \!-\! 9m^2) \MS \right\}
\\* \nonumber & & \mbox{}
        + \frac{m}{p} \g{\frac{1}{2}}(k,p,m) \ln\! 
            \left| \frac{m^2 - (p - k)^2}{m^2 - (p + k)^2} \right|
, \\
    \g{\frac{3}{2}}(k,p,m)
    & = & \frac{-1}{16m^3k^3} (m^2 \!-\! p^2 \!+\! k^2) \left\{ 
            (m^2 \!-\! p^2 \!+\! k^2)^2 - 12m^2k^2 \MS \right\}
, \\* \nonumber
    \G{\frac{3}{2}}(k,p,m)
    & = & \frac{1}{12m^2k^2} \left\{ 3(m^2 \!-\! p^2 \!+\! k^2)^2 
            - 2k^2(p^2 \!+\! 3k^2 \!+\! 15m^2) \MS \right\}
\\* \nonumber & & \mbox{}
        + \frac{m}{p} \g{\frac{3}{2}}(k,p,m) \ln\! 
            \left| \frac{m^2 - (p - k)^2}{m^2 - (p + k)^2} \right|
,
\end{eqnarray}
which are constructed such that~\mbox{$\G{h}(k,p,m)\rightarrow0$}
as~\mbox{$k\rightarrow0^{\sss+}$}\@.

\begin{figure}[p]
\begin{picture}(190,170)%
    \put(10,0){\epsfxsize=180pt\epsfbox{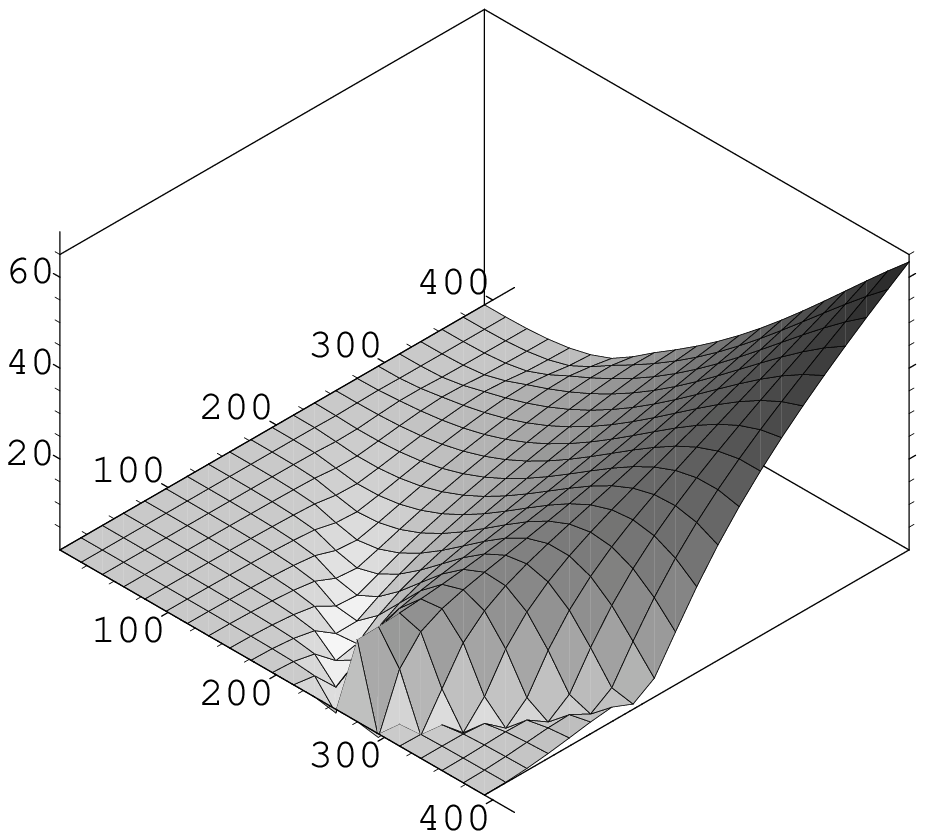}}%
    \put(0,170){\makebox(0,0)[lt]{(a)}}%
    \put(2,60){\rotatebox{90}{\scalebox{.75}{%
        $(\Delta^{\sss(h)}E_\dltM)\mbox{MeV}^{-1}$}}}%
    \put(50,25){\rotatebox{330}{\scalebox{.75}{%
        $\pFn\mbox{~MeV}^{-1}$}}}%
    \put(50,98){\rotatebox{30}{\scalebox{.75}{%
        $k\mbox{~MeV}^{-1}$}}}%
\end{picture}
\hspace*{\fill}
\begin{picture}(190,170)%
    \put(10,5){\epsfxsize=180pt\epsfbox{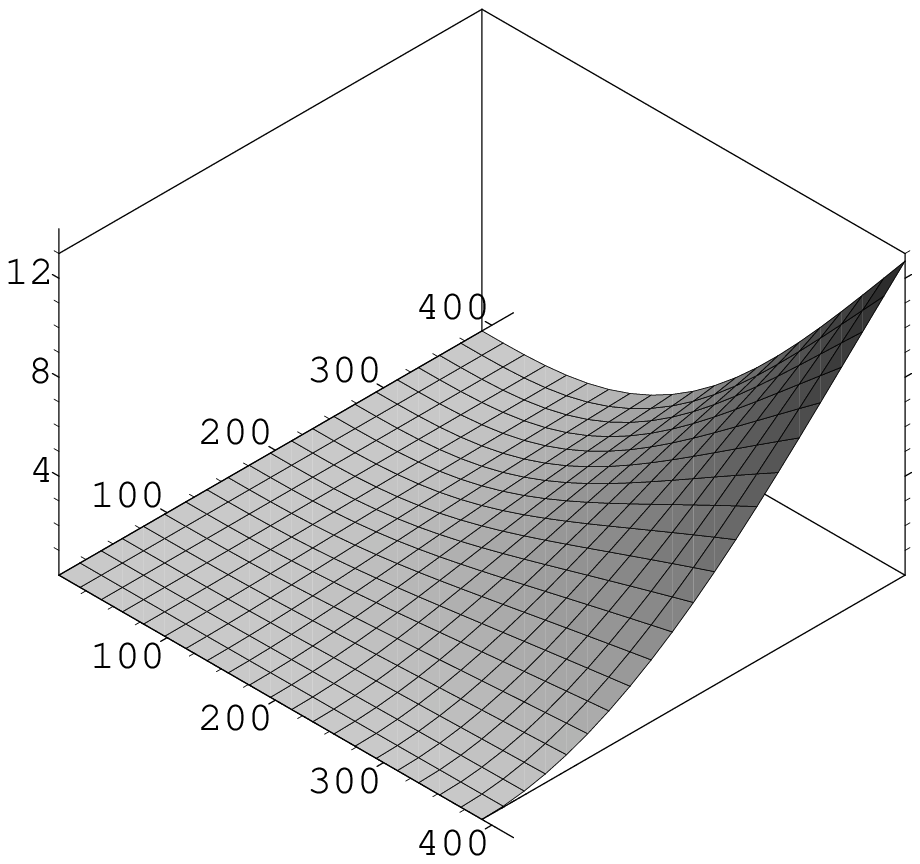}}%
    \put(0,170){\makebox(0,0)[lt]{(b)}}%
    \put(2,60){\rotatebox{90}{\scalebox{.75}{%
        $(\Delta^{\sss(h)}E_\sigSM)\mbox{MeV}^{-1}$}}}%
    \put(50,25){\rotatebox{330}{\scalebox{.75}{%
        $\pFn\mbox{~MeV}^{-1}$}}}%
    \put(50,98){\rotatebox{30}{\scalebox{.75}{%
        $k\mbox{~MeV}^{-1}$}}}%
\end{picture}
\caption{Leading-order helicity splitting of the decuplet baryon
  self-energy in nuclear matter
  \mbox{$\Delta^{\sss(h)}E=\dE{}{(1/2)}-\dE{}{(3/2)}$} as a function of
  neutron Fermi momentum~\pFn\ and baryon momentum~$k$, (a)~for the
  \dltM~baryon and (b)~for the \sigSM~baryon\@. \drafttext{\protect\\*
    \mbox{\bf[fig:NM-splits]}}}
\label{fig:NM-splits}
\end{figure}

For the negatively-charged member of each isomultiplet we find the real parts
of the self-energy shifts are
\marginal{[eq:dE-dltM]}%
\marginal{[eq:dE-sigSM]}%
\marginal{[eq:dE-xiSM]}%
\marginal{[eq:dE-omgM]}%
\begin{eqnarray}
\label{eq:dE-dltM}
    \dE{\dltM}{(h)}
    & = & \frac{1}{18\pi^2F_0^2} \left\{ 
            (3d_1)\pFp^3 \!+ (3d_1 + 3d_2)\pFn^3 \MS \right\}
\\* \nonumber & & \mbox{}
        + \frac{\C^2}{48\pi^2F_0^2} 
            \left\{ \vpf{\MS}{\MS} \!\! \right.
            \mu^3 \ln\! \left| \frac{(\mu - \pFn)^2 - k^2}{
                (\mu + \pFn)^2 - k^2} \right|
\\* \nonumber & & \mbox{} \hspace*{1in}
        + \frac{4}{3}\pFn^3 \!+ 4\mu^2\pFn 
        + \mu^2\pFn \G{h}(k, \pFn, \mu) 
        \left. \!\! \vpf{\MS}{\MS} \right\}
, \\
\label{eq:dE-sigSM}
    \dE{\sigSM}{(h)}
    & = & \frac{1}{18\pi^2F_0^2} \left\{ (3d_1 + d_3)\pFp^3 
            \!+ (3d_1 + 2d_2 + d_3 + d_4)\pFn^3 \MS \right\}
\\* \nonumber & & \mbox{}
        + \frac{\C^2}{144\pi^2F_0^2} 
            \left\{ \vpf{\MS}{\MS} \!\! \right.
            2\tmu^3 \arccos\! 
                \left( \vpf{\MS}{\MS} \!\! \right. \vc{
                \frac{\tmu^2 - \pFn^2\! + k^2}{
                    \sqrt{(\tmu^2 + \pFn\ss{}{2}\! + k^2)\ss{}{2} 
                        - 4k^2\pFn\ss{}{2} \,}}
                } \left. \!\! \vpf{\MS}{\MS} \right)
\\* \nonumber & & \mbox{} \hspace*{1in}
        + \frac{4}{3}\pFn^3 \!- 4\tmu^2\pFn 
        - \tmu^2\pFn \G{h}(k, \pFn, i\tmu) 
        \left. \!\! \vpf{\MS}{\MS} \right\}
, \\
\label{eq:dE-xiSM}
    \dE{\xiSM}{} 
    & = & \frac{1}{18\pi^2F_0^2} \left\{ (3d_1 + 2d_3)\pFp^3 
            \!+ (3d_1 + d_2 + 2d_3 + d_4)\pFn^3 \MS \right\}
, \\
\label{eq:dE-omgM}
    \dE{\omgM}{} 
    & = & \frac{1}{18\pi^2F_0^2} 
            (3d_1 + 3d_3) \left\{ \pFp^3 \!+ \pFn^3 \MS \right\} 
.
\end{eqnarray}
The helicity splitting of the~\dltM\ and \sigSM~self-energies are graphed as
functions of~\pFn\ and~$k$ in Figure~\ref{fig:NM-splits}\@.

\begin{figure}[p]
\begin{picture}(190,170)%
    \put(10,0){\epsfxsize=180pt\epsfbox{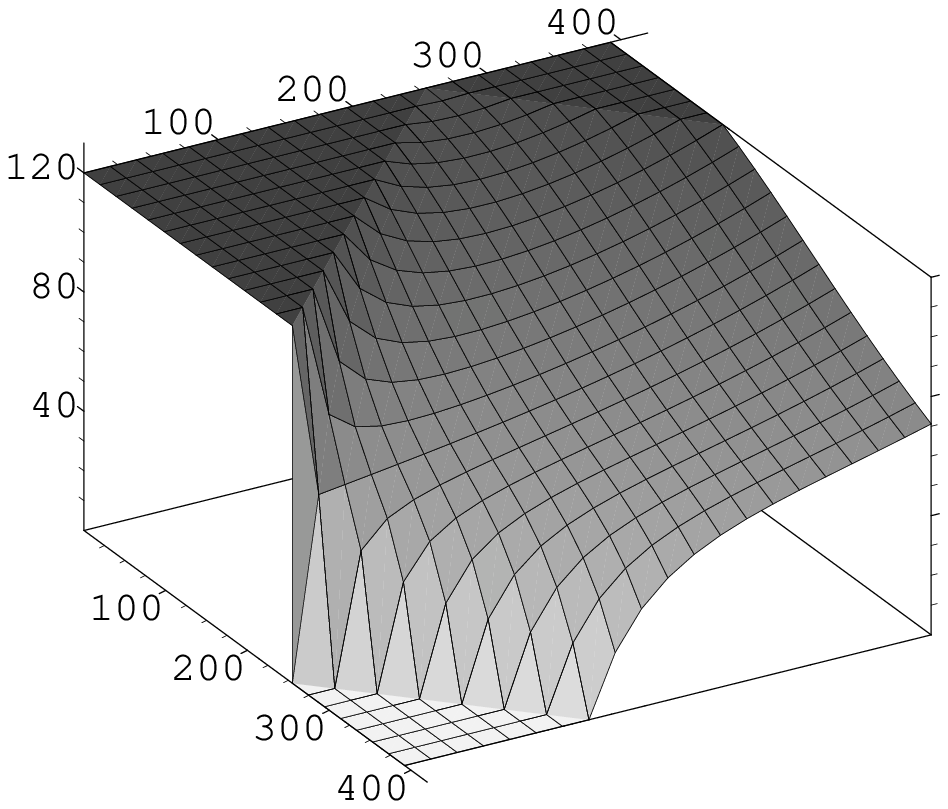}}%
    \put(0,170){\makebox(0,0)[lt]{(a)}}%
    \put(1,75){\rotatebox{90}{\scalebox{.75}{%
        $\rate{nm,\dltM}{(1/2)}\mbox{~MeV}^{-1}$}}}%
    \put(37,29){\rotatebox{321}{\scalebox{.75}{%
        $\pFn\mbox{~MeV}^{-1}$}}}%
    \put(65,153){\rotatebox{12}{\scalebox{.75}{%
        $k\mbox{~MeV}^{-1}$}}}%
\end{picture}
\hspace*{\fill}
\begin{picture}(190,170)%
    \put(10,0){\epsfxsize=180pt\epsfbox{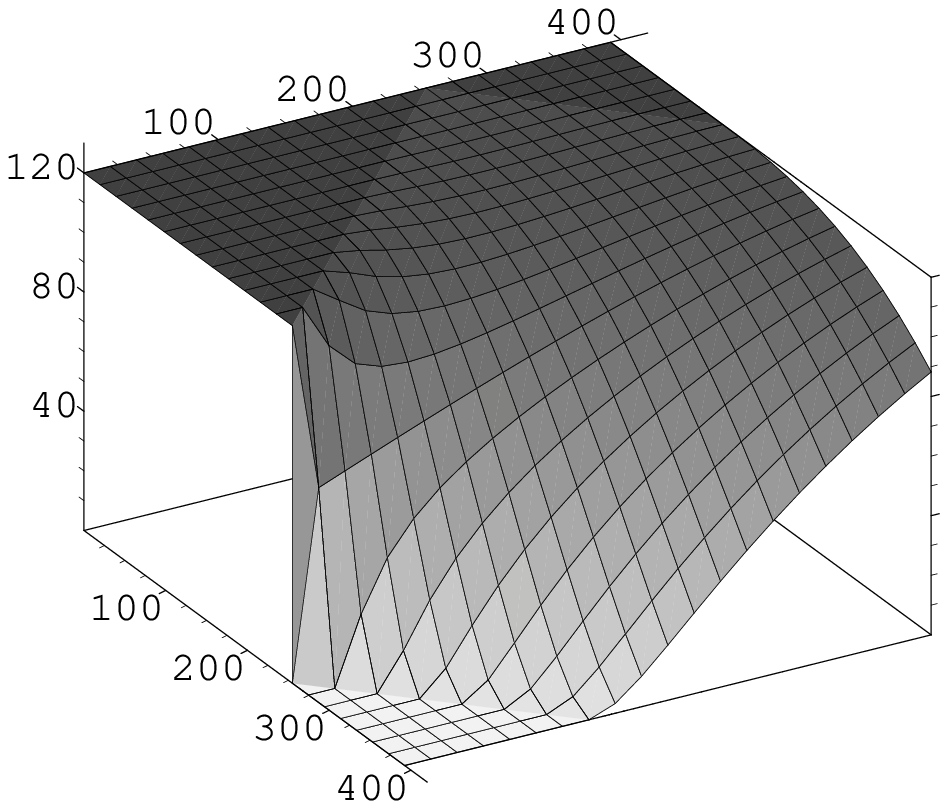}}%
    \put(0,170){\makebox(0,0)[lt]{(b)}}%
    \put(1,75){\rotatebox{90}{\scalebox{.75}{%
        $\rate{nm,\dltM}{(3/2)}\mbox{~MeV}^{-1}$}}}%
    \put(37,29){\rotatebox{321}{\scalebox{.75}{%
        $\pFn\mbox{~MeV}^{-1}$}}}%
    \put(65,153){\rotatebox{12}{\scalebox{.75}{%
        $k\mbox{~MeV}^{-1}$}}}%
\end{picture}
\caption{Leading-order \dltM~resonance widths in nuclear matter
  as functions of neutron Fermi momentum~\pFn\ and baryon momentum~$k$,
  (a)~for helicities~\mbox{$\pm\frac{1}{2}$} and
  (b)~for helicities~\mbox{$\pm\frac{3}{2}$}\@. \drafttext{\protect\\*
    \mbox{\bf[fig:NM-widths]}}}
\label{fig:NM-widths}
\end{figure}

At leading order the imaginary part of the self-energies, or resonance
widths, are shifted in nuclear matter for the $\Delta$~isomultiplet only\@.
We find the \dltM~resonance widths are
\marginal{[eq:dG-dltM]}%
\begin{eqnarray}
\label{eq:dG-dltM}
    \rate{nm,\dltM}{(h)} \!\!\!
    & = & \!\! \left\{ \MS \!\! \right. 
        \tfrac{1}{2} \left( 1 + \g{h}(k,\pFn,\mu) \right)
        \Theta\!\left( k \!+\! \pFn \!-\! \mu \right)
        \Theta\!\left( \mu^2 \!-\! (\pFn \!-\! k)^2 \right)
\\* \nonumber & & \mbox{}
        + \Theta\!\left( \mu \!-\! \pFn \!-\! k \right)
        + \Theta\!\left( k \!-\! \pFn \!-\! \mu \right)
        \left. \!\! \MS \right\} \rate{vac,\dltM}{} 
\end{eqnarray}
where~\mbox{$\rate{vac,\dltM}{}=\mu^3\C^2/12\pi{}F_0^2$} is the
\dltM~resonance width in free space at leading order\@.
Figure~\ref{fig:NM-widths} presents the resonance widths as functions
of~\pFn\ and~$k$\@.

Results for the other baryons in the decuplet can be determined from the
expressions for the negatively-charged members\@.
For the member of each isomultiplet with the most positive charge~(\dltPP,
\sigSP, and~\xiSZ) the self-energy shifts are obtained by exchanging~\pFp\ 
and~\pFn\ in
equations~\mbox{(\ref{eq:dE-dltM}--\ref{eq:dE-xiSM},~\ref{eq:dG-dltM})}\@.
The self-energy shifts of the remaining decuplet baryons are given by the
following relations:
\marginal{[eq:dE-dltP]}%
\marginal{[eq:dE-dltZ]}%
\marginal{[eq:dE-sigSZ]}%
\begin{eqnarray}
\label{eq:dE-dltP}
    \delta\E{\dltP}{} 
    & = & \tfrac{1}{3} \left(
            2 \delta\E{\dltPP}{} + \delta\E{\dltM}{} \right)
, \\
\label{eq:dE-dltZ}
    \delta\E{\dltZ}{} 
    & = & \tfrac{1}{3} \left(
            \delta\E{\dltPP}{} + 2 \delta\E{\dltM}{} \right)
, \\
\label{eq:dE-sigSZ}
    \delta\E{\sigSZ}{} 
    & = & \tfrac{1}{2} \left(
            \delta\E{\sigSP}{} + \delta\E{\sigSM}{} \right)
.
\end{eqnarray}
Clearly, in neutron-proton symmetric nuclear matter the baryons within
isomultiplets remain degenerate up to the splitting between helicity
states\@.

\ifthesisdraft\clearpage\fi%
\section{Discussion and Conclusion}%
\label{sec:NMconcl}%
\marginal{[sec:NMconcl]}%
%
\subsection{Corrections at Higher-Order}%
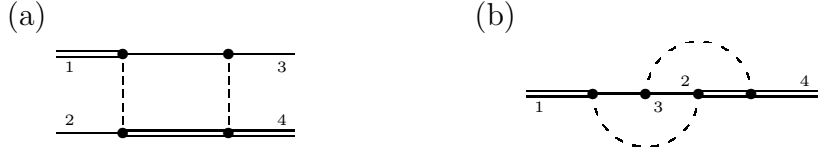
\begin{figure}[tb]
\hspace*{\fill}
\begin{picture}(90,55)%
    \put(0,55){\makebox(0,0)[rt]{(a)~}}%
    \put(3,32){\makebox(0,0)[lt]{$\sss1$}}%
    \put(0,34){\line(1,0){25}}%
    \put(0,36){\line(1,0){25}}%
    \put(25,35){\circle*{4}}%
    \put(25,35){\line(1,0){65}}%
    \put(65,35){\circle*{4}}%
    \put(87,32){\makebox(0,0)[rt]{$\sss3$}}%
    \put(3,8){\makebox(0,0)[lb]{$\sss2$}}%
    \put(0,5){\line(1,0){25}}%
    \put(25,5){\circle*{4}}%
    \put(25,4){\line(1,0){65}}%
    \put(25,6){\line(1,0){65}}%
    \put(65,5){\circle*{4}}%
    \put(87,8){\makebox(0,0)[rb]{$\sss4$}}%
    \multiput(25,8.5)(0,5){5}{\line(0,1){3}}%
    \multiput(65,8.5)(0,5){5}{\line(0,1){3}}%
\end{picture}
\hspace*{\fill}
\begin{picture}(110,55)%
    \put(0,55){\makebox(0,0)[rt]{(b)~}}%
    \put(3,17){\makebox(0,0)[lt]{$\sss1$}}%
    \put(0,19){\line(1,0){25}}%
    \put(0,21){\line(1,0){25}}%
    \put(25,20){\circle*{4}}%
    \put(25,20){\line(1,0){40}}%
    \put(45,20){\circle*{4}}%
    \put(48,17){\makebox(0,0)[lt]{$\sss3$}}%
    \put(62,23){\makebox(0,0)[rb]{$\sss2$}}%
    \put(65,20){\circle*{4}}%
    \put(65,19){\line(1,0){45}}%
    \put(65,21){\line(1,0){45}}%
    \put(85,20){\circle*{4}}%
    \put(107,23){\makebox(0,0)[rb]{$\sss4$}}%
    \savebox{\dash}(0,0){%
        \begin{picture}(0,0)%
        \put(20,-1.5){\line(0,1){3}}%
        \end{picture}}%
    \put(45,20){\rotatebox{340}{\usebox{\dash}}}%
    \put(45,20){\rotatebox{320}{\usebox{\dash}}}%
    \put(45,20){\rotatebox{300}{\usebox{\dash}}}%
    \put(45,20){\rotatebox{280}{\usebox{\dash}}}%
    \put(45,20){\rotatebox{260}{\usebox{\dash}}}%
    \put(45,20){\rotatebox{240}{\usebox{\dash}}}%
    \put(45,20){\rotatebox{220}{\usebox{\dash}}}%
    \put(45,20){\rotatebox{200}{\usebox{\dash}}}%
    \put(65,20){\rotatebox{160}{\usebox{\dash}}}%
    \put(65,20){\rotatebox{140}{\usebox{\dash}}}%
    \put(65,20){\rotatebox{120}{\usebox{\dash}}}%
    \put(65,20){\rotatebox{100}{\usebox{\dash}}}%
    \put(65,20){\rotatebox{80}{\usebox{\dash}}}%
    \put(65,20){\rotatebox{60}{\usebox{\dash}}}%
    \put(65,20){\rotatebox{40}{\usebox{\dash}}}%
    \put(65,20){\rotatebox{20}{\usebox{\dash}}}%
\end{picture}
\hspace*{\fill}
\caption{(a)~Infrared divergent baryon-meson box diagram and (b)~overlapping
  meson-loop diagram obtained from the box diagram by contracting nucleon
  lines~2 and~3\@.  The two-loop diagram in~(b) avoids the infrared
  divergence of~(a)\@. \drafttext{\protect\\*
    \mbox{\bf[fig:NM-2loop]}}}
\label{fig:NM-2loop}
\end{figure}

\label{sub:NM-HO}%
\marginal{[sub:NM-HO]}%
%
Contrary to the discussion in our prior work, the leading corrections to our
results are not from infrared divergent two-loop diagrams\@.
We previously thought diagrams with overlapping meson loops would be
afflicted with the infrared divergence that appears in baryon-meson box
diagrams, see Figure~\ref{fig:NM-2loop}\@.
That the two-loop diagrams avoid the divergence is most easily seen in terms
of time-ordered perturbation theory as discussed by
\mbox{Weinberg}~\cite{SW:nuclI,SW:nuclII}, where field propagators and
integrals over four-momenta are replaced by intermediate-state energies in 
the denominator and integrals over three-momenta\@.
In the time-ordered approach, infrared divergences arise from intermediate
states with `small' energies of order~$Q^2/m_B$; however, in loop diagrams
contributing to the proper self-energy~$\Sigma(k\ss{0}{},\vec{k}\,)$ all
intermediate states contain mesons with energies on the characteristic
scale~$Q$\@.
Consequently the two-loop diagram in Figure~\mbox{\ref{fig:NM-2loop}(b)} is
governed by equation~(\ref{eq:stat-rule}) and is suppressed relative to the
leading-order result by~\mbox{$(Q/\Lchi)^2$} and not
by~\mbox{$m_BQ/\Lambda_\chi^2\sim{}Q/\Lchi$}\@.

The leading-order results~(${\sim}Q^3$) presented in
section~\ref{sec:dEresult} are determined from the difference
\mbox{$\pse{nm}(\Delta{}m,\vec{k}\,)-\pse{vac}(\Delta{}m,\vec{k}\,)$}\@.
A priori, we would expect corrections at order~$Q^4$ from three sources:
1)~the chiral expansion of the proper self-energies~\pse{nm} and~\pse{vac}
based on the power counting of equation~(\ref{eq:stat-rule}), 2)~the
expansion of~\pse{nm} about~\E{vac}{} in equation~(\ref{eq:dE-calc}), and
3)~the \mbox{$Q$-expansion} of~\E{vac}{}
as~\mbox{$\E{vac}{}=\Delta{}m+k^2/2m_T+\mathcal{O}(Q^3)$}\@.
In the following two paragraphs we argue that the corrections from the first
two expansions do not occur until order~$Q^5$\@.
The only corrections of order~$Q^4$ arise from the third expansion and are
included by making the replacements
\marginal{[eq:mu-shift]}%
\begin{eqnarray}
\label{eq:mu-shift}
    \mu & \longrightarrow & 
        \mu + \left( \frac{\Delta m}{\mu} \right) \frac{k^2}{2 m_T}
\\* \nonumber
    \tmu & \longrightarrow &
        \tmu - \left( \frac{\Delta m}{\tmu} \right) \frac{k^2}{2 m_T}
\end{eqnarray}
in the leading-order results for the~$\Delta$ and $\Sigma^*$~isomultiplets\@.
The results for the $\Xi^*$~isomultiplet and the \omgM~baryon do not receive
corrections at order~$Q^4$\@.

The chiral expansion of the proper self-energies~\pse{nm} and~\pse{vac} would
contribute at order~$Q^4$ through one-loop diagrams obtained from
Figure~\ref{fig:NMdiags} by replacing a single vertex with one which
satisfies \mbox{$d+\frac{1}{2}n=3$}\@.
The only relevant operators, invariant under rotations,~P, and~\Gchi, are
constructed from products of the
form~\mbox{$(i\lpartial\ss{0}{})TB^\dag\Amu$} for the meson-nucleon loop
diagrams or~\mbox{$(i\lpartial\ss{0}{})TB(TB)^\dag$} for the contact
diagrams\@.
Partial integration and baryon field redefinitions can be used together to
eliminate both classes of operators in favor of terms higher-order in the
expansion of the effective Lagrangian\@.
(Such field redefinitions are reviewed in Chapter~\ref{ch:heavyK}\@.)
Consequently, there are no corrections at order~$Q^4$ from the chiral
expansion\@.
\mbox{Order-$Q^5$} corrections arise from one-loop diagrams with a single
vertex which satisfies \mbox{$d+\frac{1}{2}n=4$} and two-loop diagrams, such
as in Figure~\mbox{\ref{fig:NM-2loop}(b)}\@.

The primary corrections from the expansion of~\pse{nm} about~\E{vac}{} are
suppressed by powers of~$Q$ contributed
by~\mbox{$\partial\pse{nm}/\partial{}k\ss{0}{}$} in
equation~(\ref{eq:dE-calc})\@.
Because the \mbox{$Q$-expansion} of~\pse{nm} begins at order~$Q^2$, e.g.,\ 
through an insertion of the kinetic energy operator, we would expect the
corrections to be suppressed by~$Q$, or to contribute at order~$Q^4$\@.
However, all \mbox{$k\ss{0}{}$-depend}ence of~\pse{nm} can be moved into loop
diagrams which are order~$Q^3$ or higher by suitable redefinitions of the
decuplet baryon field\@.
Thus, \mbox{$\partial\pse{nm}/\partial{}k\ss{0}{}\sim{}Q^2$} and the primary
correction contributed by the expansion of~\pse{nm} about~\E{vac}{} is
order~$Q^5$\@.

Throughout this calculation we have made the assumption that the power
counting in equation~(\ref{eq:stat-rule}) applies\@.
However, large scattering lengths for nucleon-nucleon or nucleon-delta
interactions will require a revised power counting scheme,\footnote{We
thank M. Wise for bringing this point to our attention\@.} as seen in
references~\cite{K-S-W:NNintI,K-S-W:NNintII}\@.
Using the revised power counting for the problem considered here is
presently not tractable\@.
How these effects may be taken into account remains to be seen; the
work presented here is the best that can be done at the present time
and uses the power counting of equation~(\ref{eq:stat-rule})\@.
We hope that the discussion of some technical details, particularly
those presented in the next subsection, will be useful in further
studies of the decuplet self-energy\@.

\subsection{$\Delta$ Self-Energy Near Saturation Density}%
\label{sub:dlt-sat}%
\marginal{[sub:dlt-sat]}%
%
One important feature of the self-energy shifts of the $\Delta$~isomultiplet,
not apparent in Figure~\mbox{\ref{fig:NM-splits}(a)}, is a logarithmic
divergence when~\mbox{$\vec{k}=0$} at a `threshold' Fermi momentum~$p_F^*$
(coincident with the discontinuity of the widths)\@.
The divergence arises due to degeneracy of the $\Delta$~baryon with a pion
and a nucleon on the surface of the Fermi sea; for static baryons
\mbox{$p_F^*=\mu$}, which is just the solution of
\mbox{$m_\Delta=m_N+\sqrt{m_\pi^2+p_F^2\,}$}\@.
To the extent that the divergence complicates the analysis of the self-energy
shifts, it is unfortunate that~\mbox{$p_F^*\simeq\mbox{255~MeV}$} is so close
to the Fermi momentum associated with nuclear matter at saturation
density,~\mbox{$p_F^{(sat)}\simeq\mbox{262~MeV}$}~\cite{Wong:book}\@.
In contrast, the $\Sigma^*$~isomultiplet self-energy shifts do not have a
similar divergence because the \mbox{$\Sigma^*\rightarrow{}NK$} decay is
kinematically forbidden for any~\pFn\ and~\pFp\@.

In the \dltM~self-energy shift, equation~(\ref{eq:dE-dltM}), the divergence
manifests itself as a cancellation among quantities of order~$Q$ in the
argument of the logarithm, i.e.,~when
\marginal{[eq:Qcancel]}%
\begin{equation}
\label{eq:Qcancel}
    \left. \left( \sqrt{\ms\smash{\E{}{}^2 - m_\pi^2 \,}} - p_F 
        \right) \right|_{\E{}{} = \Delta m}
    = \mu - p_F = 0
.
\end{equation}
The divergence is resolved by keeping higher-order terms in the argument of
the logarithm, specifically the imaginary part in the expansion of~\E{}{}
about~$\Delta{}m$\@.
When the imaginary part is kept, the logarithmic divergence is replaced by a
logarithmic enhancement of the self-energy shift and the result is
order~\mbox{$Q^3\ln{Q^2}$}\@.
For Fermi momenta sufficiently below~$p_F^*$, the cancellation is not
significant and we expect equation~(\ref{eq:dE-dltM}) to provide a good
description of the \dltM~self-energy shift\@.
The effect of the cancellation becomes important when
\mbox{$\mu-p_F\sim{}Q^3$}, or approximately for
\mbox{$p_F\gtrsim(0.7)p_F^*\simeq\mbox{180~MeV}$}\@.

The logarithmically divergent term can be traced to the proper self-energy in
nuclear matter~\mbox{$\pse{nm}(\E{}{},0)$}\@.
At the threshold Fermi momentum~$p_F^*$, \pse{nm} is singular
when~\mbox{$\E{}{}=\Delta{}m$} and the expansion in
equation~(\ref{eq:dE-calc}) fails\@.
To calculate the self-energy~\E{nm}{} in the vicinity of~$p_F^*$, we
evaluate~\pse{nm} at~\E{nm}{} and expand~\mbox{$\pse{vac}(\E{vac}{},0)$}
about~\E{nm}{} (\pse{vac} is smooth in this region),
\marginal{[eq:dE-calc2]}%
\begin{eqnarray}
\label{eq:dE-calc2}
    \E{nm}{} 
    & = & \E{vac}{} + \pse{nm}(\E{nm}{},0) - \pse{vac}(\E{nm}{},0)
\\* \nonumber & & \mbox{}
        + \left( \E{nm}{} - \E{vac}{} \right) 
                \frac{\partial}{\smash{\partial k \ss{0}{}}\ms}
                    \pse{vac}(k\ss{0}{},0)
                    \left. \vpf{\ms}{\ms} \right|_{
                        k\ss{0}{} = \E{nm}{}}
        + \cdots
\end{eqnarray}
As in equation~(\ref{eq:dE-calc}), all the terms of order~$Q^3$ appear on the
first line and the second line may be neglected at leading order\@.

\begin{figure}[tb]
\centering
\begin{picture}(260,145)%
    \put(10,0){\epsfxsize=230pt\epsfbox{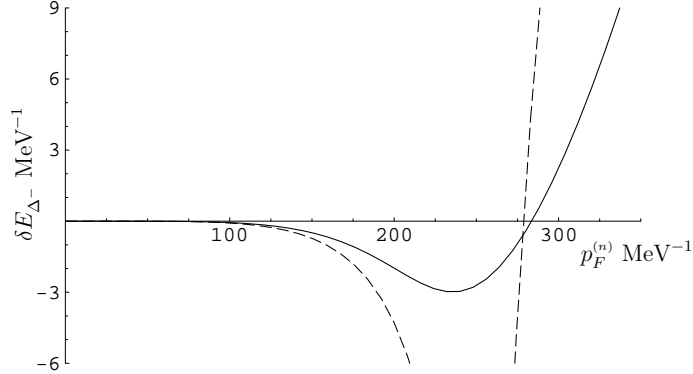}}%
    \put(2,50){\rotatebox{90}{\scalebox{.75}{%
        $\dE{\dltM}{}\mbox{~MeV}^{-1}$}}}%
    \put(260,40){\scalebox{.75}{\makebox(0,0)[rb]{%
        $\pFn\mbox{~MeV}^{-1}$}}}%
\end{picture}
\caption{Self-energy of the \dltM~baryon at rest in nuclear matter, omitting 
  contact diagrams~\mbox{($d_1=d_2=0$)}\@.  The solid curve is the numerical
  solution of equation~(\ref{eq:dE-calc2}); the dashed curve is a plot of
  equation~(\ref{eq:dE-dltM})\@.  \drafttext{\protect\\*
    \mbox{\bf[fig:log-div]}}}
\label{fig:log-div}
\end{figure}

In equation~(\ref{eq:dE-calc2}) we pay a price to determine the
self-energy~\E{nm}{} near the threshold Fermi momentum; \E{nm}{}~is the
solution of a transcendental equation and must be found numerically\@.
Figure~\ref{fig:log-div} presents a numerical calculation of the self-energy
shift~$\dE{\dltM}{}$ compared with the divergent behavior of
equation~(\ref{eq:dE-dltM})\@.
Both curves are plotted as functions of~\pFn\ with~\mbox{$\vec{k}=0$} and we
assume the coefficients of the contact terms are~\mbox{$d_1=d_2=0$}\@.
Because the contributions of the contact diagrams are effectively omitted,
the results plotted in the figure are not model-independent in the sense of
effective field theory\@.
Until the values of the coefficients~$d_1$ and~$d_2$ are determined, there is
little motivation for further numerical study of
equation~(\ref{eq:dE-calc2})\@.

Having resolved the logarithmic divergence in the real part of the
\dltM~self-energy shift, next we consider the behavior of the resonance
width~\rate{nm,\dltM}{} near~$p_F^*$\@.
The basic features of the plots in Figure~\ref{fig:NM-widths} are easily
understood in terms of the nucleon momentum in
\mbox{$\Delta\rightarrow{}N\pi$}~decay\@.
When the momentum of the \dltM~baryon is~$\vec{k}$, then up to recoil
corrections the nucleon and pion momenta in the final state are
respectively~\mbox{$\vec{k}+\vec{\mu}$} and~\mbox{$-\vec{\mu}$},
where~\mbox{$|\vec{\mu}\,|$} equals the mass parameter~$\mu$\@.
Naively, when \mbox{$|\vec{k}\,|+\mu\leq{}p_F$} all the nucleon final states
are occupied by nucleons in the Fermi sea; the \dltM~baryon becomes stable
against \mbox{$\Delta\rightarrow{}N\pi$}~decay and the width in nuclear
matter vanishes\@.
When \mbox{$p_F\leq\vbar{.3}|\vec{k}\,|-\mu\vbar{.3}$} all the nucleon states
are available for decay and the width in nuclear matter is the same as the
free-space width~\rate{vac}{}\@.
For values of~$p_F$ between those two cases, the Fermi sea partially obscures
the shell of nucleon momenta and \mbox{$0<\rate{nm,\dltM}{}<\rate{vac}{}$}\@.
When~\mbox{$\vec{k}=0$} the width transitions abruptly from~\rate{vac}{}
(for~\mbox{$p_F<\mu$}) to zero (for~\mbox{$p_F>\mu$}) creating the
discontinuities in Figure~\ref{fig:NM-widths}\@.

The pitfall in the simple analysis of the previous paragraph is determining
the momentum transfered to the intermediate-state nucleon~$\vec{\mu}$ in
terms of the free-space energy of the $\Delta$~baryon\@.
More rigorous consideration of equation~(\ref{eq:Enm-def}) shows that the
$\Delta$~resonance width is non-zero for any~$p_F$, although it may decrease
sharply near~$p_F^*$\@.
Setting~$\vec{k}=0$, \E{nm}{}~is the solution of
\marginal{[eq:Enm-calc2]}%
\begin{eqnarray}
\label{eq:Enm-calc2}
    \Delta m 
    & = & \E{}{} - \pse{nm}(\E{}{})
\\* \nonumber
    & = & \frac{\, \C^2 (m_\pi^2 - \E{}{}^2)^{\frac{3}{2}}}{24\pi^2F_0^2} 
            \left\{ \arccos\! \left( \frac{-\E{}{}}{m_\pi} \right)
                - 2 \arctan\! \left( \frac{p_F}{\sqrt{m_\pi^2-\E{}{}^2\,}} 
                    \right) \right\} + f(\E{}{},p_F)
,
\end{eqnarray}
which is analytically continued to~\mbox{$\E{}{}>m_\pi$} by
assigning~$m_\pi^2$ an infinitesimal imaginary part,
\mbox{$m_\pi^2\rightarrow{}m_\pi^2-i\epsilon$}\@.
The last term,~\mbox{$f(\E{}{},p_F)$}, is an analytic function of~\E{}{}
below the threshold for $N\pi\pi$~production from a virtual $\Delta$~baryon
and does not influence our reasoning\@.
The only real-valued solutions of equation~(\ref{eq:Enm-calc2}) satisfy
either \mbox{$\E{}{}<m_\pi$} or \mbox{$p_F>\sqrt{\E{}{}^2-m_\pi^2\,}$};
however, as~$p_F$ is decreased with~$\Delta{}m$ fixed, none of these solutions
flow continuously to the physical, complex-valued solution for small~$p_F$\@.
Starting with~\E{vac}{} at~\mbox{$p_F=0$}, as~$p_F$ is increased the physical
solution bypasses the region of real-valued solutions by passing onto a
different sheet of the Riemann surface\@.

The main result of this subsection is the logarithmic divergence
of~\dE{nm,\dltM}{} and the discontinuity of~\rate{nm,\dltM}{}, seen in
section~\ref{sec:dEresult}, can be removed by changing how the evaluation
of~\pse{nm} is handled\@.
Quantitative predictions for~\E{nm}{} near~$p_F^*$, by numeric solution of
equation~(\ref{eq:dE-calc2}) or~(\ref{eq:Enm-calc2}), require values for the
coefficients~$d_1$ and~$d_2$\@.
Independent of~$d_1$ and~$d_2$, however, we conclude that the resonance width
of the $\Delta$~isomultiplet is non-zero for any~$p_F$; the decay
channel~\mbox{$\Delta\rightarrow{}N\pi$} is never fully blocked by the
presence of the nuclear medium\@.

\subsection{Conclusion}%
\label{sub:NMconcl}%
\marginal{[sub:NMconcl]}%
%
To complete the description of the self-energy shifts of the \spin{3}
decuplet baryons in nuclear matter, the coefficients~$d_i$ of the contact
terms must still be determined\@.
In the absence of low-energy octet-decuplet scattering data, one way to
estimate the values of the coefficients may be through appealing to a larger
symmetry group, in this case the approximate \su{6}~spin-flavor symmetry\@.
Because the \spin{1} octet and \spin{3} decuplet baryons form a single
56-dimensional representation of spin-flavor~\su{6}, the \su{6}-invariant
Lagrangian determines the octet-decuplet coefficients~\mbox{$d_1$--$d_8$},
the Savage-Wise coefficients~\mbox{$c_1$--$c_6$} for octet-octet
interactions~\cite{S-W:hyperon}, and 14 coefficients for decuplet-decuplet 
interactions\@.
The \su{6}-invariant Lagrangian contains only two four-baryon contact terms
of dimension six, with coefficients~$a$ and~$b$ as defined by \mbox{Kaplan}
and \mbox{Savage}~\cite{K-S:su6}\@.
In terms of~$a$ and~$b$, the octet-decuplet coefficients in
equation~(\ref{eq:NM-Lb}) are
\marginal{[eq:SU6coeff]}%
\begin{equation}
\label{eq:SU6coeff}
    \begin{array}[t]{l@{\mathsp}l}
    d_1 = 2 a + \tfrac{5}{9} b , &
    d_5 = \tfrac{1}{9} b , \\*
    d_2 = -\tfrac{5}{9} b , &
    d_6 = -\tfrac{1}{3} b , \\*
    d_3 = -\tfrac{5}{9} b , &
    d_7 = -\tfrac{1}{9} b , \\*
    d_4 = -\tfrac{2}{9} b , &
    d_8 = -\tfrac{2}{9} b .
    \end{array}
\end{equation}
Unfortunately, neither~$a$ nor~$b$ is reliably known; therefore, for the
phenomenologically-interesting $\Delta$~isomultiplet, we have exchanged
unknowns~$d_1$ and~$d_2$ for unknowns~$a$ and~$b$\@.
What we gain from spin-flavor~\su{6} is relations between the octet-octet
coefficients~$c_i$ and the octet-decuplet coefficients~$d_i$\@.
A determination of a subset of the coefficients~$c_i$ from low-energy
octet-octet scattering would permit an estimate of the desired octet-decuplet
coefficients~$d_i$\@.

In summary, we have calculated the leading-order shift in the self-energy of
the \spin{3} decuplet baryons in nuclear matter\@.
Our work differs in two ways from earlier calculations of the
$\Delta$~isomultiplet self-energies in nuclear matter; we use chiral~\su{3}
symmetry to extend the calculation to include~$\Sigma^*$, $\Xi^*$, and~\omgM
baryons and find new momentum-independent contributions from four-baryon
operators in the effective Lagrangian\@.
We have identified quantities independent of the coefficients~$d_i$,
the helicity-splitting of the~$\Delta$ and~$\Sigma^*$ self-energy
shifts and the resonance width of the $\Delta$~isomultiplet, which are
presented in Figures~\ref{fig:NM-splits} and~\ref{fig:NM-widths}\@.
In section~\ref{sec:NMconcl} we discuss the origin and resolution of a
logarithmic divergence of the $\Delta$~self-energy near nuclear
saturation density, which is particularly relevant for future work
when the coefficients~$d_1$ and~$d_2$ are known\@.
The two major short-comings of our results are that the
coefficients~$d_i$ of the four-baryon operators have not yet been
determined and possible effects of the large nucleon-nucleon
scattering lengths have not been included\@.


%% file: heavyK.tex
\chapter[Heavy Kaon/Eta Effective Theory]{\\Heavy Kaon/Eta Effective Theory}%
\label{ch:heavyK}%
\marginal{[ch:heavyK]}%
%
In this chapter an effective field theory for treating kaon and eta
interactions in a non-relativistic framework is developed\@.
The relevant effective Lagrangians are derived to an order sufficient for
one-loop calculations of~$\pi\eta$, $\pi{}K$, and~$KK$ scattering
processes\@.
Coefficients in the effective Lagrangians at low orders are determined from a
matching calculation with \su{3}~\cpt\ and used to predict $KK$~scattering
phase shifts\@.

In section~\ref{sec:prospects} we present the motivation for developing the
non-relativistic theory, describe the three key ideas which form the
foundation, and outline the overall program to be followed\@.
Sections~\ref{sec:HKeft} and~\ref{sec:buildL} establish the elements and
principles from which the effective Lagrangian is developed then detail the
construction of the first several orders of the Lagrangian\@.
The coefficients of the effective Lagrangian in the lowest orders of the
chiral expansion are determined from a matching calculation in
section~\ref{sec:matching}\@.
A prediction of the \mbox{$s$-wave} $KK$~scattering phase shift and brief
concluding remarks on the outlook for the heavy kaon/eta effective theory are
contained in section~\ref{sec:KKphase}\@.

\ifthesisdraft\clearpage\fi%
\section{Prospectus}%
\label{sec:prospects}%
\marginal{[sec:prospects]}%
%
For the range of momenta~\mbox{$Q\lesssim{}m_\eta$} the rate of convergence
of the chiral expansion in \su{3}~chiral perturbation theory is limited by
approximately the ratio of mass scales \mbox{$(m_\eta/\Lchi)^2\sim0.3$}\@.
For processes involving only pions with momenta~\mbox{$Q\lesssim{}m_\pi$}
chiral perturabation theory based on a smaller
\mbox{$\su{2}\sub{L}\times\su{2}\sub{R}$} chiral symmetry provides an
alternative description\@.
One advantage of using \su{2}~\cpt\ for low-energy pion interactions is a
better rate of convergence; generalizing the power counting arguments in
section~\ref{sec:cpt} the expansion parameter is approximately
\mbox{$(m_\pi/m_K)^2\sim0.1$}\@.
However, compared to \su{3}~\cpt, a drawback of \su{2}~\cpt\ is the great
reduction in the variety of processes to which the theory can be applied with
only a slight reduction in the number of coefficients to be determined in the
Lagrangian\@.
Today, \su{2}~\cpt\ and \su{3}~\cpt\ are considered complementary
descriptions of pion interactions at low-energy and the relations between the
coefficients in the two \mbox{order-$Q^4$} effective Lagrangians are
known~\cite{G-L:su3}\@.

Motivated by the comparison above, we seek an effective field theory tailored
for low-energy interactions \mbox{($Q\lesssim{}m_\pi$)} of pions with kaons
or eta mesons to exploit the relatively small ratio~\mbox{$m_\pi/m_K$}\@.
One key ingredient of the effective field theory, as an expansion
in~\mbox{$1/m_K$}, is a non-relativistic treatment of the kaon and eta
degrees of freedom\@.
Because the momentum scale dictates a relativistic treatment of the pion
field, we cannot build the \mbox{$\Gchi=\su{3}\sub{L}\times\su{3}\sub{R}$}
chiral symmetry of~\QCD\ into the effective Lagrangian explicitly; flavor
rotations in~\su{3}\sub{V} would map relativistic pions into non-relativistic
kaons and vice versa\@.
However, we are able to include the smaller
\mbox{$\GchiP=\su{2}\sub{L}\times\su{2}\sub{R}$} chiral symmetry of~\QCD\ and
a description of the pion degrees of freedom based on \su{2}~\cpt\ becomes
the second key ingredient of the effective field theory\@.

The non-relativistic field theory naturally divides into a number of
\mbox{$n$-body} sectors distinguished by strangeness and the number of heavy
fields~$n$ in the initial state\@.
Sectors of the theory in which the number of heavy fields is not conserved
present a problem for the momentum expansion; the annihilation of a heavy
particle introduces pions with `hard' momenta~(${\sim}m_{K,\eta}$) in
intermediate states and the power counting scheme breaks down\@.
This problem afflicts sectors involving either $K\kbar$~pairs or the eta
meson, which is subject to pion conversion as
in~\mbox{$\eZ{}K\rightarrow{}\pi{}K$}\@.
The exception to the general rule proscribing the eta meson is for a single
eta meson interacting with any number of pions; in that case the eta meson is
protected by \mbox{$G$-parity}~\cite{G-L:eta}\@.
(When isospin-violating effects are included, the
decays~\mbox{$\eZ\rightarrow3\pi$} create new problems in the single-eta
sector\@.)
We focus on applying the effective field theory to the unaffected one- and
two-body sectors, i.e.,\ to $\pi\eZ$, $\pi{}K$,~$KK$, and the sectors 
related by charge conjugation\@.

When constructing the Lagrangian for the heavy kaon/eta effective theory, we
must introduce coefficients, or low-energy constants, which are not
determined by the imposed symmetries\@.
Even at low orders in the chiral expansion, several low-energy constants will
appear in each of the independent sectors we consider\@.
Due to the scarcity of experimental data in some sectors, the standard method
for extracting values for the low-energy constants is impractical\@.
The alternative is to pursue a matching calculation onto a better-known
theory at higher energy to determine the low-energy constants in terms of
known parameters of the high-energy theory\@.
In the regime of non-perturbative~\QCD, \su{3}~\cpt\ is the natural choice
for the `high-energy' theory\@.
An additional benefit of matching onto \su{3}~\cpt\ is that the full
\Gchi~chiral symmetry is implicitly restored to the effective Lagrangian
through constraints on the low-energy constants\@.
Because the matching calculation enhances the predictive power of the heavy
kaon/eta effective theory, we consider it the third (and final) key 
ingredient of the theory\@.
The essence of heavy kaon/eta effective theory is a reorganization of the
chiral expansion in \su{3}~\cpt, keeping order-by-order only the terms
relevant to the low-energy regime, which should result in improved
convergence\@.

The steps in the program we follow are to establish the effects of symmetry
transformations on the fields of the theory, detail the rules for
constructing a minimal effective Lagranian, build the Lagrangian for the
sectors of interest to order~$Q^4$, perform the matching calculation onto
\su{3}~\cpt, extract values for the low-energy constants of the heavy
kaon/eta effective theory, apply the results to scattering problems near
threshold, and finally look for further problems to which the theory can be
profitably applied\@.
For the $\pi{}K$~sector, this program has been started independently by
\mbox{Roessl}~\cite{AR:heavyK}; however, we find he has omitted terms from 
the effective Lagrangian at higher order~($\geq{}Q^3$)\@.
Several topics relevant to developing the effective field theory are
addressed in the discussion: field redefinitions (or use of equations of
motion); reparameterization invariance constraints on the effective
lagrangian; reconciling relativistic and non-relativistic treatments in the
matching calculations; and consequences of differences between the power
counting schemes of \su{3}~\cpt\ and of the heavy kaon/eta effective
theory\@.

To conclude a section titled ``Prospectus'' by considering future
applications for the heavy kaon/eta effective theory seems appropriate, so we
briefly skip ahead to the final step in the program\@.
While the effective Lagrangians written out for the~$\pi{}K$ and~$KK$ sectors
include isospin-violating terms for generality, the values of the associated
low-energy constants have not been determined in the matching calculations\@.
Extending the matching to determine the low-energy constants for isospin
violation and electromagnetic interactions would be straight forward and
suitable for studying processes such as
\mbox{$\pi{}K\rightarrow\gamma\pi{}K$} and
\mbox{$\gamma{}K\rightarrow\pi\pi{}K$} near threshold\@.
In another application, baryon isomultiplets from the \spin{1} octet and
\spin{3} decuplet can be coupled to the heavy kaon effective theory with the
coefficients in the Lagrangian matched to the \mbox{\Gchi-invariant} theory
in sections~\ref{sec:matter},~\ref{sec:counting}\@.
Unfortunately, in many interesting kaon-baryon systems, strangeness-exchange
reactions do not conserve the number of heavy particles,
\mbox{e.g.,~$N\kbar\rightarrow\Sigma\pi$~\cite{MS:piK1405}}, resulting in a
breakdown of the power counting scheme; however, the $NK$ and~$\Delta{}K$
sectors are protected by strangeness conservation and are compatible with the
heavy-kaon framework\@.
On a more speculative note, phenomenological models inspired by and
incorporating elements of the heavy kaon/eta effective theory may be
developed for the problematic sectors like~$K\kbar$~\cite{G-O:KKbar}\@.
For example, similar work has been done on
$\piP\piM$~atoms~\cite{GGRL:pionium}, and by integrating out $\piZ$~pairs and
allowing for a non-hermitian effective Lagrangian~\cite{E-S:pionium}\@.
Finally, if experimental data for the threshold energy region become
available in abundance, extracting the low-energy constants of the heavy
kaon/eta effective theory from the data may provide insights into
\su{3}~\cpt\@.
Knowing values for the low-energy constants would permit verification of the
constraints on the low-energy constants imposed by the matching calculation
and would provide a means of testing the \su{3}~chiral expansion\@.

\ifthesisdraft\clearpage\fi%
\section{Elements and Principles}%
\label{sec:HKeft}%
\marginal{[sec:HKeft]}%
%
The Lagrangian of heavy kaon/eta effective theory is divided into sectors by
the heavy fields present,
\marginal{[eq:L-HKeft]}%
\begin{equation}
\label{eq:L-HKeft}
    \L{}{}
    = \L{}{\pi\pi} + \L{}{\pi K} + \L{}{\pi\kbar} + \L{}{\pi\eta}
        + \L{}{\pi KK} + \cdots
\end{equation}
and the Lagrangian for each sector is subdivided as
\mbox{$\L{}{X}=\sum_j\L{j}{X}$} in a chiral expansion\@.
All the purely pionic interactions are derived from~\L{}{\pi\pi}, where
\mbox{$j=2,4,\ldots\:$}, which is identically the Lagrangian of
\su{2}~\cpt\@.
The effective Lagrangians for interactions of a single heavy field with any
number of pions are characterized by the chiral expansion
\mbox{$j=1,2,3,\ldots\:$}; the expansion of~\L{}{\pi{}KK} is similar except
that the expansion starts with~\mbox{$j=0$} due to heavy-field contact
terms\@.
In this section we briefly establish our notation and identify the 
principles used to construct the effective Lagrangians\@.

The building blocks of the heavy kaon/eta effective theory are introduced in
direct analogy to \su{3}~\cpt\@.
The three pseudo-Goldstone bosons of \su{2}~\cpt\ are introduced in the
exponential representation; the most elementary constituents
are~\mbox{$u=e^{i\Phi/2F}$} and~\mbox{$\chi=2B\Mq$} where
\marginal{[eq:u,chi-2def]}%
\begin{equation}
\label{eq:u,chi-2def}
    \Phi
    = \pi\ss{a}{} \tau\ss{a}{}
    = \left(
        \begin{array}{cc}
        \piZ & \sqrt{2}\, \piP \\*
        \sqrt{2}\, \piM & -\piZ
        \end{array} \right)
, \mathsp
    \Mq
    = \left(
        \begin{array}{cc}
        m_u & 0 \\* 0 & m_d
        \end{array} \right)
,
\end{equation}
and the parameters~$F$ and~$B$ are non-trivially related to the \su{3}~\cpt\ 
parameters~$F_0$ and~$B_0$\@.
(The relations between the parameters were obtained by \mbox{Gasser} and
\mbox{Leutwyler}~\cite{G-L:su3}; we review them in
section~\ref{sec:matching}\@.)
In the purely pionic sector, the effective Lagrangian is written in terms of
the fields~\mbox{$U=u^2$} and~$\chi$ as in section~\ref{sec:cpt}\@.
Because the kaon fields are treated as matter fields which transform in the
fundamental representation of \su{2}\sub{V}, for coupling pions to kaons we
adopt the notation used in section~\ref{sec:matter} and reintroduce
\marginal{[eq:Vmu-2def]}%
\marginal{[eq:Amu-2def]}%
\marginal{[eq:chiPM-2def]}%
\begin{eqnarray}
\label{eq:Vmu-2def}
    \Vmu & = & \tfrac{1}{2} (u^\dag \dmu u + u \dmu u^\dag) , \\
\label{eq:Amu-2def}
    \Amu & = & \tfrac{i}{2} (u^\dag \dmu u - u \dmu u^\dag) , \\
\label{eq:chiPM-2def}
    \chi_\pm & = & u^\dag \chi u^\dag \pm u \chi^\dag u .
\end{eqnarray}
Throughout this chapter this notation is reserved for the \su{2}~versions of
the fields\@.
The heavy fields are the $\eZ$~meson and the kaons, which form two
\su{2}\sub{V} doublets
\marginal{[eq:K,Kc-2def]}%
\begin{equation}
\label{eq:K,Kc-2def}
    K = \left( \begin{array}{c} \kP \\ \kZ \end{array} \right)
, \mathsp
    K_c = \left( \begin{array}{c} \kbarZ \\ -\kM \end{array} \right)
.
\end{equation} 
As non-relativistic fields, the creation and annihilation operators are
treated separately, thus we distinguish~$K^\dag$ from~$K_c$\@.

The symmetries which we explicitly build into the effective Lagrangian are
\GchiP~chiral symmetry, parity~(P), charge conjugation~(C), and
reparameterization invariance\@.
The implementation of reparameterization invariance is discussed separately
in the next subsection\@.
The transformations of the fields under \GchiP,~P, C, and hermitian
conjugation~($\dag$) are collected for reference in Table~\ref{tbl:xfm-rules}
along with the power of~$Q$, i.e.,~the chiral index, associated with each
field\@.
Under each of the transformations listed, a covariant derivative of a
field~$\Dmu{}O$ transforms in the same way as~$O$ provided, in the case of
parity, the Lorentz index of the derivative is also raised or lowered\@.
The number of low-energy constants required is restricted sector-by-sector by
imposing the constraints of hermiticity and invariance under~\GchiP\ and~P on
the effective Lagrangian\@.
Within sectors of non-zero strangeness charge conjugation does not restrict
the form of the effective Lagrangian, but determines~\L{\pi\kbar}{}
from~\L{\pi{}K}{} and~\L{\pi\kbar\kbar}{} from~\L{\pi{}KK}{}\@.

\begin{table}[tb]
\centering
\begin{tabular}{|c|c|c|c|c|c|} \hline
field & \GchiP & P & C & $\dag$ & index\MS \\* \hline \hline
$U$ & $RUL^\dag$ & $U^\dag$ & $U\ms^T$ & & 0\MS \\*
$\chi$ & $R{\chi}L^\dag$ & $\chi^\dag$ & $\chi\ms^T$ 
    & & 2\rule[-1ex]{0ex}{1ex} \\* \hline
$\Vmu$ & $H{\Vmu}H^\dag{-}({\dmu}H)H^\dag$ & $+V\ss{}{\mu}$
    & $-\Vmu\ms^T$ & $-\Vmu$ & 1\MS \\*
$\Amu$ & $H{\Amu}H^\dag$ & $-A\ss{}{\mu}$ & $\Amu\ms^T$ & $\Amu$ & 1 \\*
$\chi_+$ & $H\chi_+H^\dag$ & $+\chi_+$ & $\chi_+\ms^T$ & $\chi_+$ & 2 \\*
$\chi_-$ & $H\chi_-H^\dag$ & $-\chi_-$ & $\chi_-\ms^T$ & $-\chi_-$ & 2 \\*
$K$ & $HK$ & $-K$ & $K_c\ms^T$ & $K^\dag$ & 0 \\*
$K_c$ & $K_cH^\dag$ & $-K_c$ & $K\ms^T$ & $K_c^\dag$ & 0 \\*
$\eZ$ & $\eZ$ & $-\eZ$ & $\eZ$ & $\eZ^\dag$    
    & 0\rule[-1ex]{0ex}{1ex} \\* \hline
\end{tabular}
\caption{Transformations of fields and under \GchiP~chiral symmetry,
  parity~P, charge conjugation~C, and hermitian conjugation~$\dag$, and the
  chiral index associated with each field\@. \drafttext{\protect\\*
\mbox{\bf[tbl:xfm-rules]}}}
\label{tbl:xfm-rules}
\end{table}

\subsection{Reparameterization Invariance}%
\label{sub:RPI}%
\marginal{[sub:RPI]}%
%
By treating the kaon and eta mesons as non-relativistic (heavy) fields we
have performed an implicit rephasing, as in section~\ref{sec:counting},
compared to the relativistic counterparts~$K_r$ and~$\eta_r$,
\marginal{[eq:nrK]}
\begin{equation}
\label{eq:nrK}
    K(x) \sim \sqrt{2M\,} e^{iM v{\cdot}x} K_r(x) , \mathsp
    \eta(x) \sim \sqrt{2M^\prime\,} e^{iM^\prime v{\cdot}x} K_r(x) .
\end{equation}
Reparameterization
invariance~(RPI)~\cite{D-G-G:rpi,L-M:rpi,YQC:rpi,F-G-M:rpi} is a consequence
of arbitrariness in the choice of the velocity~$v^\mu$ used in the
rephasing\@.
The choice of a different velocity in the relations between the relativistic
and heavy fields is compensated by a shift in the residual momentum~$k^\mu$
of the heavy field\@.
Lagrangians written in terms of heavy fields rephased with different
velocities must still give the same result for physical \mbox{$S$-matrix}
elements, which implies invariance of the Lagrangian under shifts
of~$v^\mu$\@.
A consequence of RPI~symmetry in the effective Lagrangian is that
coefficients of terms at different orders in the \mbox{$Q$-expansion} are
related~\cite{L-M:rpi} and by building reparameterization invariance into an
effective Lagrangian we reduce the number of low-energy constants to be
determined\@.
In this subsection we use the kaon as a representative example; all of the
arguments apply equally well to the case of the eta meson\@.

To construct a RPI~Lagrangian it is necessary and sufficient, as shown by
\mbox{Luke} and \mbox{Manohar}~\cite{L-M:rpi}, that the rephasing velocity
and derivatives of the kaon fields appear only in the following combinations:
\marginal{[eq:RPI-V]}
\begin{eqnarray}
\label{eq:RPI-V}
    \v\ss{\mu}{}K & = & (-iMv_\mu + \Dmu)K ,
\\* \nonumber
    \v\ss{\mu}{}K^\dag & = & (+iMv_\mu + \Dmu)K^\dag ,
\end{eqnarray}
where we will refer to the operator~$\v\ss{\mu}{}$ as the RPI-covariant
derivative\@.
$\v\ss{\mu}{}$ is a valid derivative in the sense that it satisfies the chain
rule,
\marginal{[eq:RPI-chain]}
\begin{equation}
\label{eq:RPI-chain}
    \Dmu (K^\dag O K)
    = (\v\ss{\mu}{}K^\dag) O K + K^\dag(\Dmu O)K + K^\dag O (\v\ss{\mu}{}K),
\end{equation}
which permits the use of partial integration and total derivative arguments
with respect to~$\v\ss{\mu}{}$\@.
Only after building the effective Lagrangian in terms of the RPI-covariant
derivative and expanding each term in powers of the mass~$M$ do we make any
particular choice of frame, i.e.,~explicitly
setting~\mbox{$v_\mu=(1,0,0,0)$}\@.

Superficially, developing the effective Lagrangian in terms of the
RPI-covariant derivative presents a problem for power counting\@.
An infinite tower of possible terms, such \mbox{as~$K^\dag{}K$},
\mbox{$K^\dag{}\v\ss{\mu}{}\vv\ss{}{\mu}K$},
\mbox{$K^\dag{}\v\ss{\mu}{}\v\ss{\nu}{}\vv\ss{}{\mu}\vv\ss{}{\nu}K$},~\ldots,
will contribute to each order of the chiral expansion of the Lagrangian\@.
On closer inspection, the fact that RPI guarantees a particular relationship
between coefficients of terms in the effective Lagrangian assures that the
effect of adding another term with more factors of~$\v\ss{\mu}{}$ can always
be compensated by a redefinition of the coefficients already present\@.
As an explicit example we consider the effect of adding the term
\mbox{$\delta\L{}{}=b\,K^\dag\ell^\mu\v\ss{\mu}{}\v\ss{\nu}{}\vv\ss{}{\nu}K$},
where~$\ell^\mu$ represents an unspecified combination of light degrees of
freedom, to the interaction Lagrangian
\mbox{$\L{\rm{int}}{}=a\,K^\dag\ell^\mu\v\ss{\mu}{}K$}\@.
Expanding the combined result gives
\marginal{[eq:RPItest]}
\begin{eqnarray}
\label{eq:RPItest}
    \L{\rm{int}}{}+\delta\L{}{}
    & = & -iM (a-bM^2)\, K^\dag (v{\cdot}\ell) K 
        + (a-bM^2)\, K^\dag \ell^\mu \Dmu K
\\* \nonumber & & \mbox{}
        - 2bM^2\, K^\dag (v{\cdot}\ell)(v{\cdot}DK) 
        - ibM\, K^\dag (v{\cdot}\ell) \Dmu D\ss{}{\mu} K + \cdots ,
\end{eqnarray}
where we have used \mbox{$v_\mu{}v^\mu=1$}\@.
A subsequent change \mbox{of~$a$,~$a\rightarrow{}a+bM^2$,} absorbs the
contribution of~$\delta\L{}{}$ to the terms already present
in~$\L{\rm{int}}{}$\@.
The addition of another term,
\mbox{$\delta\L{}{}^\prime=c\,\v\ss{\lambda}{}K^\dag\ell^\mu%
  \v\ss{\mu}{}\v\ss{\nu}{}\vv\ss{}{\nu}\vv\ss{}{\lambda}K$}, would be
followed by a change of~$a$ and~$b$ to restore the form of the terms present
before the addition of~$\delta\L{}{}^\prime$\@.
Generally, RPI~makes the possibility of such redefinitions certain to all
orders in the expansion of the effective Lagrangian and in every sector of
the theory\@.

The procedure just proposed is an iteration of adding new terms to the
effective Lagrangian then redefining the coefficients of the terms already
present\@.
An equivalent and more practical procedure is to subtract the contribution to
terms already present in the effective Lagrangian from each new term when it
is added, for instance
\marginal{[eq:RPIsubtr]}
\begin{eqnarray}
\label{eq:RPIsubtr}
    \lefteqn{\vv\ss{}{\mu}K^\dag \v\ss{\mu}{}\v\ss{\nu}{}\vv\ss{}{\nu}K}
\\* \nonumber
    & \longrightarrow &
        \vv\ss{}{\mu}K^\dag \v\ss{\mu}{}\v\ss{\nu}{}\vv\ss{}{\nu}K
        - M^2 \left( K^\dag \v\ss{\mu}{}\vv\ss{}{\mu}K + M^2 K^\dag K \right)
\\* \nonumber & & \mbox{}
        + M^2 \left( \v\ss{\mu}{}K^\dag \vv\ss{}{\mu}K - M^2 K^\dag K \right)
        + M^4 K^\dag K .
\end{eqnarray}
RPI-covariant derivatives contracted with Lorentz indices on light degrees of
freedom are never dropped\@.
We implicitly make such subtractions from every term when we include the term
in the effective Lagrangian\@.

Having resolved the issue of infinitely many contributions to each order of
the chiral expansion of the effective Lagrangian, we turn now to determining
the proper power of~$Q$ to associate with the leading contribution of a term
in RPI~form\@.
Occurrences of RPI-covariant derivatives are divided into two classes, those
contracted with another RPI-covariant derivative and those contracted into
the light degrees of freedom\@.
In the expansion of a contracted pair of RPI-covariant derivatives, the
contribution from~$(\pm{}iMv)^2$ is cancelled by the subtracted terms; the
leading contribution to the power counting comes \mbox{from~$v{\cdot}D$} so
each contracted pair contributes~$Q^1$\@.
The leading-order contribution from expanding a RPI-covariant derivative
contracted with the light degrees of freedom is unaffected by the
subtraction; the RPI-covariant derivative is counted as one power of~$v_\mu$
or order~$Q^0$\@.
To determine the effective Lagrangian to order~$Q^D$, in terms where the
operators for the light degrees of freedom contribute~$d$ powers of~$Q$ and
carry~$n$ Lorentz indices, we must consider some terms with as many as
\marginal{[eq:RPIrule]}
\begin{equation}
\label{eq:RPIrule}
    N = n + 2 (D - d)
\end{equation}
added RPI-covariant derivatives\@.
Toward the upper limit of the necessary number of RPI~derivatives, some
relief comes from the fact that total derivative arguments will allow
replacing some terms with alternates where a derivative is moved onto the
light degrees of freedom and the alternate terms may be neglected to the
order we are working\@.

In addition to the ambiguity in the choice of the rephasing velocity~$v_\mu$,
we have freedom to choose the mass~$M$ used in the rephasing relation,
equation~(\ref{eq:nrK})\@.
This freedom was exploited in equation~(\ref{eq:Bv,Tv-def}) to avoid awkward
\mbox{$x^\mu$-depend}ent phases in the heavy baryon Lagrangian\@.
The only constraint on the choice of~$M$ is that any residual
mass~$\delta{}M$ is small~(${\lesssim}Q$) to prevent failure of the power
counting scheme\@.
In the context of heavy kaon/eta theory, we perform the rephasing of the kaon
and eta fields with the masses~$\Mbar_K$ and~$\Mbar_\eta$ respectively,
where~$\Mbar_K$ and~$\Mbar_\eta$ are defined as the meson masses in the
\su{2}~chiral limit \mbox{$m_{u,d}\rightarrow0$} and $m_s$~finite\@.
With this choice, the residual masses of the~$K$ and~$\eta$ mesons at
tree-level are generated through interactions with the field~$\chi_+$ and are
included perturbatively\@.
Because the residual masses vanish in the \su{2}~chiral limit, by definition,
the effective Lagrangian does not include any explicit residual mass
term~\mbox{$-\delta{}M\,K^\dag{}K$}; in this sense our choice for the
rephasing mass is the `natural' choice\@.

\subsection{Use of Field Redefinitions}%
\label{sub:redefs}%
\marginal{[sub:redefs]}%
%
Redefinitions of the fields of the theory can be used in two ways to
potentially improve or simplify the effective Lagrangian\@.
The first class of field redefinitions eliminates terms from the effective
Lagrangian which are proportional to a classical equation of motion; this
class of field redefinitions, combined with total derivative arguments, is a
powerful method for reducing the number of unknown coefficients required\@.
The technique is well-documented in the
literature~\cite{DP:eom,HG:eom,CA:eom,F-S:eom} for simplifications of
effective Lagrangians and we have nothing to add aside from identifying the
general form of the terms we eliminate by this method\@.
Terms with an explicit factor of~$\Dmu{}A\ss{}{\mu}$,
\mbox{$\v\ss{\mu}{}\vv\ss{}{\mu}K$}, \mbox{$\v\ss{\mu}{}\vv\ss{}{\mu}\eta$},
or the hermitian conjugates can be combined with contributions from other
terms in the effective Lagrangian to form complete equation of motion terms
and are subsequently eliminated\@.

The second class of field redefinitions allows the elimination of all
time-like derivatives of the heavy fields from the interaction
Lagrangian~\cite{SW:nuclII,L-S:nrqcd,C-R-S:RelCorr}\@.
The principle is closely related to the elimination of equation of motion
terms, and if all equation of motion terms have already been eliminated by
redefinitions of the first class then further redefinitions from the second
class will permit a re-expression of the effective Lagrangian but not the
further elimination of any low-energy constants present\@.
Instead, the motive for considering field redefinitions in the second class
is to simplify the application of the effective Lagrangian by replacing
time-like derivatives of the heavy fields with spatial derivatives\@.
Spatial derivatives permit a more transparent power counting for~$Q$ in loop
diagrams than is possible when time-like derivatives occur in vertices\@.

Considering the free field theory of a (complex) heavy scalar is sufficient
to illustrate the general method and highlight conclusions relevant to how
the matching calculation between relativistic and non-relativistic theories
is performed\@.
The free-field Lagrangian for the heavy scalar, after rephasing as in
equation~(\ref{eq:nrK}), becomes
\marginal{[eq:Lphi0]}
\begin{equation}
\label{eq:Lphi0}
    \L{\phi_0}{} = \phi_0^\dag \left[ iv{\cdot}\partial
        + \frac{\nabla_\perp^2+(iv{\cdot}\partial)^2}{2m} \right] \phi_0
\end{equation}
where the spatial derivative~$\vec{\nabla}_\perp$ is defined
by~\mbox{$\partial_\mu\partial^\mu=(v{\cdot}\partial)^2-\nabla_\perp^2$}\@.
After substituting a field redefinition \mbox{$\phi_0=\phi_1+\delta\phi$}, we
find the well-known result that
\mbox{$\delta\phi=\tfrac{-1}{4m}iv{\cdot}\partial\phi$} cancels the time-like
derivatives from terms at order~$1/m$ and gives
\marginal{[eq:Lphi1]}
\begin{eqnarray}
\label{eq:Lphi1}
    \L{\phi_1}{} & = & \phi_1^\dag \left[ iv{\cdot}\partial 
        + \frac{\nabla_\perp^2}{2m} - \frac{3(iv{\cdot}\partial)^3}{16m^2}
        - \frac{(iv{\cdot}\partial)\nabla_\perp^2}{4m^2}
        + \mathcal{O}(1/m^3) \right] \phi_1 .
\end{eqnarray}
(A slightly different approach is taken in
references~\cite{L-S:nrqcd,C-R-S:RelCorr}\@.)
Substituting a second field redefinition
\mbox{$\phi_1=\phi_2+\delta\phi^\prime$}, dependence on time-like derivatives
is canceled in~\L{\phi_2}{} through order~$1/m^2$ by
\mbox{$\delta\phi^\prime=(4\nabla_\perp^2+3(iv{\cdot}\partial)^2)%
  \phi/32m^2$}\@.
Repeating this process, time-like derivatives of~$\phi$ can be eliminated to
any arbitrary order in an expansion in terms of~$(\partial/m)$\@.

Our original contribution to this discussion is the observation that the
infinite series of field redefinitions can be formally summed and a
relatively simple expression for the field redefinition
\mbox{$\phi_0=\mathcal{F}[\tilde{\phi}]$} can be obtained\@.
The approach is based on substituting a series representation
for~\mbox{$\mathcal{F}[\tilde{\phi}]$}, and deriving recursion relations from
the constraint that all time derivatives cancel from~\L{\tilde{\phi}}{}\@.
Here we present a brief plausibility argument for the inverse solution
\mbox{$\tilde{\phi}=\mathcal{F}^{-1}[\phi_0]$}; the actual recursion
relations generated are presented in Appendix~\ref{ch:recur} with a short
description of their solution\@.
The first step is to complete the square for the time-like derivative
appearing in equation~(\ref{eq:Lphi0});
\marginal{[eq:redef-1]}
\begin{eqnarray}
\label{eq:redef-1}
    \L{\phi_0}{}
    & = & \tfrac{1}{2m} \phi_0^\dag \left[\MS (iv{\cdot}\partial + m)^2
            - m^2 + \nabla_\perp^2 \right] \phi_0
\\* \nonumber
    & = & \tfrac{1}{2m} \phi_0^\dag \left[\MS iv{\cdot}\partial + m 
                - \sqrt{\ms\smash{m^2-\nabla_\perp^2}\,} \right]
            \left[\MS iv{\cdot}\partial + m 
                + \sqrt{\ms\smash{m^2-\nabla_\perp^2}\,} \right] \phi_0 .
\end{eqnarray}
Second, observing \mbox{$(\sqrt{m^2-\nabla_\perp^2\,}-m)$} is the
relativistic kinetic energy operator~$\widehat{K}$, we recognize the
first expression in brackets is the kernel of the desired Lagrangian
for~$\tilde{\phi}$,
\mbox{$\L{\tilde{\phi}}{}=\tilde{\phi}^\dag(iv{\cdot}\partial%
    -\widehat{K})\tilde{\phi}$}\@.
The second term in brackets is \mbox{$(iv{\cdot}\partial+\widehat{K}+2m)$}
which we divide between~$\phi_0$ and~$\phi_0^\dag$ and finally use to
identify~$\tilde{\phi}$;
\marginal{[eq:redef-2]}
\begin{eqnarray}
\label{eq:redef-2}
    \L{\phi_0}{} \!\! & = & \!\! 
        \left[\MS\sqrt{\frac{-iv{\cdot}\partial+\widehat{K}+2m}{2m}\,}\  
            \phi_0^\dag\right]\!\!\left(iv{\cdot}\partial-\widehat{K}\right)
            \!\!\left[\MS\sqrt{\frac{iv{\cdot}\partial+\widehat{K}+2m}{2m}\,}
            \ \phi_0\right]
\\* \nonumber & &
    \longrightarrow \ \ \tilde{\phi}
        = \sqrt{(iv{\cdot}\partial+\widehat{K}+2m)/2m\,}\ \phi_0
        = \mathcal{F}^{-1}[\phi_0] .
\end{eqnarray}

The explicit form for the redefinition resolves some differences between the
perturbative field theories of the heavy fields~$\phi_0$
and~$\tilde{\phi}$\@.
Developing the field propagator from the complete free-field Lagrangian in
equation~(\ref{eq:Lphi0}) results in two poles in the propagator, at
\mbox{$iv{\cdot}\partial=\pm{}(m+\widehat{K})$}\@.
Only one pole is relevant for the non-relativistic treatment and the second
pole is eliminated differently in the~$\phi_0$ and~$\tilde{\phi}$ theories\@.
The second pole is completely absent from the Lagrangian~\L{\tilde{\phi}}{};
the disappearance of the second pole is traced to a singularity of the field
redefinition~$\mathcal{F}[\tilde\phi]$ which causes~$\tilde{\phi}$ to vanish
at the location of the second pole\@.
In the~$\phi_0$ theory, the pole is removed by treating the
operator~$(iv{\cdot}\partial)^2/2m$ in~\L{\phi_0}{} as a perturbative
correction to the \mbox{$\phi_0$~two-point} function; the second pole does 
not appear at any finite order\@.
The second difference between the two non-relativistic theories is the
normalization of one-particle states relative to the normalization of the
relativistic one-particle state~\cite{AM:nrqcd,L-S:nrqcd}, which becomes
relevant when we pursue the matching calculation between the non-relativistic
heavy kaon/eta theory and fully relativistic~\cpt\@.
We determine the relative normalization by evaluating
\mbox{$\tilde{\phi}=\mathcal{F}^{-1}[\phi_0]$} at the location of the
relevant pole, \mbox{$(iv{\cdot}\partial)=\widehat{K}=E-m$}, which gives
\mbox{$\sqrt{2m\,}\tilde{\phi}=\sqrt{2E\,}\phi_0$}\@.
Which factor is included in matching calculations between relativistic and
non-relativistic theories depends on whether relativistic corrections to the
propagator are included as insertions of~$(iv{\cdot}\partial)^2/2m$ in
the~$\phi_0$ theory or as insertions
of~\mbox{$\nabla_\perp^4/8m^3+\nabla_\perp^6/16m^5+\cdots$} in
the~$\tilde{\phi}$ theory\@.

As stated above, the motivation for eliminating time-like derivatives of
heavy fields is to simplify the power counting of loop diagrams involving
both light- and heavy-field propagators\@.
However, the field redefinitions required to eliminate time-like derivatives
are not reparameterization invariant\@.
We consider the question open as to whether RPI can be reformulated as a
principle of the redefined $\tilde{\phi}$~scalar field theory\@.
Without explicit RPI~symmetry of the $\tilde{\phi}$~effective Lagrangian, 
the relations between coefficients of different orders in the
\mbox{$Q$-expansion} are not guaranteed\@.
As a consequence, we forego using further \mbox{non-RPI} field redefinitions
and treat the heavy kaon and eta fields in analogy with the $\phi_0$~scalar
field theory described in this subsection\@.

\ifthesisdraft\clearpage\fi%
\section{The Effective Lagrangians}%
\label{sec:buildL}%
\marginal{[sec:buildL]}%
%
The approach we take to developing the effective Lagrangians is based on the
work of \mbox{Fearing} and \mbox{Scherer}~\cite{F-S:orderQ6} extending
\L{\cpt}{} to order~$Q^6$\@.
We begin by addressing two points where we use a different approach,
specifically when performing the \su{2}\sub{V} contractions and when using
total derivative arguments to eliminate terms\@.
After discussing those issues we give a brief summary of the overall
procedure then present the explicit construction of the effective
Lagrangians~\L{}{\pi{}K}, \L{}{\pi\eta}, and~\L{}{\pi{}KK}.

Given a product of field operators, \mbox{Fearing} and \mbox{Scherer} form
\mbox{\Gchi-invariant} contractions by taking traces of all possible
permutations of the matrix fields\@.
(\mbox{Roessl} applies the same method for~\L{}{\pi{}K} by introducing a
matrix field~$KK^\dag$ for the kaons~\cite{AR:heavyK}\@.)
Then trace relations, such as
\marginal{[eq:trace-rel]}%
\begin{eqnarray}
\label{trace-rel}
    \lefteqn{\tr{ABC}+\tr{ACB}} 
\\* \nonumber
    & = & \tr{AB}\tr{C} + \tr{CA}\tr{B} + \tr{BC}\tr{A} - \tr{A}\tr{B}\tr{C}
\end{eqnarray}
for~$2\times2$ matrices, can be used to eliminate some of the resulting terms
in favor of others in the set\@.
For the relatively simple case of~\su{2}\sub{V}, we prefer to form a minimal
set of contractions by appealing to the algebra for addition of angular
momenta\@.
We prefer the second method because 1)~it permits a clean separation between
isospin-conserving and isospin-violating terms, 2)~the connection between
generated terms and physical processes is more meaningful and direct, and
3)~the method is more familiar\@.

A general matrix field~$O$ transforms under~\su{2}\sub{V} as
\mbox{$\bf0\oplus1$}, where the~$\bf0$ component is identified as~$\tr{O}$\@.
To separate out the~$\bf1$ component of a matrix field, we define the 
notation
\marginal{[eq:hat-def]} 
\begin{equation}
    \widehat{O} = O - \tfrac{1}{2}\tr{O} .
\end{equation}
Since~\tr{\Amu} and~\tr{\Vmu} both vanish, these fields transform in just 
the $\bf1$~representation\@.
This notation is particularly helpful when applied to~$\chi_\pm$; we find
that~$\tr{\chi_+}$ and~$\hat{\chi}_-$ are proportional to~$\hat{m}$,
but~$\tr{\chi_-}$ and~$\hat{\chi}_+$ are proportional to~\mbox{$(m_u-m_d)$}
and give isospin-violating terms\@.

As a simple illustration of the method, the product of
fields~\mbox{$KK^\dag{}\Amu\chi_-$} transforms as
\mbox{$\bf\frac{1}{2}\otimes\frac{1}{2}\otimes1\otimes(0\oplus1)$} which
contains three singlet representations; the resulting set of contracted forms
is \mbox{$\tr{\hat{\chi}_-\Amu}K^\dag{}K$},
\mbox{$\tr{\chi_-}K^\dag{}\Amu{}K$}, and
\mbox{$K^\dag[\hat{\chi}_-,\Amu]K$}\@.
We choose to take \mbox{$K^\dag[\hat{\chi}_-,\Amu]K$} instead of the equally
suitable form \mbox{$K^\dag\hat{\chi}_-\Amu{}K$} because the commutator
projects out just the $\bf1$ component of the product
\mbox{$(\hat{\chi}_-\times\Amu)$}\@.
Also, we can distribute any covariant derivatives which are present after
performing all \su{2}\sub{V} contractions since they do not change the
\su{2}\sub{V} transformation of the field on which they operate\@.
(However, antisymmetric expressions such as~$[\Amu,A\ss{}{\mu}]$ must be kept
until derivatives have been distributed across the fields\@.)

The second point to discuss is how to select terms which are to be eliminated
by adding total derivatives to the effective Lagrangian\@.
Clearly, the set of selected terms should contain as few terms proportional
to an equation of motion (EoM~terms) as possible, since such terms will be
eliminated anyway by subsequent field redefinitions.
\mbox{Fearing} and \mbox{Scherer}~\cite{F-S:orderQ6} take an approach driven
by intuition, and apply rules such as \textit{all terms where more than half
  of the derivatives act on a single field may be eliminated,} and use
Lorentz index exchange arguments to preserve as many EoM terms as possible
from the remaining set\@.
Because our power counting scheme permits many more derivatives at any order
than appear in \L{\cpt}{}, the intuitive approach is vulnerable to questions
of whether every possible total derivative was used to eliminate a term and
whether the set of terms retained contains the most EoM terms possible\@.

We present an algorithm which reduces the use of total derivative arguments
to linear algebra and assures that no EoM terms are eliminated at the expense
of keeping another term (and low-energy constant) in the effective
Lagrangian\@.
We view the set of~$m$ terms prior to any eliminations as the basis of a
\mbox{$m$-dimen}sional vector space~$\Omega$, and the set of coefficients of
those terms in the effective Lagrangian represents an arbitrary vector
in~$\Omega$\@.
Also, each total derivative, when expanded by the chain rule, is represented
by a vector in~$\Omega$; we define~$\omega$ to be the \mbox{$n$-dimen}sional
sub-space of~$\Omega$ spanned by the complete set of total derivatives\@.
The key ideas are that the vector representing the most general effective
Lagrangian is arbitrary up to the addition of any vector from~$\omega$ and
any set of \mbox{$m-n$}~terms which spans the complementary
sub-space~\mbox{$\Omega/\omega$} is sufficient for the most general effective
Lagrangian\@.

We construct a matrix~$P$ from the column vectors of a complete set of total
derivatives, not necessarily linearly independent\@.
A complete set of total derivatives for terms containing $k$~derivatives can
always be generated by explicitly taking the derivatives of each possible
term with $k-1$~derivatives\@.
For any matrix~$O$ we define the operation NS[$O$] which returns a matrix
whose columns span the (right) null space of~$O$\@.
Then, trivially, the columns of the \mbox{$(m-n)\times{}m$}~matrix NS[$P^T$]
span the desired sub-space~\mbox{$\Omega/\omega$}\@.
A particular choice of $m-n$ terms from the original set also
spans~\mbox{$\Omega/\omega$} if the matrix~$R$ constructed from the column
vectors representing the terms satisfies the relation
\mbox{$\mbox{det}[R^T\cdot\mbox{NS}[P^T]]\neq0$}\@.
Thus we can guarantee a complete set of total derivatives and test any
particular choice of terms for suitability as a basis
of~\mbox{$\Omega/\omega$}\@.
The remaining hurdle is to establish a means for identifying which choices of
terms contain the maximum number of EoM terms; because the number of
acceptable sets grows combinatorially with the number of derivatives and
fields in each term, exhaustive testing of the possible sets is
impractical\@.

By definition the columns of NS[$O$] represent the linear combinations of the
columns of~$O$ which vanish\@.
The key observation is if we construct a matrix~$Q$ from the rows of
NS[$P^T$] which correspond to the EoM terms, then the columns of NS[$Q$]
represent all independent linear combinations of the columns of NS[$P^T$]
with vanishing projection onto the EoM terms\@.
The result is the columns of the matrix \mbox{NS[$P^T$]$\cdot$NS[$Q$]} span
the sub-space of~$\Omega/\omega$ which is orthogonal to all of the EoM terms,
and the number of columns is exactly the minimum number of \mbox{non-EoM}
terms which must be included in the effective Lagrangian\@.

One considerable advantage of the algorithm just described is that the
computations, from developing the complete sets of terms and total
derivatives to selecting a single optimal basis of terms to include in the
effective Lagrangian, are easily programmed with software for symbolic
mathematics\@.
We also note that for very large sets of terms, the problem can be broken
into two smaller problems by dividing the full set of terms
considered~\mbox{$\{t_j\}_{1\leq{}j\leq{}m}$} into disjoint sets
\mbox{$\{t_j+t_j^\dag\}$} and \mbox{$\{i(t_j-t_j^\dag)\}$} and applying the
algorithm to each smaller set\@.
Finally, if the full set of terms forms a hierarchy of `better' terms and
`lesser' terms (beyond the distinction between EoM and \mbox{non-EoM} terms),
the steps for removing a maximal set of EoM terms from the basis
of~$\Omega/\omega$ can be repeated for each level in the hierarchy to assure
the minimum number of terms must be selected from each of the successive
`lesser' sets of terms\@.

In summary, we list the steps of the procedure, based on \mbox{Fearing} and
\mbox{Scherer}~\cite{F-S:orderQ6}, which we apply to construct the minimal
effective Lagrangians~\L{}{\pi{}K} and~\L{}{\pi{}KK} order by order in~$Q$\@.
\begin{enumerate}
\item Identify all simple (uncontracted) products of fields and derivatives
  consistent with parity and of the proper power of~$Q$\@.
\item For each product, form a minimal set of forms contracted over
  \su{2}\sub{V} indices in analogy to addition of angular momenta\@.
\item For each contracted form, distribute the derivatives over the fields
  and contract Lorentz indices all possible ways, keeping in mind that
  derivatives of~$\chi_\pm$ are unnecessary and~\mbox{$\Dmu{}A\ss{\nu}{}$}
  and all multiple derivatives are implicitly symmetrized\@.
\item For sets of terms with many derivatives, the power counting may
  indicate that some contractions of Lorentz indices yield terms higher order
  in~$Q$ than the order to which we are working; drop all such terms\@.
\item Replace each pair of terms~$t$ and~$t^\dag$ with the hermitian
  combinations \mbox{$(t+t^\dag)$} and \mbox{$i(t-t^\dag)$}\@.
\item For each set of terms, apply the total derivative algorithm to select
  a basis set containing the most EoM terms possible\@.
\item Eliminate EoM terms, proportional to either~$\Dmu{}A\ss{}{\mu}$,
  $\v\ss{\mu}{}\vv\ss{}{\mu}K$, or $\v\ss{\mu}{}\vv\ss{}{\mu}K^\dag$ through
  use of implicit field redefinitions\@.
\item Expand RPI-covariant derivatives of the kaon fields as
  \mbox{$\v\ss{\mu}{}=\pm{}i\Mbar_Kv_\mu+\Dmu$} then set
  \mbox{$v_\mu=(1,0,0,0)$}\@.
\end{enumerate}
In practice, after identifying a product of fields in step~(1.), we work out
all the terms generated by adding covariant derivatives to the product of
fields up to the highest relevant order in~$Q$ before moving on to the next
product of fields\@.

\subsection{The $\pi{}K$ and $\pi\eta$ Sectors}%
\label{sub:LpiK}%
\marginal{[sub:LpiK]}%
%
In the interest of transparency, so that the process and results can be
independently verified and because some knowledge of intermediate results is
necessary to extend the Lagrangian to higher order, we present details on
some of the steps in the construction of the effective Lagrangian\@.
After the discussion related to each product of fields, we list the RPI forms
of the terms which are kept, together with the coefficient associated with
each\@.
The particular coefficient we assign to a given term depends on the leading
power of~$Q$ contributed by the term after the RPI-covariant derivatives are
expanded\@.

The only product of fields permitted at order~$Q^0$ is~\mbox{$K^\dag{}K$}\@.
Because covariant derivatives can be freely moved from one field to the
other, sophisticated total derivative arguments are unnecessary regardless of
the number of derivatives distributed on the kaon fields; in each order of
the \mbox{$Q$-expansion}, only a single term of this form needs to be
retained\@.
For no derivatives or two derivatives, the terms we keep
are~\mbox{$K^\dag{}K$} and~\mbox{$\v\ss{\mu}{}K^\dag\vv\ss{}{\mu}K$}\@.
By definition, the coefficient of the mass term is the square of the mass
used in the heavy field rephasing, i.e.,~the mass of the kaon in the chiral
limit \mbox{$m_{u,d}\rightarrow0$}\@.
For four or more derivatives, we can always move all derivatives except a
single contracted pair off the kaon field leaving only the EoM term which is
subsequently eliminated\@.
The only combination of terms generated from the product~\mbox{$K^\dag{}K$}
(at any order) is
\begin{displaymath}
    \v\ss{\mu}{}K^\dag \vv\ss{}{\mu}K - \Mbar_K^2 (K^\dag K) .
\end{displaymath}

No product of fields at order~$Q^1$ is consistent with RPI and parity
invariance\@.
At order~$Q^2$ two products of field operators must be taken into
account,~\mbox{$KK^\dag\chi_+$} and~\mbox{$KK^\dag\Amu{}A\ss{\nu}{}$}\@.

For the field product~\mbox{$KK^\dag\chi_+$} we use the contracted forms
\begin{displaymath}
    \tr{\chi_+} K^\dag K \ \ (\coeff{c}{K}{3}),\mathsp
    K^\dag \hat{\chi}_+ K \ \ (\coeff{c}{K}{4}).
\end{displaymath}
Because~$\chi_+$ is effectively `transparent' to covariant derivatives
(since~$\Dmu\chi_+$ is equivalent to alternate terms we consider elsewhere),
we can freely move covariant derivatives between the kaon fields just as with
the product~\mbox{$K^\dag{}K$}\@.
Consequently, for any number of derivatives distributed on this form, we can
always reduce the set of terms to a single EoM term proportional to
\mbox{$\v\ss{\mu}{}\vv\ss{}{\mu}K$}\@.

For the field product~\mbox{$KK^\dag\Amu{}A\ss{\nu}{}$} we choose the
contracted forms \mbox{$\tr{\Amu{}A\ss{\nu}{}}K^\dag{}K$} and
\mbox{$K^\dag[\Amu,A\ss{\nu}{}]K$}, and up to order~$Q^4$ in each case we can
distribute up to six derivatives over the fields\@.
First we consider the form \mbox{$\tr{\Amu{}A\ss{\nu}{}}K^\dag{}K$}; adding
two derivatives, we must keep a total of seven terms of which five may be EoM
terms; adding four derivatives, we must keep a total of nine terms at
order~$Q^4$ of which eight may be EoM terms; adding six derivatives, we must
keep a total of four terms at order~$Q^4$ of which three may be EoM terms\@.
The terms we select after all eliminations are
\begin{eqnarray*}
    & \tr{A\ss{\mu}{} A\ss{}{\mu}} K^\dag K \ \ (\coeff{c}{K}{2}), & \\
    & \tr{A\ss{}{\mu} A\ss{}{\nu}} \v\ss{\mu}{}K^\dag \v\ss{\nu}{}K 
        \ \ (\coeff{c}{K}{1}), & \\
    & \tr{D\ss{\mu}{}A\ss{\nu}{} D\ss{}{\mu}A\ss{}{\nu}} K^\dag K 
        \ \ (\coeff{e}{K}{3}), & \\
    & \tr{A\ss{}{\mu} D\ss{}{\nu}A\ss{}{\lambda}} 
        \v\ss{\lambda}{}K^\dag \v\ss{\mu}{}\v\ss{\nu}{}K + \mbox{h.c.}    
        \ \ (\coeff{e}{K}{2}), & \\
    & \tr{D\ss{}{\mu}A\ss{}{\nu} D\ss{}{\lambda}A\ss{}{\kappa}}
        \v\ss{\mu}{}\v\ss{\nu}{}K^\dag \v\ss{\lambda}{}\v\ss{\kappa}{}K
        \ \ (\coeff{e}{K}{1}). &
\end{eqnarray*}
The second form \mbox{$K^\dag[\Amu,A\ss{\nu}{}]K$} vanishes unless
derivatives are added; adding two derivatives, we must keep three terms of
which two may be EoM terms; adding four derivatives, we must keep six terms
of which five may be EoM terms; adding six derivatives, we only need to keep
two EoM terms\@.
The terms we select after all eliminations are
\begin{eqnarray*}
    & \v\ss{\mu}{}K^\dag [A\ss{}{\mu},A\ss{}{\nu}] \v\ss{\nu}{}K
        \ \ (\coeff{d}{K}{2}), & \\
    & \v\ss{\mu}{}K^\dag [A\ss{}{\mu},D\ss{}{\nu}A\ss{}{\lambda}]
        \v\ss{\nu}{}\v\ss{\lambda}{}K + \mbox{h.c.} 
        \ \ (\coeff{d}{K}{1}). &
\end{eqnarray*}

The only product of fields consistent with RPI and parity invariance at
order~$Q^3$ is \mbox{$KK^\dag\Amu\chi_-$}\@.
The product results in three contracted forms,
\mbox{$\tr{\hat{\chi}_-\Amu}K^\dag{}K$}, \mbox{$K^\dag[\hat{\chi}_-,\Amu]K$},
and \mbox{$\tr{\chi_-}K^\dag\Amu{}K$}, which up to order~$Q^4$ can be
combined with either one or three derivatives\@.
Because all three of the contracted forms are a product of the same four
distinguishable fields (no exchange symmetry), the total derivative arguments
are exactly the same in each case; with one derivative we must keep two terms
of which one may be an EoM term; with three derivatives we may keep only two
EoM terms\@.
The three \mbox{non-EoM} terms which we elect to keep are
\begin{eqnarray*}
    & \tr{\hat{\chi}_- A\ss{}{\mu}} K^\dag \v\ss{\mu}{}K + \mbox{h.c.}
        \ \ (\coeff{d}{K}{3}), & \\
    & i K^\dag [\hat{\chi}_-,A\ss{}{\mu}] \v\ss{\mu}{}K + \mbox{h.c.}
        \ \ (\coeff{d}{K}{4}), & \\
    & \tr{\chi_-} K^\dag A\ss{}{\mu} \v\ss{\mu}{}K + \mbox{h.c.}
        \ \ (\coeff{d}{K}{5}). &
\end{eqnarray*}

At order~$Q^4$ four products of field operators are taken into account:
\mbox{$KK^\dag\chi_+\chi_+$}, \mbox{$KK^\dag\chi_-\chi_-$},
\mbox{$KK^\dag\Amu{}A\ss{\nu}{}\chi_+$}, and
\mbox{$KK^\dag\Amu{}A\ss{\nu}{}A\ss{\lambda}{}A\ss{\kappa}{}$}\@.
The first two products will only appear without any additional derivatives,
and are both contracted the same way under \su{2}\sub{V}\@.
When coupling the representations of the two matrix fields, only the
symmetric combinations
\mbox{$\bf[(0\oplus1)\otimes(0\oplus1)]_S=0\oplus0\oplus1_S\oplus2$} will
appear resulting in three \su{2}\sub{V} contractions for each product\@.
The six terms generated in this fashion are
\begin{displaymath}
    \begin{array}[t]{c@{\mathsp}c}
    \tr{\chi_+}^2 K^\dag K \ \ (\coeff{e}{K}{12}), &
    \tr{\hat{\chi}_-\hat{\chi}_-} K^\dag K \ \ (\coeff{e}{K}{13}), \\
    \tr{\chi_+} K^\dag \hat{\chi}_+ K \ \ (\coeff{e}{K}{18}), &
    \tr{\chi_-} K^\dag \hat{\chi}_- K \ \ (\coeff{e}{K}{19}), \\
    \tr{\hat{\chi}_+\hat{\chi}_+} K^\dag K \ \ (\coeff{e}{K}{20}), &
    \tr{\chi_-}^2 K^\dag K \ \ (\coeff{e}{K}{21}).
    \end{array}
\end{displaymath}

The product of fields \mbox{$KK^\dag\Amu{}A\ss{\nu}{}\chi_+$} permits six
distinct \su{2}\sub{V} contractions and under the
exchange~\mbox{$\Amu\leftrightarrow{}A\ss{\nu}{}$} three are symmetric and
three are antisymmetric\@.
For the six contracted forms, the power counting permits the addition of two
derivatives\@.
The three antisymmetric forms are
\mbox{$\tr{\hat{\chi}_+\Amu{}A\ss{\nu}{}}K^\dag{}K$},
\mbox{$\tr{\chi_+}K^\dag[\Amu,A\ss{\nu}{}]K$}, and
\mbox{$K^\dag[\hat{\chi}_+,[\Amu,A\ss{\nu}{}]]K$} which all vanish if no
derivatives are included\@.
In each of the three cases, when two derivatives are included, the symmetry
allows only a single term of order~$Q^4$ which can be eliminated by total
derivative arguments\@.
The three symmetric contracted forms are
\mbox{$\tr{\chi_+}\tr{\Amu{}A\ss{\nu}{}}K^\dag{}K$},
\mbox{$\tr{\Amu{}A\ss{\nu}{}}K^\dag\hat{\chi}_+K$}, and
\mbox{$K^\dag(\Amu\hat{\chi}_+A\ss{\nu}{}+A\ss{\nu}{}\hat{\chi}_+\Amu)K$}\@.
With the addition of two derivatives to each of the symmetric forms, we are
required to keep only a single \mbox{non-EoM} term\@.
The three symmetric forms result in a total of six terms in the
effective Lagrangian at order~$Q^4$,
\begin{displaymath}
    \begin{array}[t]{c@{\mathsp}c}
    \tr{\chi_+} \tr{\Amu A\ss{}{\mu}} K^\dag K \ \ (\coeff{e}{K}{11}), &
    \tr{\chi_+} \tr{A\ss{}{\mu} A\ss{}{\nu}} 
        \v\ss{\mu}{}K^\dag \v\ss{\nu}{}K \ \ (\coeff{e}{K}{10}), \\
    \tr{\Amu A\ss{}{\mu}} K^\dag \hat{\chi}_+ K \ \ (\coeff{e}{K}{17}), &
    \tr{A\ss{}{\mu} A\ss{}{\nu}} \v\ss{\mu}{}K^\dag 
        \hat{\chi}_+ \v\ss{\nu}{}K \ \ (\coeff{e}{K}{16}), \\
    K^\dag \Amu \hat{\chi}_+ A\ss{}{\mu} K \ \ (\coeff{e}{K}{15}), &
    \v\ss{\mu}{}K^\dag (A\ss{}{\mu} \hat{\chi}_+ A\ss{}{\nu}
        + A\ss{}{\nu} \hat{\chi}_+ A\ss{}{\mu}) \v\ss{\nu}{}K 
        \ \ (\coeff{e}{K}{14}).
    \end{array}
\end{displaymath}

The final product of fields,
\mbox{$KK^\dag\Amu{}A\ss{\nu}{}A\ss{\lambda}{}A\ss{\kappa}{}$}, transforms as
\mbox{$\bf\frac{1}{2}\otimes\frac{1}{2}\otimes1\otimes1\otimes1\otimes1$}
which contains nine singlet representations\@.
Three of the nine \su{2}\sub{V} contractions correspond to
\mbox{$\tr{\Amu{}A\ss{\nu}{}}\tr{A\ss{\lambda}{}A\ss{\kappa}{}}K^\dag{}K$}
under distinct permutations of Lorentz indices; the remaining six contracted
forms are distinct permutations among the Lorentz indices of
\mbox{$\tr{\Amu{}A\ss{\nu}{}}K^\dag[A\ss{\lambda}{},A\ss{\kappa}{}]K$}\@.
In each of the two cases, the power counting permits distributing up to four
derivatives on the fields, but the power counting also prohibits any EoM 
terms in the set of terms which can contribute at order~$Q^4$\@.
First we consider the second, commutator-type form\@.
The contributions of order~$Q^4$ from this term vanish if either no
derivatives or four derivatives are distributed on the fields; in the case of
adding two derivatives, a single term of order~$Q^4$ must be kept\@.
The term we choose to keep is
\begin{displaymath}
    i \tr{A\ss{\mu}{} A\ss{}{\nu}} K^\dag [A\ss{}{\mu},A\ss{}{\lambda}]
        \v\ss{\nu}{}\v\ss{\lambda}{}K + \mbox{h.c.} \ \ (\coeff{e}{K}{9}).
\end{displaymath}
For the double-trace contracted form, there are two ways to contract the
Lorentz indices without adding derivatives of the fields\@.
In addition, with two derivatives on the fields we must keep two terms at
order~$Q^4$; with four derivatives we must keep a single term at
order~$Q^4$\@.
The net contribution to the effective Lagrangian is five terms; we select the
following terms for this set:
\begin{eqnarray*}
    & \tr{A\ss{\mu}{} A\ss{}{\mu}} \tr{A\ss{\nu}{} A\ss{}{\nu}} K^\dag K
        \ \ (\coeff{e}{K}{7}), & \\
    & \tr{A\ss{\mu}{} A\ss{\nu}{}} \tr{A\ss{}{\mu} A\ss{}{\nu}} K^\dag K
        \ \ (\coeff{e}{K}{8}), & \\
    & \tr{A\ss{\mu}{} A\ss{}{\mu}} \tr{A\ss{}{\nu} A\ss{}{\lambda}}
        \v\ss{\nu}{}K^\dag \v\ss{\lambda}{}K \ \ (\coeff{e}{K}{5}), & \\
    & \tr{A\ss{\mu}{} A\ss{}{\nu}} \tr{A\ss{}{\mu} A\ss{}{\lambda}}
        \v\ss{\nu}{}K^\dag \v\ss{\lambda}{}K \ \ (\coeff{e}{K}{6}), & \\
    & \tr{A\ss{}{\mu} A\ss{}{\nu}} \tr{A\ss{}{\lambda} A\ss{}{\kappa}}
        \v\ss{\mu}{}\v\ss{\nu}{}K^\dag \v\ss{\lambda}{}\v\ss{\kappa}{}K
        \ \ (\coeff{e}{K}{4}). &
\end{eqnarray*}

We collect the terms of~\L{}{\pi{}K} listed above and present the complete
results for the chiral expansion \mbox{$\L{}{\pi{}K}=\sum_j\L{j}{\pi{}K}$} to
order~$Q^4$\@.
We expand the RPI-covariant derivatives to separate the terms which
contribute with different powers of~$Q$\@.
We also explicitly symmetrize the multiple derivatives and derivatives
of~\Amu\ wherever necessary and absorb some constant factors into the
coefficients where convenient\@.
Finally, all of the isospin-violating terms are collected together at the end
of each equation, offset with parentheses\@.
\marginal{[eq:Lpk1] [eq:Lpk2] [eq:Lpk3] [eq:Lpk4]} 
\begin{eqnarray}
\label{eq:Lpk1}
    \L{1}{\pi K} 
    & = & i K^\dag \D\ss{0}{}K
\\
\label{eq:Lpk2}
    \Mbar_K \L{2}{\pi K} 
    & = & \tfrac{1}{2} \D\ss{\mu}{}K^\dag D\ss{}{\mu}K
        + \coeff{c}{K}{1} \tr{A\ss{}{0} A\ss{}{0}} K^\dag K
        + \coeff{c}{K}{2} \tr{A\ss{\mu}{} A\ss{}{\mu}} K^\dag K
\\* \nonumber & & \mbox{}
        + \coeff{c}{K}{3} \tr{\chi_+} K^\dag K
        \left(\MS + \coeff{c}{K}{4} K^\dag \hat{\chi}_+ K \right)
\\
\label{eq:Lpk3}
    \Mbar_K^2 \L{3}{\pi K} 
    & = & i\coeff{c}{K}{1} \tr{A\ss{}{0} A\ss{}{\mu}}
            (K^\dag \D\ss{\mu}{}K - \mbox{h.c.} )
        + i\coeff{d}{K}{1} K^\dag [A\ss{}{0},\D\ss{0}{}A\ss{}{0}] K
\\* \nonumber & & \mbox{}
        + i\coeff{d}{K}{2} (K^\dag [A\ss{}{0},A\ss{}{\mu}] \D\ss{\mu}{}K
            + \mbox{h.c.} )
        + i\coeff{d}{K}{3} \tr{\hat{\chi}_- A\ss{}{0}} K^\dag K
\\* \nonumber & & \mbox{}
        + \coeff{d}{K}{4} K^\dag [\hat{\chi}_-,A\ss{}{0}] K
        \left(\MS + i\coeff{d}{K}{5} \tr{\chi_-} K^\dag A\ss{}{0} K \right) 
\\
\label{eq:Lpk4}
    \Mbar_K^3 \L{4}{\pi K} 
    & = & \coeff{c}{K}{1} \tr{A\ss{}{\mu} A\ss{}{\nu}}
            \D\ss{\mu}{}K^\dag \D\ss{\nu}{}K
\\* \nonumber & & \mbox{}
        - \tfrac{1}{2}\coeff{d}{K}{1} \left\{\MS K^\dag 
            ( [A\ss{}{\mu},\D\ss{0}{}A\ss{}{0}]
                + [A\ss{}{0},\D\ss{0}{}A\ss{}{\mu}]
                + [A\ss{}{0},D\ss{}{\mu}A\ss{}{0}] ) \D\ss{\mu}{}K
            - \mbox{h.c.} \right\}
\\* \nonumber & & \mbox{}
        + \coeff{d}{K}{2} \D\ss{\mu}{}K^\dag [A\ss{}{\mu},A\ss{}{\nu}]
            \D\ss{\nu}{}K
        - \tfrac{1}{2}\coeff{d}{K}{3} \tr{\hat{\chi}_- A\ss{}{\mu}}
            ( K^\dag \D\ss{\mu}{}K - \mbox{h.c.} )
\\ \nonumber & & \mbox{}
        + \tfrac{i}{2}\coeff{d}{K}{4} ( K^\dag
            [\hat{\chi}_-,A\ss{}{\mu}] \D\ss{\mu}{}K - \mbox{h.c.} )
        + \coeff{e}{K}{1} \tr{\D\ss{0}{}A\ss{}{0} \D\ss{0}{}A\ss{}{0}}
            K^\dag K
\\ \nonumber & & \mbox{}
        + \coeff{e}{K}{2} \tr{A\ss{}{\mu} \D\ss{0}{}A\ss{}{0}} (
            K^\dag \D\ss{\mu}{}K + \mbox{h.c.} )
        + \coeff{e}{K}{4} \tr{A\ss{}{0} A\ss{}{0}} 
            \tr{A\ss{}{0} A\ss{}{0}} K^\dag K
\\ \nonumber & & \mbox{}
        + \coeff{e}{K}{3} \tr{\D\ss{\mu}{}A\ss{\nu}{}
            (D\ss{}{\mu}A\ss{}{\nu} + D\ss{}{\nu}A\ss{}{\mu})} K^\dag K
        + \coeff{e}{K}{5} \tr{A\ss{}{0} A\ss{}{0}} 
            \tr{A\ss{\mu}{} A\ss{}{\mu}} K^\dag K
\\ \nonumber & & \mbox{}
        + \coeff{e}{K}{6} \tr{A\ss{}{0} A\ss{\mu}{}} 
            \tr{A\ss{}{0} A\ss{}{\mu}} K^\dag K 
        + \coeff{e}{K}{7} \tr{A\ss{\mu}{} A\ss{}{\mu}} 
            \tr{A\ss{\nu}{} A\ss{}{\nu}} K^\dag K
\\ \nonumber & & \mbox{}
        + \coeff{e}{K}{8} \tr{A\ss{\mu}{} A\ss{\nu}{}} 
            \tr{A\ss{}{\mu} A\ss{}{\nu}} K^\dag K
        + i\coeff{e}{K}{9} \tr{A\ss{}{0} A\ss{\mu}{}} K^\dag
            [A\ss{}{0},A\ss{}{\mu}] K
\\ \nonumber & & \mbox{}
        + \coeff{e}{K}{10} \tr{\chi_+} \tr{A\ss{}{0} A\ss{}{0}} K^\dag K
        + \coeff{e}{K}{11} \tr{\chi_+} \tr{A\ss{\mu}{} A\ss{}{\mu}} K^\dag K
\\ \nonumber & & \mbox{}
        + \coeff{e}{K}{12} \tr{\chi_+}^2 K^\dag K
        + \coeff{e}{K}{13} \tr{\hat{\chi}_- \hat{\chi}_-} K^\dag K
\\ \nonumber & & \left( \vc{\vpf{\ms}{\ms}} \mbox{} 
        - \tfrac{1}{2}\coeff{d}{K}{5} \tr{\chi_-} ( 
            K^\dag A\ss{}{\mu} \D\ss{\mu}{}K - \mbox{.h.c} ) \right.
\\ \nonumber & & \mbox{\ \ \ } 
        + \coeff{e}{K}{14} K^\dag A\ss{}{0} \hat{\chi}_+ A\ss{}{0} K
        + \coeff{e}{K}{15} K^\dag A\ss{\mu}{} \hat{\chi}_+ A\ss{}{\mu} K
        + \coeff{e}{K}{16} \tr{A\ss{}{0} A\ss{}{0}} K^\dag \hat{\chi}_+ K
\\* \nonumber & & \mbox{\ \ \ } 
        + \coeff{e}{K}{17} \tr{A\ss{\mu}{} A\ss{}{\mu}} K^\dag \hat{\chi}_+ K
        + \coeff{e}{K}{18} \tr{\chi_+} K^\dag \hat{\chi}_+ K
        + \coeff{e}{K}{19} \tr{\chi_-} K^\dag \hat{\chi}_- K
\\* \nonumber & & \left. \vc{\vpf{\ms}{\ms}} \mbox{\ \ \ }  
        + \coeff{e}{K}{20} \tr{\hat{\chi}_+ \hat{\chi}_+} K^\dag K
        + \coeff{e}{K}{21} \tr{\chi_-}^2 K^\dag K \right)
\end{eqnarray}

This procedure does not need to be repeated for the $\pi\eta$~sector\@.
The construction of~\L{}{\pi\eta} is identical to the construction
of~\L{}{\pi{}K} with three exceptions: the eta transforms as a singlet under
\su{2}\sub{V}; all isospin-violating terms must be omitted as discussed in
section~\ref{sec:prospects}; and the effective Lagrangian within the
$\pi\eta$~sector is invariant under charge conjugation\@.
From the results for~\L{}{\pi{}K} we construct~\L{}{\pi\eta} directly by
simply dropping all terms which violate isospin conservation or are not
proportional to the general form \mbox{$(D^mK^\dag{}D^nK)$},
replacing~$\Mbar_K$ by~$\Mbar_\eta$ and each occurrence of
\mbox{$(D^mK^\dag{}D^nK)$} by \mbox{$(\partial^m\eta^\dag{}\partial^n\eta)$},
and finally checking the remaining terms for charge conjugation invariance.
\marginal{[eq:Lpe1] [eq:Lpe2] [eq:Lpe3] [eq:Lpe4]} 
\begin{eqnarray}
\label{eq:Lpe1}
    \L{1}{\pi\eta} 
    & = & i \eta^\dag \partial\ss{0}{}\eta
\\
\label{eq:Lpe2}
    \Mbar_\eta \L{2}{\pi\eta} 
    & = & \tfrac{1}{2} \partial\ss{\mu}{}\eta^\dag \partial\ss{}{\mu}\eta
        + \coeff{c}{\eta}{1} \tr{A\ss{}{0} A\ss{}{0}} \eta^\dag \eta
\\* \nonumber & & \mbox{}
        + \coeff{c}{\eta}{2} \tr{A\ss{\mu}{} A\ss{}{\mu}} \eta^\dag \eta
        + \coeff{c}{\eta}{3} \tr{\chi_+} \eta^\dag \eta
\\
\label{eq:Lpe3}
    \Mbar_\eta^2 \L{3}{\pi\eta} 
    & = & i\coeff{c}{\eta}{1} \tr{A\ss{}{0} A\ss{}{\mu}}
            (\eta^\dag \dmu\eta - \mbox{h.c.} )
        + \coeff{d}{\eta}{} \tr{\hat{\chi}_- A\ss{}{0}} \eta^\dag \eta
\\
\label{eq:Lpe4}
    \Mbar_\eta^3 \L{4}{\pi\eta} 
    & = & \coeff{c}{\eta}{1} \tr{A\ss{}{\mu} A\ss{}{\nu}}
            \dmu\eta^\dag \dnu\eta
        - \tfrac{1}{2}\coeff{d}{\eta}{} \tr{\hat{\chi}_- A\ss{}{\mu}}
            ( \eta^\dag \dmu\eta - \mbox{h.c.} )
\\* \nonumber & & \mbox{}
        + \coeff{e}{\eta}{1} \tr{\D\ss{0}{}A\ss{}{0} \D\ss{0}{}A\ss{}{0}}
            \eta^\dag \eta
        + \coeff{e}{\eta}{2} \tr{A\ss{}{\mu} \D\ss{0}{}A\ss{}{0}}
            (\eta^\dag \dmu\eta + \mbox{h.c.} )
\\* \nonumber & & \mbox{}
        + \coeff{e}{\eta}{3} \tr{\D\ss{\mu}{}A\ss{\nu}{}
            (D\ss{}{\mu}A\ss{}{\nu} + D\ss{}{\nu}A\ss{}{\mu})} \eta^\dag \eta
        + \coeff{e}{\eta}{4} \tr{A\ss{}{0} A\ss{}{0}} 
            \tr{A\ss{}{0} A\ss{}{0}} \eta^\dag \eta
\\ \nonumber & & \mbox{}
        + \coeff{e}{\eta}{5} \tr{A\ss{}{0} A\ss{}{0}} 
            \tr{A\ss{\mu}{} A\ss{}{\mu}} \eta^\dag \eta
        + \coeff{e}{\eta}{6} \tr{A\ss{}{0} A\ss{\mu}{}} 
            \tr{A\ss{}{0} A\ss{}{\mu}} \eta^\dag \eta 
\\* \nonumber & & \mbox{}
        + \coeff{e}{\eta}{7} \tr{A\ss{\mu}{} A\ss{}{\mu}} 
            \tr{A\ss{\nu}{} A\ss{}{\nu}} \eta^\dag \eta 
        + \coeff{e}{\eta}{8} \tr{A\ss{\mu}{} A\ss{\nu}{}} 
            \tr{A\ss{}{\mu} A\ss{}{\nu}} \eta^\dag \eta 
\\* \nonumber & & \mbox{}
        + \coeff{e}{\eta}{10} \tr{\chi_+} \tr{A\ss{}{0} A\ss{}{0}} 
            \eta^\dag \eta
        + \coeff{e}{\eta}{11} \tr{\chi_+} \tr{A\ss{\mu}{} A\ss{}{\mu}} 
            \eta^\dag \eta
\\* \nonumber & & \mbox{}
        + \coeff{e}{\eta}{12} \tr{\chi_+}^2 \eta^\dag \eta
        + \coeff{e}{\eta}{13} \tr{\hat{\chi}_- \hat{\chi}_-} \eta^\dag \eta
\end{eqnarray}

\subsection{The $\pi{}KK$ Sector}%
\label{sub:LpiKK}%
\marginal{[sub:LpiKK]}%
%
In the $\pi{}KK$ sector, the full expression for the effective Lagrangian to
order~$Q^4$ is not necessary to treat \mbox{$2\leftrightarrow2$} scattering
processes until very high order\@.
For instance terms involving four factors of~\Amu\ only appear with at least
two closed pion loops and will be suppressed by at least a factor of~$Q^6$\@.
We develop the full effective Lagrangian to order~$Q^2$; at orders~$Q^3$
and~$Q^4$ we consider only those terms which give rise to four-kaon contact
terms\@.
To distinguish contact terms from pion interactions, we note that an
expansion of quantities in the number of pion fields gives
\mbox{$\Amu\simeq-\dmu\pi/2F$}, \mbox{$\hat{\chi}_-\simeq-4i\hat{m}B\pi/F$},
and \mbox{$\tr{\chi_-}\simeq-4i(m_u-m_d)B\piZ/F$}\@.

Also, we choose to couple the kaon annihilation operators together as
\mbox{$K\epsilon{}K$} and \mbox{$K\epsilon\tau_jK$},
where~\mbox{$\epsilon=i\tau_2$}, before contracting with the other fields
present for the advantage of greater symmetry under exchanges, antisymmetric
and symmetric respectively\@.
Coupling four kaon operators together in this way gives six representations
under \su{2}\sub{V}\@.
The three representations involving an antisymmetric contraction of fields
are simple products of the two forms above and transform
as~{$\bf0\oplus1\oplus1$}; the remaining three combinations of fields are
\mbox{$(K\epsilon\tau_jK)^\dag(K\epsilon\tau_jK)$} (a~$\bf0$),
\mbox{$i\epsilon_{jkl}(K\epsilon\tau_jK)^\dag(K\epsilon\tau_kK)$} (a~$\bf1$),
and \mbox{$\left((K\epsilon\tau_jK)^\dag(K\epsilon\tau_kK)%
    +(K\epsilon\tau_kK)^\dag(K\epsilon\tau_jK)\right)$} (a~$\bf2$), where use
of~$\epsilon_{jkl}$, the three-dimensional \mbox{Levi-Civita} symbol,
is always clear from context\@.
For brevity of notation we define the combinations
\marginal{[eq:4Kto0] [eq:4Kto1]}
\begin{eqnarray}
\label{eq:4Kto0}
    \KKKK{K_1}{K_2}{K_3}{K_4} 
    & = & (K_1\epsilon\tau_jK_2)^\dag(K_3\epsilon\tau_jK_4)
, \\
\label{eq:4Kto1}
    \KKOKK{K_1}{K_2}{O}{K_3}{K_4} 
    & = & i\epsilon_{jkl} (K_1\epsilon\tau_jK_2)^\dag 
            \tr{\tau_kO} (K_3\epsilon\tau_lK_4)
\end{eqnarray}
where~$K_j$ represent general fields in the $\bf\frac{1}{2}$~representation
of \su{2}\sub{V} and $O$ is any \su{2}\sub{V} matrix operator\@.

Without any factors of~\Amu\ or $\chi_\pm$ there are two contracted forms of
four kaon field operators, \mbox{$(K\epsilon{}K)^\dag(K\epsilon{}K)$} and
\mbox{$\KKKK{K}{K}{K}{K}$}\@.
To determine all contact operators up to order~$Q^4$ requires distributing up
to eight RPI-covariant derivatives on each form\@.
(Many with several derivatives contribute only at much higher order in tree
diagrams; however, we include them for completeness regardless\@.)
The algorithm for making optimal use of total derivative arguments serves us
very well in this endeavor; we need to deal with as many as 221 candidate
terms restricted by 179 linearly independent total derivatives\@.
We consider the symmetric contracted form first and count the terms left
after total derivative arguments; one term with no derivatives is necessary;
adding two derivatives, we must keep three terms of which two are EoM terms;
adding four derivatives, we must keep nine terms of which seven are EoM
terms; adding six derivatives, 20 terms are required of which 18 are EoM
terms; finally adding eight derivatives, 42 terms are required of which 39
are EoM terms\@.
After all redundant terms are eliminated, we select the following set of
terms:
\begin{eqnarray*}
    & \begin{array}[t]{c@{\mathsp}c}
    \KKKK{K}{K}{K}{K} \ \ (\coeff{a}{KK}{}), &
    \KKKK{K}{\v\ss{\mu}{}K}{K}{\vv\ss{}{\mu}K} \ \ (\coeff{b}{KK}{}), \\
    \KKKK{\v\ss{\mu}{}K}{\vv\ss{}{\mu}K}{\v\ss{\nu}{}K}{\vv\ss{}{\nu}K}
        \ \ (\coeff{c}{KK}{1}), &
    \KKKK{\v\ss{\mu}{}K}{\v\ss{\nu}{}K}{\vv\ss{}{\mu}K}{\vv\ss{}{\nu}K}
        \ \ (\coeff{c}{KK}{2}), \\
    \KKKK{\v\ss{\mu}{}K}{\v\ss{\nu}{}\v\ss{\lambda}{}K}%
        {\vv\ss{}{\mu}K}{\vv\ss{}{\nu}\vv\ss{}{\lambda}K} 
        \ \ (\coeff{d}{KK}{1}), &
    \KKKK{\v\ss{\nu}{}K}{\v\ss{\mu}{}\v\ss{\lambda}{}K}%
        {\vv\ss{}{\lambda}K}{\vv\ss{}{\mu}\vv\ss{}{\nu}K} 
        \ \ (\coeff{d}{KK}{2}), 
    \end{array} & \\
    & \KKKK{\v\ss{\mu}{}\v\ss{\nu}{}K}{\v\ss{\lambda}{}\v\ss{\kappa}{}K}%
        {\vv\ss{}{\mu}\vv\ss{}{\nu}K}{\vv\ss{}{\lambda}\vv\ss{}{\kappa}K}
        \ \ (\coeff{e}{KK}{1}), & \\
    & \KKKK{\v\ss{\mu}{}\v\ss{\lambda}{}K}{\v\ss{\nu}{}\vv\ss{}{\lambda}K}%
        {\v\ss{\kappa}{}\vv\ss{}{\mu}K}{\vv\ss{}{\kappa}\vv\ss{}{\nu}K}
        \ \ (\coeff{e}{KK}{2}), & \\
    & \KKKK{\v\ss{\mu}{}\v\ss{\nu}{}K}{\vv\ss{}{\mu}\vv\ss{}{\nu}K}%
        {\v\ss{\lambda}{}\v\ss{\kappa}{}K}{\vv\ss{}{\lambda}\vv\ss{}{\kappa}K}
        \ \ (\coeff{e}{KK}{3}). &
\end{eqnarray*}
For the antisymmetric contracted form the term with no derivatives vanishes;
adding two derivatives, we must keep a single \mbox{non-EoM} term; adding
four derivatives, we must keep four terms of which three are EoM terms;
adding six derivatives, twelve terms must be retained of which ten are EoM
terms; adding eight derivatives, 28 terms must be retained of which 26 are
EoM terms\@.
After all redundant terms are eliminated, we select the following set of
terms:
\begin{eqnarray*}
    & (K\epsilon \v\ss{\mu}{}K)^\dag (K\epsilon \vv\ss{}{\mu}K) 
        \ \ (\coeff{c}{KK}{3}), & \\
    & (K\epsilon \v\ss{\mu}{}K)^\dag 
        (\v\ss{\nu}{}K\epsilon \vv\ss{}{\mu}\vv\ss{}{\nu}) + \mbox{h.c.}
        \ \ (\coeff{d}{KK}{4}), & \\
    & (\v\ss{\nu}{}K\epsilon \v\ss{\mu}{}\v\ss{\lambda}{}K)^\dag
        (\vv\ss{}{\lambda}K\epsilon \vv\ss{}{\mu}\vv\ss{}{\nu}K)
        \ \ (\coeff{d}{KK}{3}), & \\
    & (\v\ss{\nu}{}K\epsilon \v\ss{\mu}{}\vv\ss{}{\nu}K)^\dag
        (\v\ss{\lambda}{}K\epsilon \vv\ss{}{\mu}\vv\ss{}{\lambda}K)
        \ \ (\coeff{e}{KK}{5}), & \\
    & (\v\ss{\mu}{}\v\ss{\nu}{}K\epsilon 
            \v\ss{\lambda}{}\v\ss{\kappa}{}K)^\dag
        (\vv\ss{}{\mu}\vv\ss{}{\nu}K\epsilon
            \vv\ss{}{\lambda}\vv\ss{}{\kappa}K) \ \ (\coeff{e}{KK}{4}), & \\
    & (\v\ss{\nu}{}K\epsilon \v\ss{\mu}{}\vv\ss{}{\nu}K)^\dag
        (\v\ss{\lambda}{}\v\ss{\kappa}{}K\epsilon
            \vv\ss{}{\mu}\vv\ss{}{\lambda}\vv\ss{}{\kappa}K) . & 
\end{eqnarray*}
The final term in the above list does not contribute until order~$Q^5$, so we
do not list an associated coefficient for that term\@.

The next product of fields we consider is~$KKK^\dag{}K^\dag\chi_+$ which
results in five contracted forms and each form must be considered with up to
four derivatives\@.
Because~$\chi_+$ is transparent to covariant derivatives, three of the
contracted forms yield terms analogous to a product of
fields~$KKK^\dag{}K^\dag$\@.
The contracted forms \tr{\chi_+}\KKKK{K}{K}{K}{K} and
\KKOKK{K}{K}{\chi_+}{K}{K} give four terms each after all redundant terms are
eliminated in analogy to~\KKKK{K}{K}{K}{K};
\begin{displaymath}
    \begin{array}[t]{c@{\mathsp}c}
    \tr{\chi_+}\KKKK{K}{K}{K}{K} \ \ (\coeff{c}{KK}{10}), &
    \KKOKK{K}{K}{\chi_+}{K}{K} \ \ (\coeff{c}{KK}{11}), \\
    \tr{\chi_+}\KKKK{K}{\v\ss{\mu}{}K}{K}{\vv\ss{}{\mu}K}
        \ \ (\coeff{d}{KK}{5}), &
    \KKOKK{K}{\v\ss{\mu}{}K}{\chi_+}{K}{\vv\ss{}{\mu}K}
        \ \ (\coeff{d}{KK}{6}), \\
    \tr{\chi_+}\KKKK{\v\ss{\mu}{}K}{\vv\ss{}{\mu}K}{\v\ss{\nu}{}K}%
        {\vv\ss{}{\nu}K} \ \ (\coeff{e}{KK}{6}), &
    \KKOKK{\v\ss{\mu}{}K}{\vv\ss{}{\mu}K}{\chi_+}{\v\ss{\nu}{}K}%
        {\vv\ss{}{\nu}K} \ \ (\coeff{e}{KK}{11}), \\
    \tr{\chi_+}\KKKK{\v\ss{\mu}{}K}{\v\ss{\nu}{}K}{\vv\ss{}{\mu}K}%
        {\vv\ss{}{\nu}K} \ \ (\coeff{e}{KK}{7}), &
    \KKOKK{\v\ss{\mu}{}K}{\v\ss{\nu}{}K}{\chi_+}{\vv\ss{}{\mu}K}%
        {\vv\ss{}{\nu}K} \ \ (\coeff{e}{KK}{12}).
    \end{array}
\end{displaymath}
The contracted form \mbox{$\tr{\chi_+}(K\epsilon{}K)^\dag(K\epsilon{}K)$}
gives two terms after all redundant terms are eliminated in analogy to
\mbox{$(K\epsilon{}K)^\dag(K\epsilon{}K)$};
\begin{eqnarray*}
    & \tr{\chi_+}(K\epsilon \v\ss{\mu}{}K)^\dag (K\epsilon \vv\ss{}{\mu}K)
        \ \ (\coeff{e}{KK}{8}), & \\
    & \tr{\chi_+}(K\epsilon \v\ss{\mu}{}K)^\dag
        (\v\ss{\nu}{}K\epsilon \vv\ss{}{\mu}\vv\ss{}{\nu}K) . &
\end{eqnarray*}
Again, the final term in the list contributes only beyond order~$Q^4$\@.
The final two contracted forms for the product $KKK^\dag{}K^\dag\chi_+$ are
the hermitian conjugate pair
\mbox{$(K\epsilon{}K)^\dag(K\epsilon\hat{\chi}_+K)$} and
\mbox{$(K\epsilon\hat{\chi}_+K)^\dag(K\epsilon{}K)$}\@.
Each of the forms vanishes without derivatives; adding two derivatives, in
each case only one EoM term is required; adding four derivatives, four EoM
terms are required\@.
These final two contracted pairs do not result in any necessary terms in the
effective Lagrangian\@.

The only remaining terms we will consider at order~$Q^4$ are derived from the
field product \mbox{$KKK^\dag{}K^\dag\chi_+\chi_+$} without any
derivatives\@.
Considering the \su{2}\sub{V} couplings gives eight distinct contractions,
but four vanish due to antisymmetry\@.
The remaining four terms contributing to the effective Lagrangian at
order~$Q^4$ are
\begin{displaymath}
    \begin{array}[t]{c@{\mathsp}c}
    \tr{\chi_+}^2 \KKKK{K}{K}{K}{K} \ \ (\coeff{e}{KK}{9}), &
    \tr{\chi_+} \KKOKK{K}{K}{\hat{\chi}_+}{K}{K} 
        \ \ (\coeff{e}{KK}{10}), \\ 
    \tr{\hat{\chi}_+\hat{\chi}_+} \KKKK{K}{K}{K}{K} 
        \ \ (\coeff{e}{KK}{13}), & 
    (K\epsilon\hat{\chi}_+K)^\dag(K\epsilon\hat{\chi}_+K) 
        \ \ (\coeff{e}{KK}{14}).
    \end{array}
\end{displaymath}

The final product of fields in the $\pi{}KK$~sector is
\mbox{$KKK^\dag{}K^\dag\Amu{}A\ss{\nu}{}$} and to derive the Lagrangian to
order~$Q^2$ we only need to consider adding up to two derivatives\@.
The \su{2}\sub{V} contractions yield six independent forms, two of which are
related by hermitian conjugation\@.
Without adding any derivatives to the product of fields, four of the
contracted forms vanish due to an antisymmetry under an exchange of fields\@.
Adding two derivatives to each of these four contracted forms gives terms
which only appear at higher order than~$Q^2$; we list the terms derived with
the power of~$Q$ at which it would first contribute\@.
From the hermitian-conjugate pair of contracted forms we get two EoM terms
and
\begin{eqnarray*}
    & (K\epsilon \v\ss{\mu}{}K)^\dag (K\epsilon [A\ss{}{\mu},A\ss{}{\nu}] 
        \v\ss{\nu}{}K) \ \ ({\sim} Q^3,\mbox{~twice}) , & \\
    & (K\epsilon \v\ss{\mu}{}K)^\dag (K\epsilon [A\ss{\nu}{},
        D\ss{}{\mu}A\ss{}{\nu}] K) \ \ ({\sim} Q^4,\mbox{~twice}) ; &
\end{eqnarray*}
each form contributing both as~$(O+O^\dag)$ and as~$i(O-O^\dag)$\@.
The other two contracted forms with a pair of fields antisymmetric under
exchange give two EoM terms and the higher-order terms
\begin{eqnarray*}
    & \KKOKK{K}{\v\ss{\mu}{}K}{A\ss{}{\mu}A\ss{}{\nu}}{K}{\v\ss{\nu}{}K}
        \ \ ({\sim} Q^3) , & \\
    & \tr{A\ss{\mu}{} A\ss{}{\mu}} (K\epsilon \v\ss{\nu}{}K)^\dag
        (K\epsilon \vv\ss{}{\nu}K) \ \ ({\sim} Q^4) , & \\
    & \tr{A\ss{}{\mu} A\ss{}{\nu}} (K\epsilon \v\ss{\mu}{}K)^\dag
        (K\epsilon \v\ss{\nu}{}K) \ \ ({\sim} Q^4) . &
\end{eqnarray*}
The two terms without any derivatives which do contribute to~\L{2}{\pi{}KK}
are
\begin{displaymath}
    \tr{A\ss{\mu}{} A\ss{}{\mu}} \KKKK{K}{K}{K}{K} 
        \ \ (\coeff{c}{KK}{6}) , \mathsp
    (K\epsilon A\ss{\mu}{}K)^\dag (K\epsilon A\ss{}{\mu}K) 
        \ \ (\coeff{c}{KK}{9}) .
\end{displaymath}
When two derivatives are distributed on either form, we must retain eleven
terms of which five may be EoM terms\@.
Of the twelve terms appearing in the effective Lagrangian, the only four
contributing to~\L{2}{\pi{}KK} are
\begin{eqnarray*}
    & \tr{A\ss{}{\mu} A\ss{}{\nu}} \KKKK{K}{\v\ss{\mu}{}K}{K}{\v\ss{\nu}{}K}
        \ \ (\coeff{c}{KK}{4}) , & \\
    & \tr{A\ss{}{\mu} A\ss{}{\nu}} \KKKK{K}{K}{\v\ss{\mu}{}K}{\v\ss{\nu}{}K}
        + \mbox{h.c.} \ \ (\coeff{c}{KK}{5}) , & \\
    & (K\epsilon A\ss{}{\mu}K)^\dag (\v\ss{\mu}{}K\epsilon
        A\ss{}{\nu}\v\ss{\nu}{}K) + \mbox{h.c.} \ \ (\coeff{c}{KK}{7}) ,&\\
    & (K\epsilon A\ss{}{\mu}\v\ss{\mu}{}K)^\dag
            (K\epsilon A\ss{}{\nu}\v\ss{\nu}{}K)
        + (K\epsilon A\ss{}{\mu}\v\ss{\nu}{}K)^\dag
            (K\epsilon A\ss{}{\nu}\v\ss{\mu}{}K) \ \ (\coeff{c}{KK}{8}) . &
\end{eqnarray*}
From the product of fields~$KKK^\dag{}K^\dag\Amu{}A\ss{\nu}{}$, the remaining
eight terms derived are
\begin{eqnarray*}
    & i\tr{A\ss{}{\mu} A\ss{}{\nu}} \KKKK{K}{K}{\v\ss{\mu}{}K}{\v\ss{\nu}{}K}
        + \mbox{h.c.} \ \ ({\sim} Q^3) , & \\
    & i(K\epsilon A\ss{}{\mu}K)^\dag (\v\ss{\mu}{}K\epsilon
        A\ss{}{\nu}\v\ss{\nu}{}K) + \mbox{h.c.} \ \ ({\sim} Q^3) , & \\
    & \tr{A\ss{\mu}{} A\ss{}{\mu}} \KKKK{K}{K}{\v\ss{\nu}{}K}{\vv\ss{}{\nu}K}
        \ \ ({\sim} Q^3,\mbox{~twice}) , & \\
    & (K\epsilon A\ss{\mu}{}K)^\dag (\v\ss{\nu}{}K\epsilon
        A\ss{}{\mu}\vv\ss{}{\nu}K) \ \ ({\sim} Q^3,\mbox{twice}) , & \\
    & \tr{\D\ss{\mu}{}A\ss{\nu}{}D\ss{}{\mu}A\ss{}{\nu}} \KKKK{K}{K}{K}{K}
        \ \ ({\sim} Q^4) , & \\
    & (K\epsilon \D\ss{\mu}{}A\ss{\nu}{}K)^\dag 
        (K\epsilon D\ss{}{\mu}A\ss{}{\nu}K) \ \ ({\sim} Q^4) . & 
\end{eqnarray*}

We collect the terms listed above and present the results for~\L{0}{\pi{}KK}
through~\L{3}{\pi{}KK}\@.
The contact terms of~\L{4}{\pi{}KK} are easily derived from the enumerated
terms above; we do not present the fully-expanded form of~\L{4}{\pi{}KK}\@.
\marginal{[eq:Lpkk0] [eq:Lpkk1] [eq:Lpkk2] [eq:Lpkk3]}
\begin{eqnarray}
\label{eq:Lpkk0}
    \Mbar_K^2 \L{0}{\pi KK}
    & = & \coeff{a}{KK}{} \KKKK{K}{K}{K}{K}
\\
\label{eq:Lpkk1}
    \Mbar_K^3 \L{1}{\pi KK}
    & = & i \coeff{b}{KK}{} \{ \KKKK{K}{K}{K}{\D\ss{0}{}K} - \mbox{h.c.} \}
\\
\label{eq:Lpkk2}
    \Mbar_K^4 \L{2}{\pi KK}
    & = & \coeff{b}{KK}{} \KKKK{K}{\D\ss{\mu}{}K}{K}{D\ss{}{\mu}K}
        + \coeff{c}{KK}{3} (K\epsilon \D\ss{\mu}{}K)^\dag
            (K\epsilon D\ss{}{\mu}K)
\\* \nonumber & & \mbox{}
        + (\coeff{c}{KK}{1} + 2\coeff{c}{KK}{2})
            \KKKK{K}{\D\ss{0}{}K}{K}{\D\ss{0}{}K}
\\* \nonumber & & \mbox{}
        - \coeff{c}{KK}{2} \{ \KKKK{K}{K}{\D\ss{0}{}K}{\D\ss{0}{}K}
            + \mbox{h.c.} \}
\\* \nonumber & & \mbox{}
        + (\coeff{c}{KK}{4} + \coeff{c}{KK}{5})
            \tr{A\ss{0}{} A\ss{0}{}} \KKKK{K}{K}{K}{K}
        + \coeff{c}{KK}{6}
            \tr{A\ss{\mu}{} A\ss{}{\mu}} \KKKK{K}{K}{K}{K}        
\\* \nonumber & & \mbox{}
        + (\coeff{c}{KK}{7} + \coeff{c}{KK}{8})
            (K\epsilon A\ss{0}{}K)^\dag (K\epsilon A\ss{0}{}K)
        + \coeff{c}{KK}{9} 
            (K\epsilon A\ss{\mu}{}K)^\dag (K\epsilon A\ss{}{\mu}K)
\\* \nonumber & & \mbox{}
        + \coeff{c}{KK}{10} \tr{\chi_+} \KKKK{K}{K}{K}{K}
        \left(\MS +\coeff{c}{KK}{11}\KKOKK{K}{K}{\hat{\chi}_+}{K}{K} \right)
\\
\label{eq:Lpkk3}
    \Mbar_K^5 \L{3}{\pi KK}
    & = & \tfrac{i}{2} \coeff{c}{KK}{1} 
            \{ \KKKK{K}{\D\ss{0}{}K}{\D\ss{\mu}{}K}{D\ss{}{\mu}K}
                - \mbox{h.c.} \}
\\* \nonumber & & \mbox{}
        + 2i \coeff{c}{KK}{2} 
            \{ \KKKK{K}{\D\ss{\mu}{}K}{\D\ss{0}{}K}{D\ss{}{\mu}K}
                - \mbox{h.c.} \}
\\* \nonumber & & \mbox{}
        + i (\coeff{d}{KK}{1} + \coeff{d}{KK}{2}) \left\{\MS \right.
            \KKKK{K}{K}{\D\ss{0}{}K}{\D\ss{0}{}\D\ss{0}{}K}
            - \KKKK{K}{\D\ss{0}{}K}{K}{\D\ss{0}{}\D\ss{0}{}K}
\\* \nonumber & & \mbox{} \hspace*{1.5in}
            -2 \KKKK{K}{\D\ss{0}{}K}{\D\ss{0}{}K}{\D\ss{0}{}K}
            - \mbox{h.c.} \left.\MS \right\}
\\ \nonumber & & \mbox{}
        + i \coeff{d}{KK}{3} \{ (K\epsilon \D\ss{0}{}K)^\dag
            (K\epsilon \D\ss{0}{}\D\ss{0}{}K) - \mbox{h.c.} \}
\\ \nonumber & & \mbox{}
        + i \coeff{d}{KK}{4} \left\{\MS \right.
            (K\epsilon \D\ss{\mu}{}K)^\dag
                (K\epsilon \D\ss{0}{}D\ss{}{\mu}K)
            + (K\epsilon \D\ss{\mu}{}K)^\dag
                (K\epsilon D\ss{}{\mu}\D\ss{0}{}K) 
\\* \nonumber & & \mbox{} \hspace*{.75in}
            + 2 (K\epsilon \D\ss{\mu}{}K)^\dag
                (\D\ss{0}{}K\epsilon D\ss{}{\mu}K) 
            - \mbox{h.c.} \left.\MS \right\}
\\* \nonumber & & \mbox{}
        + i \coeff{d}{KK}{5} \tr{\chi_+} 
            \{ \KKKK{K}{K}{K}{\D\ss{0}{}K} - \mbox{h.c.} \}
\\* \nonumber & & \mbox{}
        \left(\MS + i \coeff{d}{KK}{6} \{ 
            \KKOKK{K}{K}{\hat{\chi}_+}{K}{\D\ss{0}{}K} 
                - \mbox{h.c.} \} \right)
\\* \nonumber & & \mbox{}+\mbox{other terms involving \Amu}
\end{eqnarray}
Having worked out the forms of the effective Lagrangians, we turn in the next
section to a matching calculation to determine the coefficients which appear
in terms of the parameters of the `high-energy' theory,~\su{3}~\cpt\@.

\ifthesisdraft\clearpage\fi%
\section{Matching Calculations}%
\label{sec:matching}%
\marginal{[sec:matching]}%
%
A topic which needs to be addressed before setting up the matching
calculations between the heavy kaon/eta theory and \su{3}~chiral perturbation
theory is the difference between the expansions of the two theories\@.
The heavy kaon/eta effective theory is an expansion in~$Q/M$, where~$Q$
represents the generic scale of~$m_\pi$ and external momenta~$|\vec{p}\,|$,
and~$M$ represents the scale of heavy masses~$\Mbar_{K,\eta}$\@.
In contrast, \su{3}~\cpt\ is an expansion in~$(Q/\Lambda)^2$ where~$Q$
represents the scale of external momenta and the full set of pseudo-Goldstone
boson masses\@.
To reconcile the two effective field theories in the low-energy
regime~\mbox{$|\vec{p}\,|\lesssim{}m_\pi$}, we must recognize three distinct
mass scales~$Q$, $M$, and~$\Lambda$ which represent respectively external
momenta and~$m_\pi$, $\Mbar_K$ and~$\Mbar_\eta$, and the chiral symmetry
breaking scale\@.

Distinguishing the scales in~\cpt\ is trivial\@.
We compute scattering amplitudes to a given order in~$1/\Lambda$, then in
each order separate the expression by powers of~\mbox{$m_sB_0\sim{}M^2$}\@.
In contrast, the role of~$\Lambda$ in the heavy kaon/eta effective theory is
buried in the coefficients of the effective theory, just as the coefficients
of~\cpt\ are implicitly unknown functions of~$\Lambda_\QCD$ and the heavy
quark masses~\cite{HG:book}\@.
From within the heavy kaon/eta theory there is no way to determine the
dependence of the coefficients on~$\Lambda$ via scaling arguments as in
section~\ref{sec:cpt} because the dependence is obscured by the intermediate
scale~$M$\@.
The matching calculation provides a direct way to establish the relationship
between the coefficients and the chiral symmetry breaking scale\@.

We match the theories by equating on-shell scattering amplitudes and the
locations and residues of poles in the heavy field propagators of the two
theories\@.
(Off the mass-shell, the scattering amplitudes depend, in an unphysical way,
on how the meson fields are defined and matching would require including all
of the EoM terms eliminated by field redefinitions\@.)
We choose to match amplitudes with a relativistic normalization; so a factor
of~$\sqrt{\MS2\Mbar_{K,\eta}}$ is included for each external heavy-particle
state in the heavy kaon/eta theory, consistent with the discussion in
subsection~\ref{sub:redefs}\@.
For the amplitudes calculated in~\cpt\ we perform a non-relativistic
expansion, making the
replacement~\mbox{$p\ss{}{0}\rightarrow(\Mbar_{K,\eta}+k\ss{}{0})$}\@.
The expansions for an arbitrary amplitude~$\A{}{}$ of mass-dimension~$d$,
in~\cpt\ and in heavy kaon/eta theory respectively, are (symbolically)
\marginal{[eq:A-cpt] [eq:A-NR]}
\begin{eqnarray}
\label{eq:A-cpt}
    \A{\cpt}{}
    & = & M^d \sum_{j,k} \bar{\alpha}_{2j,k} 
            \frac{Q^{2j}M^{2(k-j)}}{\Lambda^{2k}}
    = M^d \sum_j \frac{Q^{2j}}{M^{2j}} 
            \sum_k \bar{\alpha}_{2j,k} \frac{M^{2k}}{\Lambda^{2k}}
, \\
\label{eq:A-NR}
    \A{NR}{} 
    & = & M^d \sum_j \alpha_j \frac{Q^j}{M^j}
,
\end{eqnarray}
where the \mbox{$(1/\Lambda)$-expansion} of~\cpt\ has been reorganized as an
expansion in~$M/\Lambda$ within an expansion in~$Q/M$\@.
Equating the expansions in powers of~$Q$ determines the combinations of terms
\mbox{$\alpha_jQ^j=\sum_k\bar{\alpha}_{j,k}Q^j(M^2/\Lambda^2)^k$} for
even~$j$ and \mbox{$\alpha_jQ^j=0$} for odd~$j$\@.
Solving for the coefficients in the heavy kaon/eta effective theory, we find
that matching gives the coefficients as an expansion in powers
of~$(M/\Lambda)^2$\@.
This behavior is seen explicitly in the results for matching~$F$ and~$B$ to
the parameters of \su{3}~\cpt\ by comparing amplitudes for pion
interactions\@.
The matching relations were derived by \mbox{Gasser} and
\mbox{Leutwyler}~\cite{G-L:su3};
\marginal{[eq:matchF] [eq:matchB]}
\begin{eqnarray}
\label{eq:matchF}
    F & = & F_0 \left\{ 1 
            -\frac{\Mbar_K^2}{32\pi^2F_0^2} \ln\frac{\Mbar_K^2}{\mu^2}
            +8L_4^r \frac{\Mbar_K^2}{F_0^2} 
            +\mathcal{O}(M^4/\Lambda^4) \right\}
, \\
\label{eq:matchB}
    B & = & B_0 \left\{ 1
            -\frac{\Mbar_\eta^2}{96\pi^2F_0^2} \ln\frac{\Mbar_\eta^2}{\mu^2} 
            -16(L_4^r-2L_6^r) \frac{\Mbar_K^2}{F_0^2} 
            +\mathcal{O}(M^4/\Lambda^4) \right\}
.
\end{eqnarray}
The leading-order results from~\cpt\ determine coefficents of the
effective Lagrangians~\L{j\leq3}{(X)} to leading order in~$(M/\Lambda)^2$
(with the exception of coefficients for terms with time-like derivatives)\@.
Next-to-leading order results from~\cpt\ determine the same coefficients at 
the next order in~$(M/\Lambda)^2$ and the leading order of a further set of 
coefficients from the heavy kaon/eta theory\@.

We present the results for the leading order matching calculation then apply
the results to make a prediction for~$KK$ scattering phase shifts in the next
section\@.
The parameters of~\L{}{\pi{}K} are, up to corrections of
order~$(M/\Lambda)^2$, 
\marginal{[eq:matchPK]}
\begin{equation}
\label{eq:matchPK}
    \Mbar_K^2 = m_s B_0 ,\mathsp
    \coeff{c}{K}{1} = 0 , \mathsp
    \coeff{c}{K}{2} = -1/4 , \mathsp
    \coeff{c}{K}{3} = -1/16 ,
\end{equation}
and~\mbox{$\coeff{d}{K}{1}=\coeff{d}{K}{2}=\coeff{d}{K}{4}=0$}.
The parameters appearing in~\L{}{\pi\eta} are, up to corrections of
order~$(M/\Lambda)^2$,
\marginal{[eq:matchPE]}
\begin{equation}
\label{eq:matchPE}
    \Mbar_\eta^2 = \tfrac{4}{3} m_s B_0 , \mathsp
    \coeff{c}{\eta}{1} 
        = \coeff{c}{\eta}{2} = 0 , \mathsp
    \coeff{c}{\eta}{3} = -1/16 .
\end{equation}
The coefficients~\coeff{d}{K}{3} and~\coeff{d}{\eta}{} are not determined
until matching at next-to-leading order in~\cpt\@.
For~\L{}{\pi{}KK} we only perform the matching for the coefficients of the
four-kaon contact terms and find the leading order results are
\marginal{[eq:matchPKK]}
\begin{equation}
\label{eq:matchPKK}
    \coeff{a}{KK}{} = \frac{-m_s B_0}{16 F_0^2} , \mathsp
    \coeff{b}{KK}{} = \frac{-m_s B_0}{8 F_0^2} , \mathsp
    \coeff{c}{KK}{3} = 0, \mathsp
    \coeff{c}{KK}{10} = \frac{m_s B_0}{128 F_0^2} ,
\end{equation}
with corrections of order~$(M/\Lambda)^4$\@.

In effect, the heavy kaon/eta effective theory becomes a dual expansion
in~$Q/M$ and~$(M/\Lambda)^2$ through the matching calculation\@.
Applying the theory to physical processes requires establishing a relative
weight to the two expansion parameters\@.
The relative weight dictates to what order of the expansion in powers
of~$(M/\Lambda)^2$ one must work for a consistent result to a particular
order in~$Q/M$\@.
Equivalently we introduce a common expansion parameter~$\xi$ and assign~$Q/M$
and~$M/\Lambda$ each a characteristic power of~$\xi$\@.
The ratio of masses \mbox{$m_\pi/M_K\simeq0.3$} and the typical size of
\su{3}~symmetry breaking effects~\cite{G-L:su3} suggest assigning the ratios
of mass scales \mbox{$Q/M$} and \mbox{$(M/\Lambda)^2$} equal powers of the
parameter~$\xi$, so we take \mbox{$\xi\sim{}Q/M\sim(M/\Lambda)^2\sim0.3$}\@.

\ifthesisdraft\clearpage\fi%
\section{Application: KK Scattering}%
\label{sec:KKphase}%
\marginal{[sec:KKphase]}%
%
From the effective Lagrangians~\L{}{\pi{}K} and~\L{}{\pi{}KK} we calculate
the~$KK$ scattering amplitudes in the isospin~\mbox{I=0,1} channels to
order~$Q^2$\@.
Using the results of the matching calculations of section~\ref{sec:matching},
we make a prediction for the~$KK$ \mbox{$s$-wave} scattering phase shift in
the~\mbox{I=1} channel\@.
We plot a comparison of the leading order results of the heavy kaon/eta
theory and \su{3}~\cpt\ and comment on the usefulness of this approach\@.

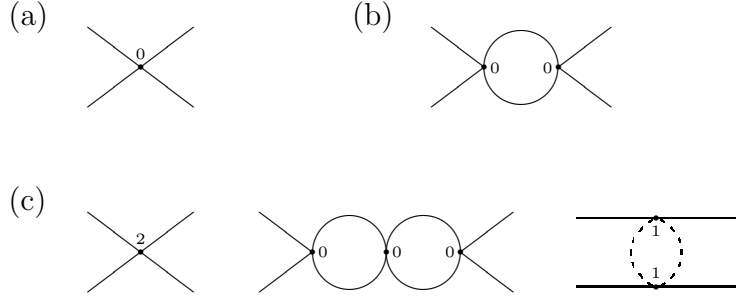
\begin{figure}[tb]
\hspace*{\fill}
\begin{picture}(300,120)%
    \put(0,120){\makebox(0,0)[lt]{(a)~}}%
    \put(30,80){\line(4,3){40}}%
    \put(30,110){\line(4,-3){40}}%
    \put(50,95){\circle*{2}}%
    \put(50,100){\makebox(0,0){$\sss0$}}%
    \put(130,120){\makebox(0,0)[lt]{(b)~}}%
    \put(180,95){\line(-4,-3){20}}%
    \put(180,95){\line(-4,3){20}}%
    \put(180,95){\circle*{2}}%
    \put(184,95){\makebox(0,0){$\sss0$}}%
    \put(194,95){\circle{29}}%
    \put(208,95){\circle*{2}}%
    \put(204,95){\makebox(0,0){$\sss0$}}%
    \put(208,95){\line(4,-3){20}}%
    \put(208,95){\line(4,3){20}}%
    \put(0,50){\makebox(0,0)[lt]{(c)~}}%
    \put(30,10){\line(4,3){40}}%
    \put(30,40){\line(4,-3){40}}%
    \put(50,25){\circle*{2}}%
    \put(50,30){\makebox(0,0){$\sss2$}}%
    \put(115,25){\line(-4,-3){20}}%
    \put(115,25){\line(-4,3){20}}%
    \put(115,25){\circle*{2}}%
    \put(119,25){\makebox(0,0){$\sss0$}}%
    \put(129,25){\circle{29}}%
    \put(143,25){\circle*{2}}%
    \put(147,25){\makebox(0,0){$\sss0$}}%
    \put(157,25){\circle{29}}%
    \put(171,25){\circle*{2}}%
    \put(167,25){\makebox(0,0){$\sss0$}}%
    \put(171,25){\line(4,-3){20}}%
    \put(171,25){\line(4,3){20}}%
    \put(215,12){\line(1,0){60}}%
    \put(245,12){\circle*{2}}%
    \put(245,17){\makebox(0,0){$\sss1$}}%
    \put(215,38){\line(1,0){60}}%
    \put(245,38){\circle*{2}}%
    \put(245,33){\makebox(0,0){$\sss1$}}%
    \savebox{\dash}(0,0){%
        \begin{picture}(0,0)%
        \put(15,-1){\line(0,1){2}}%
        \end{picture}}%
    \put(238.5,25){\rotatebox{-54}{\usebox{\dash}}}%
    \put(239,25){\rotatebox{-36}{\usebox{\dash}}}%
    \put(239.5,25){\rotatebox{-18}{\usebox{\dash}}}%
    \put(239.5,25){\rotatebox{0}{\usebox{\dash}}}%
    \put(239.5,25){\rotatebox{18}{\usebox{\dash}}}%
    \put(239,25){\rotatebox{36}{\usebox{\dash}}}%
    \put(238.5,25){\rotatebox{54}{\usebox{\dash}}}%
    \put(251.5,25){\rotatebox{126}{\usebox{\dash}}}%
    \put(251,25){\rotatebox{144}{\usebox{\dash}}}%
    \put(250.5,25){\rotatebox{162}{\usebox{\dash}}}%
    \put(250.5,25){\rotatebox{180}{\usebox{\dash}}}%
    \put(250.5,25){\rotatebox{198}{\usebox{\dash}}}%
    \put(251,25){\rotatebox{216}{\usebox{\dash}}}%
    \put(251.5,25){\rotatebox{234}{\usebox{\dash}}}%
\end{picture}
\hspace*{\fill}
\caption{Feynman diagrams contributing to the~$KK$ scattering amplitude in
  the heavy kaon/eta effective theory; (a)~order~$Q^0$ tree diagram,
  (b)~order~$Q$ kaon-bubble diagram, (c)~order~$Q^2$ tree, double-bubble, and
  pion exchange diagrams\@. \drafttext{\protect\\*
    \mbox{\bf[fig:KKdiags]}}}
\label{fig:KKdiags}
\end{figure}

Initially we calculate the~$KK$ scattering amplitude as an expansion in
powers of~$Q/M$, treating all coefficients as
intrinsically~$\mathcal{O}(\xi^0)$\@.
The leading-order contribution to the scattering amplitude is order~$Q^0$ and
arises from the tree diagram with an \mbox{\coeff{a}{KK}{}-vertex}\@.
Because time derivatives on external legs contribute~$Q^2$, the effective
Lagrangian results in no contribution from tree diagrams at order~$Q$\@.
At order~$Q^2$ we include tree diagrams with a vertex from~\L{1}{\pi{}KK}
or~\L{2}{\pi{}KK} and which give one of the coefficients~\coeff{b}{KK}{},
\coeff{c}{KK}{3}, or~\coeff{c}{KK}{10}\@.
Because the two-body sector admits nearly-infrared-divergent behavior, the
appropriate power counting for loop diagrams is given in
equation~(\ref{eq:NR-rule})\@.
The power counting shows that kaon-bubble diagrams are suppressed by a single
power of~$Q$, so to order~$Q^2$ we take double- and single-bubble diagrams
with only the \coeff{a}{KK}{}-vertex included\@.
Finally, one-pion loop diagrams are suppressed by only the powers of~$Q$
generated at the vertices, which means double pion exchange is included with
vertices from the \mbox{order-$Q$} effective Lagrangian~\L{1}{\pi{}K}\@.
The set of relevant diagrams is illustrated in Figure~\ref{fig:KKdiags}\@.

Integrations over loop energies~$dq\ss{0}{}$ are performed by contour
integration and the resulting momentum integrals~$d\vec{q}$ are evaluated in
dimensional regularization\@.
The scattering amplitudes are calculated in the center-of-mass frame and the
\mbox{order-$Q^2$} equation of motion,
\mbox{$2\Mbar_KE_k=k^2-16\coeff{c}{K}{3}\hat{m}B$}, gives the kaon energy of
the external states\@.
The results for~$\A{NR}{\rm(I)}$ are
\marginal{[eq:KK-A0] [eq:KK-A1]}
\begin{eqnarray}
\label{eq:KK-A0}
    i\A{NR}{(0)}(k)
    & = & 32i \coeff{c}{KK}{3} \frac{k^2 \cos\theta}{\MS\Mbar_K^2}
, \\
\label{eq:KK-A1}
    i\A{NR}{(1)}(k)
    & = & 32i \coeff{a}{KK}{} \left\{ 1
        + i \coeff{a}{KK}{} \sqrt{ \frac{2E_k}{\MS \pi^2 \Mbar_K} \:}
        - \frac{(\coeff{a}{KK}{})^2 2E_k}{\MS \pi^2 \Mbar_K} \right\}
\\* \nonumber & & \mbox{}
        + 64i \coeff{b}{KK}{} \frac{E_k}{\MS\Mbar_K}
        + 256i \coeff{c}{KK}{10} \frac{\hat{m}B}{\MS\Mbar_K^2}
        - i \frac{ m_\pi^2 \Mbar_K^2 }{\MS 16 \pi^2 F^4 }
            \ln \frac{m_\pi^2}{\MS\mu^2}
.
\end{eqnarray}
From the matching calculation we determined~\mbox{$\coeff{c}{KK}{3}\simeq0$};
consequently, the (I=0) \mbox{$p$-wave} scattering amplitude and phase shift
vanish to the order we are working\@.
Including the \mbox{$\xi$-scaling} of the coefficients, the leading
contribution to~\A{NR}{(1)} comes from the \mbox{order-$Q^0$} tree diagram 
and counts as order~$\xi$\@.
The next terms in the phenomenological expansion in~$\xi$, of
equation~(\ref{eq:KK-A1}), are the \mbox{order-$Q$} kaon-bubble diagram 
and the \mbox{order-$Q^2$} tree diagram\@.
To work consistently at this order would require extending the matching of
the coefficient in the leading-order result, \coeff{a}{KK}{}, by two orders
in~$(M/\Lambda)^2$\@.

\begin{figure}[tb]
\hspace*{\fill}
\begin{picture}(190,130)%
    \put(10,25){\epsfxsize=150pt\epsfbox{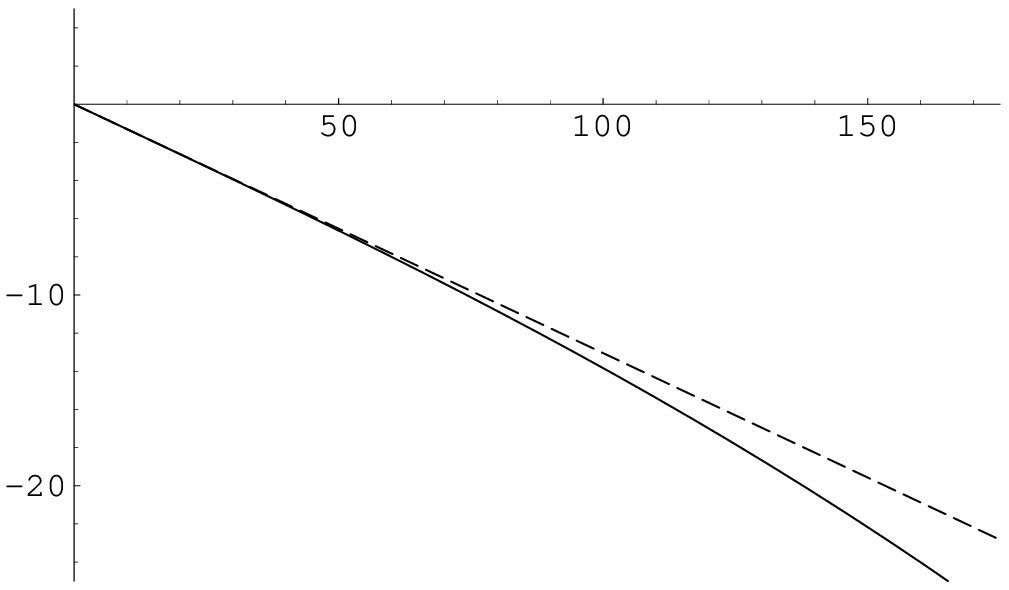}}%
    \put(0,120){\makebox(0,0)[lt]{(a)}}%
    \put(0,50){\rotatebox{90}{\scalebox{.75}{$\delta_0(k)$}}}%
    \put(90,105){\scalebox{.75}{$k\mbox{~MeV}^{-1}$}}%
\end{picture}
\hspace*{\fill}
\begin{picture}(190,130)%
    \put(10,20){\epsfxsize=150pt\epsfbox{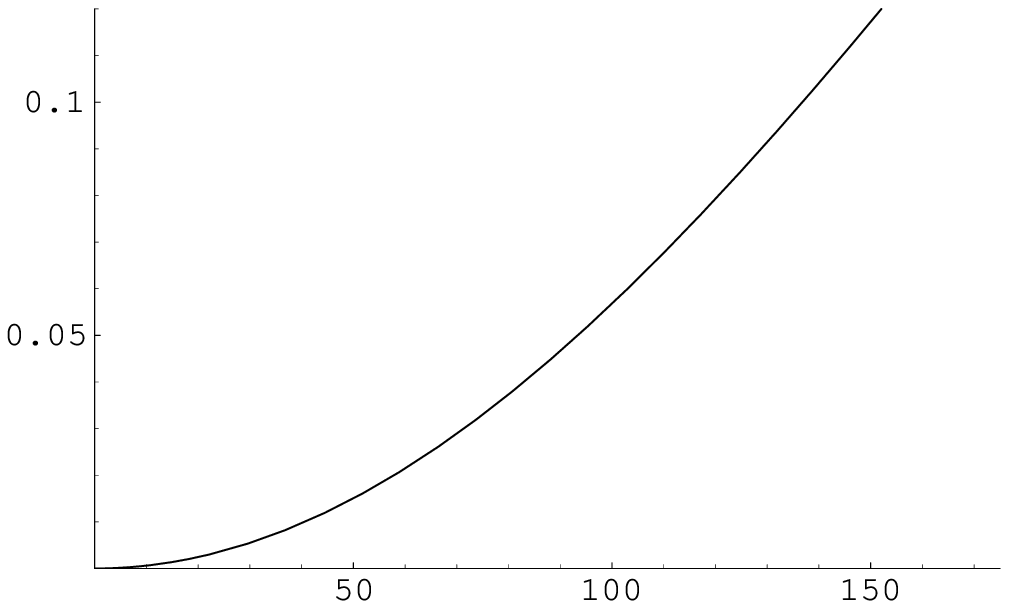}}%
    \put(0,120){\makebox(0,0)[lt]{(b)}}%
    \put(0,45){\rotatebox{90}{\scalebox{.75}{%
        $\Delta[\delta_0]/|\delta_0|$}}}%
    \put(90,5){\scalebox{.75}{$k\mbox{~MeV}^{-1}$}}%
\end{picture}
\hspace*{\fill}
\caption{(a)~Leading-order calculations of the $KK$ $s$-wave (I=1) scattering
  phase shift~$\delta_0(k)$ in degrees as a function of center-of-mass
  momentum~$k$; the dashed line is heavy kaon/eta theory, the solid line
  is~\su{3}~\cpt\@. (b)~Relative error between the results
  \mbox{$|\delta_0^{NR}-\delta_0^\cpt|/|\delta_0^\cpt|$}\@.
  \drafttext{\protect\\*
    \mbox{\bf[fig:KKshift]}}}
\label{fig:KKshift}
\end{figure}

For comparison, the result for the scattering amplitude in~\cpt\ is
\marginal{[eq:A1-cpt]}
\begin{equation}
\label{eq:A1-cpt}
    i\A{\cpt}{(1)} = \frac{-1}{3F_0^2} \left[\MS
        8 M_K^2 + 12 k^2 - 2(m_s+\hat{m})B_0 \right] .
\end{equation}
The leading-order results for the \mbox{$s$-wave} phase shift in both the
heavy kaon/eta theory and \su{3}~\cpt\ are presented together in
Figure~\ref{fig:KKshift}\@.
The \mbox{$s$-wave} $KK$~scattering length determined from the phase
shift plotted in the figure is
\marginal{[eq:scatt-len]}
\begin{equation}
\label{eq:scatt-len}
    a^{(1)} = 0.45 \times 10^{-13} \mbox{~cm},
\end{equation}
which is consistent with a repulsive \mbox{$K$-$K$~inter}action potential\@.
In both cases the leading corrections are suppressed
by~\mbox{${\sim}(M/\Lambda)^2\sim0.3$}\@.
In light of the fact that the heavy kaon/eta theory is determined directly
from the scattering amplitudes of~\cpt, the agreement of the two expansions
to within the expected~30\% corrections is not surprising\@.

The calculation of the \mbox{$s$-wave} phase shift in both \cpt\ and the
heavy kaon/eta theory has illustrated a fundamental point relevant to the
utility of the matching calculation\@.
The motivation for developing the heavy kaon/eta theory was to achieve a
better expansion in the threshold regime by virtue of the better ratio of
scales \mbox{$(Q/M)^2\sim.1$} versus \mbox{$(M/\Lambda)^2\sim.3$}\@.
In effect, the price for the improved convergence was an increase in the
number of low-energy constants, i.e.,\ coefficients in the Lagrangian, to 
be determined\@.
The initial proposal was made to determine the unknowns of the low-energy
theory by performing a matching calculation onto a theory with fewer unknown
parameters, \su{3}~\cpt\@.
However, by matching onto \cpt\ we guarantee that the heavy kaon/eta theory
is limited to converging no better than the theory to which it is matched\@.
Indeed, we find that the matching calculation explicitly reintroduces the 
mass-scale ratio~$(M/\Lambda)^2$ that we attempted to circumvent\@.
The matching calculation appears to be both the savior and the Achilles' heel
of the heavy kaon/eta effective theory\@.

For a wide variety of processes, including $KK$ scattering, the heavy
kaon/eta effective theory does not provide a computational advantage over
\cpt, and the additional work required to carry out the matching calculation
cannot be justified\@.
However, the outlook for the heavy kaon/eta theory or for the use of 
matching calculations is not entirely bleak\@.
The approach is useful is two senses that were suggested in
section~\ref{sec:prospects}\@.
The first is that independent of the matching calculation, when data are
available for a direct empirical determination of the low-energy constants,
then taking advantage of the better rate of convergence in the
non-relativistic theory will be possible\@.
For a short-term perspective, the available data on
$\pi{}K$~scattering~\cite{J-N:piKdata,K-S:piKdata} may be useful in this
way\@.

The second way in which the non-relativistic effective theory may be useful
is calculating quantities for which the relativistic formalism of~\cpt\ 
is awkward\@.
An example where a non-relativistic treatment has proven beneficial is
calculating the properties of bound-state
systems~\cite{C-L:nrqed,L-M:bound,K-S-W:deuteron,AGGR:bound}\@.
In applications of this sort, reviving the matching calculation approach has
merit because the motive for using the non-relativistic effective theory is
not solely to improve the convergence of the expansion\@.
Since bound state formation always involves summing an infinite number
of Feynman diagrams, the power counting considered here cannot be
directly applicable for these problems\@.
Two cases should be considered regarding bound states in the heavy
kaon/eta effective theory\@.
First, the effective field theory cannot be applied to deeply bound
systems because they lie outside the range of the momentum
expansion\@.
Second, for shallow bound states, the power counting scheme needs to
reflect the fine-tuning which is implicit in the associated large
scattering length\@.
In systems with a shallow bound state, the matching calculation will
not provide an estimate of coefficients of the terms responsible for
binding the system, but may permit the determination of coefficients
which contribute perturbatively to properties of the bound state\@.
The interesting possibilities for strong-interaction bound states, $K\kbar$
and $\kbar{}N$, both suffer kaon annihilation and are not suitable
candidates\@.
Another possibility is to calculate strong-interaction perturbations to a
Coulomb bound state with a kaon, for instance $\piM\kP$ or $\kM\Sigma^+$, 
or kaon electromagnetic form-factors via $e^-\kP$\@.

In conclusion, we have presented the foundations and laid out the principles
for an effective field theory to describe the interactions of pions with
non-relativistic kaons and eta mesons\@.
The effective Lagrangians for few-body sectors were constructed explicitly 
for the first several orders in a chiral expansion\@.
Much work remains to be done both in applying the theory to systems of
interest and in generalizing the theory to include electromagnetic
interactions, couplings to baryons, and the study of isospin violation\@.


%% file: codes.tex
\chapter[Symbolic Expansion of \L{\cpt}{}]%
    {\\Symbolic Expansion of \L{\cpt}{}}%
\label{ch:codes}%
\marginal{[ch:codes]}%
%
This appendix presents \mbox{\textit{Mathematica}}\footnotemark\ 
routines~\cite{Wolfram} for the symbolic expansion of the \su{3}~chiral
lagrangian in terms of the embedded chiral~\su{2} fields in the sincere hope
that they will prove useful to others\@.
\footnotetext{\mbox{\textit{Mathematica}} is a registered trademark of
  Wolfram Research, Inc.}
The source code is divided into two files, \verb@lagrangian.math@ on
page~\pageref{pgA:lagrangian.math} which specifies the sequence of steps in
expanding and saving the desired parts of the lagrangian and
\verb@definitions.math@ on
pages~\mbox{\pageref{pgA:definitions.math}--\pageref{pgB:definitions.math}}
which defines the functions called to perform the expansion and simplify the
results\@.
Currently, the chiral Lagrangian~\L{2}{} in equation~(\ref{eq:Lcpt2}) is
expanded out to terms including six boson fields; \L{4}{}~in
equation~(\ref{eq:Lcpt4}) is expanded to four boson fields\@.
The source code can be applied to higher-order parts of the chiral Lagrangian
or to terms involving more bosons by adding transformation rules to the
\verb@trace@ routine in \verb@definitions.math@ at line 42\@.

The routines are of very limited use without a brief description of the
symbols appearing in the input and output files\@.
In addition to the obvious parameters \verb@F@, \verb@B@, and
\verb@L1@--\verb@L8@ defined as in section~\ref{sec:cpt}, the quark masses
are included as
\begin{displaymath}
    \verb@mhat@ \rightarrow \hat{m} = \tfrac{1}{2} (m_u + m_d), \mathsp
    \verb@mdiff@ \rightarrow (m_u - m_d), \mathsp
    \verb@ms@ \rightarrow m_s.
\end{displaymath}
The correspondences for the boson fields are
\begin{displaymath}
    \verb@pion@ \rightarrow \pi\ss{a}{} \tau\ss{a}{}, \mathsp
    \verb@kaon@ \rightarrow K, \mathsp
    \verb@kbar@ \rightarrow \kbar, \mathsp
    \verb@eta@ \rightarrow \eta.
\end{displaymath}
Each occurrence of a boson field is multiplied by a tag, \verb@boson@,
which simplifies expanding the lagrangian and separating the
lagrangian into parts by the number of bosons appearing\@.
Also, two matrices are given symbolic names in the input files; \verb@one@
refers to the $2\times2$~identity matrix and \verb@tau3@ refers to the Pauli
matrix~\mbox{$\tau\ss{3}{}=\sigma_z$}, which appears in isospin violating
terms\@.
Finally, the following notation appears in the output:
\begin{itemize}
\item derivatives, 
    $\verb@d[A,x]@ \rightarrow \lpartial\ss{x}{}A$, \\
    $\verb@d[pion,mu]@ \rightarrow (\dmu\pi\ss{a}{})\tau\ss{a}{}$, \mathsp
    $\verb@d[kaon,nu]@ \rightarrow {\dnu}K$ 
\item isovector products, 
    $\verb@v[A,B]@ \rightarrow \tfrac{1}{2}\tr{AB}$, \\
    $\verb@v[pion,pion]@ \rightarrow 
        \tfrac{1}{2} \pi\ss{a}{}\pi\ss{b}{} \tr{\tau\ss{a}{}\tau\ss{b}{}} 
    = \pi\cdot\pi$, \mathsp
    $\verb@v[d[pion,mu],tau3]@ \rightarrow \dmu\piZ$
\item isospinor contractions,
    $\verb@m[A,B,C]@ \rightarrow 
        A\ss{\alpha}{}B\ss{\alpha\beta}{}C\ss{\beta}{}$, \\
    $\verb@m[kbar,d[kaon,mu]]@ \rightarrow \kbar{\dmu}K$, \mathsp
    $\verb@m[kbar,pion,kaon]@ \rightarrow \pi\cdot(\kbar{\tau}K)$
\item commutators,
    $\verb@c[A,B]@ \rightarrow [A,B]$, \\
    $\verb@v[c[pion,d[pion,mu]],tau3]@ \rightarrow 
        \pi\ss{a}{} (\dmu\pi\ss{b}{}) (2i\epsilon\ss{ab3}{})
    = 2i\, (\pi\times\dmu\pi)\cdot\tau\ss{3}{}$, \\
    $\verb@m[kbar,c[d[pion,mu],d[pion,nu]],kaon]@ \rightarrow 
        2i\, (\dmu\pi\times\dnu\pi)\cdot(\kbar{\tau}K)$
\end{itemize}

The output is written to the file \verb@lagrangian.save@ and breaks the
lagrangian into pieces of manageable size\@.
The different parts of the lagrangian are named~\verb@L@\textit{XyZ}, where
\textit{X}~is the chiral order as in~\L{2}{} versus~\L{4}{}, \textit{Z}~is
the number of boson fields appearing in the terms, and \textit{y}~is a letter
which indicates the number of factors of~\Mq\ occurring in place of
derivatives\@.
A few select lines of \verb@lagrangian.save@ are given below to illustrate\@.
\input{Lcpt-out}
\noindent The translation of the lines above is
\begin{eqnarray*}
    \L{2a2}{}
    & = & \tfrac{1}{2}\, \dmu\pi \cdot \dum\pi
        + \dmu\kbar {\dum}K + \tfrac{1}{2}\, \dmu\eta \dum\eta
, \\*
    \L{2b2}{}
    & = & - B_0\hat{m}\, \pi \cdot \pi - B_0 (m_s + \hat{m})\, {\kbar}K
        - \tfrac{1}{3} B_0 (2m_s + \hat{m})\, \eta^2 
.
\end{eqnarray*}

\clearpage%
\label{pgA:lagrangian.math}%
\input{Lcpt-in}%
\label{pgB:lagrangian.math}%

\clearpage%
\label{pgA:definitions.math}%
\input{defines}%
\label{pgB:definitions.math}%


%% file: Lcpt-out.tex
\begin{spacing}{1}
\footnotesize
\begin{verbatim}
L2a2 = d[eta, mu]^2/2 + m[d[kbar, mu], d[kaon, mu]] +
     v[d[pion, mu], d[pion, mu]]/2

L2b2 = eta^2*(-(B*mhat)/3 - (2*B*ms)/3) + (-(B*mhat) - B*ms)*m[kbar, kaon] -
     B*mhat*v[pion, pion]
\end{verbatim}
\end{spacing}

%% file: Lcpt-in.tex
\begin{spacing}{1}
\footnotesize
\begin{verbatim}
(******************** lagrangian.math ************************)
(* symbolically expands the SU(3) ChPT lagrangian in terms of the      *)
(*   embedded SU(2) fields, out to 6 bosons for L2 and 4 bosons for L4 *)
<< "definitions.math";  timetag["Session started:"];

mdiff=0;  (* drops all isospin violating terms, remove line to keep them *)
X=Xd=2*B*{{mhat*one+mdiff*tau3/2,0},{0,ms}};
P=boson*Sqrt[2]*{{pion/Sqrt[2]+eta*one/Sqrt[6],kaon},{kbar,-2*eta/Sqrt[6]}};
U=matrixseries[Exp[(I/F)*#]&,P,{boson,6}];    Um=derivative[U,mu];
Ud=matrixseries[Exp[(-I/F)*#]&,P,{boson,6}];  Udm=derivative[Ud,mu];
timetag["Defined the basics:"];

(* Calculate L2a *)
UmUdm=product[dot,Um,Udm,{boson,6}];  L2a=trace[UmUdm]*(F/2)^2;
{L2a2,L2a4,L2a6}=Map[reduce[Coefficient[L2a,boson,#]]&,{2,4,6}];
L2aN=Expand[reduce[L2a/.boson->1]-(L2a2+L2a4+L2a6)];
Save["lagrangian.save",L2a2,L2a4,L2a6,L2aN];
{Um,Udm,UmUdm}=Map[truncate[#,{boson,4}]&,{Um,Udm,UmUdm}];
Clear[L2a,L2a2,L2a4,L2a6,L2aN];  timetag["Calculated L2a:"];
(* Calculate L2b *)
XUd=dot[X,Ud];  UXd=dot[U,Xd];  L2b=trace[XUd+UXd]*(F/2)^2;
{L2b0,L2b2,L2b4,L2b6}=Map[reduce[Coefficient[L2b,boson,#]]&,{0,2,4,6}];
L2bN=Expand[reduce[L2b/.boson->1]-(L2b0+L2b2+L2b4+L2b6)];
Save["lagrangian.save",L2b0,L2b2,L2b4,L2b6,L2bN];
{XUd,UXd}=Map[truncate[#,{boson,4}]&,{XUd,UXd}];
Clear[U,Ud,L2b,L2b0,L2b2,L2b4,L2b6,L2bN];  timetag["Calculated L2b:"];
(* Calculate L4a *)   TUmUdm=trace[UmUdm];
TUmUdn=trace[product[dot,Um,(Udm/.mu->nu),{boson,4}]];
L4a=(L1*product[Times,TUmUdm,(TUmUdm/.mu->nu),{boson,4}]
    +L2*product[Times,TUmUdn,TUmUdn,{boson,4}]
    +L3*trace[product[dot,UmUdm,(UmUdm/.mu->nu),{boson,4}]] );
L4a4=reduce[Coefficient[L4a,boson,4]]; L4aN=Expand[reduce[L4a/.boson->1]-L4a4];
Save["lagrangian.save",L4a4,L4aN];
Clear[Um,Udm,TUmUdn,L4a,L4a4,L4aN];  timetag["Calculated L4a:"];
(* Calculate L4b *)   TpXU=trace[XUd+UXd];
L4b=(L4*product[Times,TUmUdm,TpXU,{boson,4}]
    +L5*trace[product[dot,UmUdm,(XUd+UXd),{boson,4}]] );
{L4b2,L4b4}=Map[reduce[Coefficient[L4b,boson,#]]&,{2,4}];
L4bN=Expand[reduce[L4b/.boson->1]-(L4b2+L4b4)];
Save["lagrangian.save",L4b2,L4b4,L4bN];
Clear[UmUdm,TUmUdm,L4b,L4b2,L4b4,L4bN];  timetag["Calculated L4b:"];
(* Calculate L4c *)   TmXU=trace[XUd-UXd];
TpXUXU=trace[product[dot,XUd,XUd,{boson,4}]+product[dot,UXd,UXd,{boson,4}] ];
L4c=(L6*product[Times,TpXU,TpXU,{boson,4}]
    +L7*product[Times,TmXU,TmXU,{boson,4}]+L8*TpXUXU );
{L4c0,L4c2,L4c4} = Map[reduce[Coefficient[L4c,boson,#]]&,{0,2,4}];
L4cN = Expand[reduce[L4c/.boson->1]-(L4c0+L4c2+L4c4)];
Save["lagrangian.save",L4c0,L4c2,L4c4,L4cN];  timetag["Calculated L4c:"];
(********************   end of file   ************************)
\end{verbatim}
\end{spacing}

%% file: defines.tex
\begin{spacing}{1}
\footnotesize
\begin{verbatim}
(************************ definitions.math ************************)
(* defines functions called by "lagrangian.math" *)

(* time-stamp printing function *)
timetag[msg_]:=(Print[];  timenew=SessionTime[];
    If[untimed, (timeorigin=timeold=timenew; untimed=False); ];
    Print[ StringTake[msg<>spaces,20],PaddedForm[(timenew-timeold),{9,3}],
        PaddedForm[(timenew-timeorigin),{9,3}],PaddedForm[TimeUsed[],{9,3}] ];
    timeold=timenew );
untimed=True;  spaces="                      ";

(* catagories of symbols which are used in simplifying rules *)
fields={pion,kaon,kbar,eta};  ident={{one,0},{0,1}};
structs={pion,kaon,kbar,one,tau3};  bras={kbar};  kets={kaon};

(* functions called directly from "lagrangian.math" *)
truncate[exp_,xpn_]:=Expand[Normal[Series[exp,{First[xpn],0,Last[xpn]}]]];
derivative[exp_,var_]:=(D[exp,var,NonConstants->fields] 
    //. Dot[___,0,___]->0 //. Literal[D][f_,v_,___]->d[f,v] );
dot[x_,y_]:=(distribute[{
    {Dot[x[[1,1]],y[[1,1]]]+Dot[x[[1,2]],y[[2,1]]],
        Dot[x[[1,1]],y[[1,2]]]+x[[1,2]]*y[[2,2]]},
    {Dot[x[[2,1]],y[[1,1]]]+x[[2,2]]*y[[2,1]],
        Dot[x[[2,1]],y[[1,2]]]+x[[2,2]]*y[[2,2]]} }]);
product[mult_,x_,y_,xpn_]:=(Module[{i,j,coeff,result=0},(
    coeff=Coefficient[#1,First[xpn],#2]&;
    For[i=0,i<=Last[xpn],i++, For[j=0,j<=i,j++,
        result+=mult[coeff[x,j],coeff[y,(i-j)]]*First[xpn]^i; ];];
    Return[Expand[result]] );]);
matrixseries[func_,matr_,xpn_]:=(Module[{x,i=0,m={truncate[matr,xpn]}},(
    While[Union[Flatten[Last[m]]]!={0},
        (AppendTo[m,product[dot,Last[m],First[m],xpn]]; i++); ];
    Return[ReleaseHold[Expand[ truncate[func[x],{x,0,i}]
        +Limit[func[x],x->0]*(Hold[ident]-1) ] /. x^n_.:>Hold[m[[n]]] ]]);]);
trace[exp_]:=(Expand[ExpandAll[tr[exp[[1,1]]]+tr[exp[[2,2]]]] 
    //.{tr[x_?(FreeQ[#,any[structs]]&)]->x,
        tr[x:(_Plus|_Times)]:>Map[tr,x], tr[x_^n_Integer]->tr[x]^n } 
    /. tr[x_Dot]:>isoscalars[Apply[List,x]]
    //.{tr[one]->2, tr[tau3|pion|d[pion,_]]->0, tr[x_,y_]->2*v[x,y],
        tr[x_,a__,x_]->v[x,x]*tr[a], tr[a___,x_,x_,b___]->v[x,x]*tr[a,b],
        m[a__,x_,x_,b__]->v[x,x]*m[a,b] }
    /. {tr[x_,y_,z_]->v[c[x,y],z], tr[x_,y_,x_,y_,z_]->2*v[x,y]*v[c[x,y],z],
        tr[x_,y_,z_,w_]->2*v[x,y]*v[z,w]-2*v[x,z]*v[y,w]+2*v[x,w]*v[y,z],
        m[a_,x_,y_,b_]->v[x,y]*m[a,b]+m[a,c[x,y],b]/2,
        m[a_,x_,y_,z_,b_]->v[c[x,y],z]*m[a,b]/2
            +v[x,y]*m[a,z,b]-v[x,z]*m[a,y,b]+v[y,z]*m[a,x,b],
        m[a_,x_,y_,x_,y_,b_]->v[x,y]*m[a,c[x,y],b]
            +2*v[x,y]*v[x,y]*m[a,b]-v[x,x]*v[y,y]*m[a,b] }
    /. {v[tau3,tau3]->1, m[_,c[x_,x_],_]->0,
        v[c[x_,y_],z_]:>0/;((x===y)||(x===z)||(y===z)) }]);
reduce[exp_]:=(Module[{terms,forms,freeq=FreeQ[#,any[fields]]&},(
    terms=Map[commonform,Flatten[{
        Replace[ExpandAll[exp],x_Plus:>Apply[List,x]] }]];
    forms=Union[DeleteCases[terms,_?freeq,{1}] //. _?freeq*x_->x];
    Return[Collect[Apply[Plus,terms],forms]] );]);

(* functions used internally to simplify and reduce expressions *)
any[list_]:=Apply[Alternatives,list];
indexed[form_]:=(MatchQ[form,( any[structs]|d[any[structs],__]
    |Dot[_?(FreeQ[#,any[bras]]&),__]|Dot[__,_?(FreeQ[#,any[kets]]&)] )]); 
distribute[exp_]:=(ExpandAll[exp] //. Dot[___,0,___]->0 
    //.{Dot[w___,x_Plus,y___]:>Distribute[Dot[w,x,y],Plus],
        Dot[w___,a_*(x_?indexed),y___]->a*Dot[w,x,y],
        Dot[w___,one,y___]->Dot[w,y] });
isoscalars[list_]:=(Module[{olist,break=Map[First,Position[list,any[bras]]]},(
    If[break=={},Return[Apply[tr,list]];];
    olist=RotateLeft[list,First[break]-1];
    break=Append[break-First[break]+1,Length[olist]+1];
    Return[Product[ Apply[m,Take[olist,{break[[i]],break[[i+1]]-1}]],
        {i,1,Length[break]-1} ]]);]);
commonform[exp_]:=(Module[{vars,swaps,array},(
    array={1,exp} //. List[a_,b_?(FreeQ[#,any[fields]]&)*x_]->{a*b,x}
    //.{{a__,x_*y_,z___}->{a,x,y,z}, {a__,x_^n_Integer,y___}->{a,x^(n-1),x,y}};
    array=Apply[Union[Flatten[Outer[Times,##]]]&,Map[Union, Map[List,array]
    /. {{m[a_,c[x_,y_],b_]}->{m[a,c[x,y],b],-m[a,c[y,x],b]},
        {v[c[x_,y_],z_]}->{v[c[x,y],z],-v[c[y,x],z],
            v[c[y,z],x],-v[c[z,y],x],v[c[z,x],y],-v[c[x,z],y]},
        {v[x_,y_]}->{v[x,y],v[y,x]} }
    /. {m[a_,c[tau3,x_],b_]->-m[a,c[x,tau3],b], v[tau3,x_]->v[x,tau3],
        (v[c[tau3,x_],y_]|v[c[y_,tau3],x_])->v[c[x,y],tau3] }]];
    vars=Union[Cases[exp,(d[_,x_]->x),Infinity]];
    swaps=Map[Thread[Rule[vars,#]]&,Permutations[vars],{1}];
    Return[First[Union[Flatten[Map[(#/.swaps)&,array]]]]]); ]);
(************************    end of file    ************************)
\end{verbatim}
\end{spacing}

%% file: recur.tex
\chapter[Recursion Relations for \redefname]%
    {\\Recursion Relations for \redefname}%
\label{ch:recur}%
\marginal{[ch:recur]}%
%
We start with the free-field Lagrangian for a (complex) heavy scalar field,
\begin{displaymath}
    \L{\phi}{}
    = \phi^\dag \left[ \vpf{\MS}{\MS} iv{\cdot}\partial
        + \frac{\nabla_\perp^2 + (iv{\cdot}\partial)^2}{2m} \right] \phi
\end{displaymath}
where~$v^\mu$ is the scalar-field velocity
and~\mbox{$\dmu\dum=(v{\cdot}\partial)^2-\nabla_\perp^2$}\@.
We seek a redefinition of the scalar
field~\mbox{$\phi=\mathcal{F}[\tilde{\phi}]$} which eliminates secondary
time-like derivatives of the heavy scalar in favor of higher powers of the
space-like derivative~$\nabla_\perp^2$,
\begin{displaymath}
    \L{\tilde{\phi}}{}
    = \tilde{\phi}^\dag \left[ \vpf{\MS}{\MS} iv{\cdot}\partial
        + m \sum_{j=1}^\infty g_j \left( \frac{\nabla_\perp^2}{m^2} \right)^j 
        \right] \tilde{\phi}.
\end{displaymath}
The general field redefinition we consider here is
\begin{displaymath}
    \mathcal{F}[\tilde{\phi}]
    = \sum_{j=0}^\infty \sum_{k=0}^\infty 
            f^j_k \left( \frac{iv{\cdot}\partial}{m} \right)^j 
            \left( \frac{\nabla_\perp^2}{m^2} \right)^k \tilde{\phi},
\end{displaymath}
and constraints on the form of~\L{\tilde{\phi}}{} give recursion relations
for the coefficients~$f^j_k$\@.
Using integration by parts, we find the recursion relations are
\begin{eqnarray*}
\lefteqn{
    (f^0_0)^2 = 1, \mathsp
    f^1_0 = -\tfrac{1}{4} f^0_0, \mathsp
    f^0_1 = -\tfrac{1}{2} f^1_0, } \hspace*{.4in}
\\
    - f^0_0 f^{j+1}_0 
    & = & \tfrac{1}{4} f^0_0 f^j_0 
        + \tfrac{1}{4} \sum_{p=0}^{j-1} \left(
            f^p_0 f^{j-p}_0  + 2 f^{p+1}_0 f^{j-p}_0 \right)
, \\
    - f^0_0 f^j_{k+1}
    & = & - \tfrac{1}{2} f^j_0 f^0_{k+1}
        + \tfrac{1}{4} \sum_{p=0}^{j-1} \sum_{q=0}^k \left(
            f^p_q f^{j-p-1}_{k-q+1} + 2 f^{p+1}_q f^{j-p-1}_{k-q+1} 
                + f^{p+1}_q f^{j-p}_{k-q} \right)
\\* & & \mbox{}
        + \tfrac{1}{4} \sum_{p=0}^{j-1} \left(
            f^p_{k+1} f^{j-p-1}_0 + 2 f^p_{k+1} f^{j-p}_0 \right)
        + \tfrac{1}{2} \sum_{q=0}^k \left(
            f^0_q f^{j+1}_{k-q} + f^0_{q+1} f^j_{k-q} \right)
, \\
    - f^0_0 f^0_{k+1} 
    & = & \tfrac{1}{4} (f^0_0 f^1_k + f^1_0 f^0_k) 
        + \tfrac{1}{4} \sum_{q=0}^{k-1} \left(
            f^0_q f^1_{k-q} + f^1_q f^0_{k-q} + 2 f^0_{q+1} f^0_{k-q} \right)
, \\
    g_{k+1} 
    & = & \tfrac{1}{2} \sum_{q=0}^k f^0_q f^0_{k-q}
,
\end{eqnarray*}
and determine the coefficients in the
sequence~\mbox{$f^0_0,f^1_0,f^0_1,f^2_0,f^1_1,f^0_2,f^3_0,\ldots$}

Solving for~\mbox{$\mathcal{F}[\tilde{\phi}]$} is simplified by recognizing
that the coefficients~$g_j$ must reproduce the kinetic energy and well-known
relativistic corrections;
\mbox{$\L{\tilde{\phi}}{}%
  =\tilde{\phi}^\dag(iv{\cdot}\partial-\hat{K})\tilde{\phi}$} where
\begin{eqnarray*}
    \hat{K}
    & = & \sqrt{m^2-\nabla_\perp^2\,} - m
\\* 
    & = & - \frac{\nabla_\perp^2}{2m} - \frac{\nabla_\perp^4}{8m^3}
        - \frac{\nabla_\perp^6}{16m^5} - \cdots
\end{eqnarray*}
We postulate that the field redefinition is a function of one of the
combinations~\mbox{($iv{\cdot}\partial\pm\nabla_\perp^2$)}
or~\mbox{($iv{\cdot}\partial\pm\hat{K}$)} only, then verify that only the
form~\mbox{($iv{\cdot}\partial+\hat{K}$)} satisfies the recursion
relations\@.
We construct the solution through trial and error, guided by comparing
expansions of the trial function with the recursion relations above,
and find
\begin{displaymath}
    \mathcal{F}[\tilde{\phi}]
    = \pm \left[ 1 + \frac{ iv{\cdot}\partial + \hat{K} }{2m} 
        \right]^{-\frac{1}{2}} \tilde{\phi}.
\end{displaymath}
Inspired by hindsight we present a formal, and much shorter, derivation of
this result in Chapter~\ref{ch:heavyK}\@.